\documentclass[fleqn,usenatbib]{mnras}

\usepackage{mathptmx}
\usepackage[T1]{fontenc}
\usepackage{ae,aecompl}

\usepackage{graphicx}	% Including figure files
\usepackage{savesym}
\usepackage{amsmath}
\savesymbol{iint}
\savesymbol{iiint}
\usepackage{txfonts}

\restoresymbol{TXF}{iint}
\restoresymbol{TXF} {iiint}

\def\la{\raise.5ex\hbox{$<$}\kern-.8em\lower 1mm\hbox{$\sim$}}
\def\ma{\raise.5ex\hbox{$>$}\kern-.8em\lower 1mm\hbox{$\sim$}}

\def\kms{$\rm km\, s^{-1}$~}
\def\cm3{cm$^{-3}$~}

\def\Vs{$V_{\rm s}$~}
\def\n0{$n_{\rm 0}$}
\def\B0{$B_{\rm 0}$}
\def\ne{$n_{\rm e}$~}
\def\Ne{$n_{e}$~}
\def\Te{$T_{\rm e}$~}

\def\ergcs{$\rm erg\, cm^{-2}\, s^{-1}$}

\def\L12{$L_{12\mu m}$~}
\def\F12{$F_{12\mu m}$~}

\def\Hb{H${\beta}$~}
\def\Hbc{H${\beta}_c$~}
\def\Ha{H${\alpha}$~}

\def\dx{$\Delta$X~}

\title[Ionised gas outflow in the Circinus galaxy]{The ionised gas outflow in the Circinus galaxy: kinematics and physical conditions}

\author[Fonseca-Faria, M. A. et al.]{
M. A. Fonseca-Faria$^{1}$,\thanks{E-mail: marcosfonsecafaria@gmail.com}
A. Rodr\'iguez-Ardila$^{1,2}$,
M. Contini$^{3}$, V. Reynaldi$^{4}$
\\
$^{1}$ INPE - Instituto Nacional de Pesquisas Espaciais, Av. dos Astronautas, CEP 12227-010, S\~ao Jos\'e dos Campos - SP, Brazil\\
$^{2}$ Laborat\'orio Nacional de Astrof\'{\i}sica - Rua dos Estados Unidos 154, Bairro das Na\c{c}\~oes . CEP 37504-364, Itajub\'a, MG, Brazil\\
$^{3}$School of Physics and Astronomy, Tel Aviv University, Tel Aviv 69978, Israel.\\
$^{4}$Facultad de Ciencias Astron\'omicas y Geof{\'{\i}}sicas - Universidad Nacional de La Plata, La Plata, Argentina
}

\date{Accepted XXX. Received YYY; in original form ZZZ}

\pubyear{2015}

\begin{document}
\label{firstpage}
\pagerange{\pageref{firstpage}--\pageref{lastpage}}
\maketitle

\begin{abstract}
We employ MUSE/VLT data to study the ionised and highly ionised gas phases of the feedback in Circinus, 
the closest Seyfert~2 galaxy to us. The analysis of the nebular emission allowed us to detect a remarkable high-ionisation gas outflow  beyond the galaxy plane  
traced by the coronal lines [Fe\,{\sc vii}]~$\lambda$6089 and [Fe\,{\sc x}]~$\lambda$6374, extending up to 700~pc and 350~pc NW from the nucleus, respectively. This is the first time that the [Fe\,{\sc x}] emission is observed at such distances from the central engine in an AGN. The gas kinematics reveals expanding gas shells with velocities of a few hundred ~km\,s$^{-1}$, 
spatially coincident with prominent hard X-ray emission detected by Chandra. Density and temperature sensitive line 
ratios show that the extended high-ionisation gas is characterized by a temperature  reaching 25000~K and an
electron density $> 10^2$ \cm3. We found that local gas excitation by shocks produced by the passage of a
radio jet leads to the spectacular
high-ionisation emission in this object. This hypothesis is fully supported by photoionisation models that accounts for the combined effects of the central engine and shocks. They reproduce the observed emission line spectrum at different locations inside and outside of the NW ionisation cone. The energetic outflow produced by the radio jet is  spatially located close to an  extended molecular outflow recently reported 
using ALMA which suggests 
that they both represent different phases of the same feedback process acting on the AGN.  

\end{abstract}


%%%%%%%%%%%%%%%%%%%%%%%%%%%%%%%%%%%%%%%%%%%%%%%%%%

%%%%%%%%%%%%%%%%% BODY OF PAPER %%%%%%%%%%%%%%%

\section{Introduction}

Outflows  from active galactic nuclei (AGN) expel part of the gas surrounding the accretion disc transporting energy and 
matter to different regions of the host galaxy. Starting from the broad line region (BLR), they cross the narrow line 
region (NLR) and the galaxy disc, reaching the outermost regions at tens of kpc from the galaxy centre. 
The energy carried out by the outflowing gas is transmitted to different environments and distances throughout the galaxy,
leaving its signature into the different medium phases: ionised, molecular and neutral \citep{wada_2018}. 
Classic examples of  AGN where outflows have been detected in the three phases are the Seyfert~2 galaxies 
IC\,5063 \citep{dasyra_2015, dasyra_2016, oosterloo_2017} and NGC\,5643 \citep{garcia+20}.
Outflows  from AGN are directly related to the feedback modes currently identified in these sources. They can be 
of radiative or kinematic origin. Radiative feedback (or quasar feedback mode) results from winds generated by radiation 
pressure very close to the optically thick accretion disc. They are usually associated to luminous AGN   
approaching the Eddington limit \citep{valentini_2020}, 
$L_{Edd} \cong 1.3 \times 10^{38} (M_{BH}) /M \odot$~erg\,s$^{-1}$ \citep{frank_2002}). These winds can leave the BRL 
region and propagate over to large distances  at galactic scales. 
Kinematic feedback (or jet feedback) is associated to the kinetic energy transported by 
the radio jet as in radio galaxies. In this case, the jets may extend over dozens 
of kpc away from the galaxy core. The mechanical contribution of the jet to the host galaxy is of little relevance 
\citep{morganti_2015} as it affects mostly the intergalactic medium. In contrast, radio-quiet objects show weaker jets, 
of sizes less than a few kpc. They  mechanically contribute to the  outflows within the host galaxy.

Recently, with the use of integral field unit (IFU) data from MUSE/VLT (Multi Unit Spectroscopic Explorer),  it was 
identified for the first time the highly ionised component of the outflowing gas in the Circinus galaxy. 
Earlier, this component was  only studied by  means of  the [\ion{O}{iii}] line data \citep{mingozzi_2019}.  
\citet{ardila_2020} reported the  discovery of an extended  high ionisation gas outflow  in Circinus by means of 
the [\ion{Fe}{vii}]~$\lambda6087$ line, detected up to a distance of 700 pc from the central source. 
To explain the structure 
of the  outflow, the authors propose the existence of an expanding shell of gas formed by the interaction of the  radio 
jet with the interstellar gas.

The choice of the Circinus galaxy for the present investigation is not by chance.  Circinus is the closest 
Seyfert~2, located at a distance of
$4.2 \pm 0.8 $~Mpc \citep{freeman_1977}, with a projected scale of 1'' = 20~pc. It is hosted by a spiral 
galaxy of SAb morphology and an inclination angle of approximately 65$^{\rm o}$. 
The   supermassive black hole lurking at the centre of the galaxy has a mass of $ 1.7 \pm 0.3 \times 10^{6} $ $ \textup {M}{\odot}$ \citep{greenhill_2003}. An outstanding cone-shaped outflow  extending to kpc distances  is easily visible in
the [\ion{O}{iii}]~$\lambda$5007 line \citep{muller_2006}. The global star formation rate has been measured 
between 3 and 8 M$\odot$~yr$^{-1}$ \citep{for_2012} and in the most central regions ($ <100 \text {~pc}$) 
the star formation rate is even smaller  \citep[0.1{} M$\odot$~yr$^{-1}$,][]{esquej_2014}.
The  observations  show  a molecular gas outflow from Circinus with velocities consistent with those
of the ionised gas outflow \citep{zs_2016}. However, no conclusive  evidence exists so far that both are produced 
by the same mechanism. There is also evidence of a radio jet with a PA (position angle) of $295^{\circ} \pm  5^{\circ} $ 
\citep{el_1998b}.
The  nuclear spectrum displays prominent high ionisation lines including [\ion{Fe}{vii}], [\ion{Fe}{x}] and 
[\ion{Fe}{xiv}] in the optical \citep{oliva_1999}, [S {\sc ix}], [\ion{Si}{vi}] and [\ion{Ca}{viii}] in the NIR 
and [\ion{Ne}{v}] in the MIR \citep{storchi_1999}.

In this work we carry out a detailed analysis of the physical conditions of the highly-ionised gas (HIG) component 
in Circinus, its kinematics and ionisation structure. Our aim is to explore scenarios that best explain the 
spectacular extension of that component, which reaches a projected linear extension of 700~pc from the AGN. 
In Sect.~2 we describe the observations, the
data treatment and the emission gas distribution and morphology. Sect.~3 deals with  the physical conditions of 
the gas  and in Sect.~4 we study the stellar and gas kinematics. The outflow properties are derived in Sect~5. 
The photoionisation models used to reproduce the emission line spectra detected at several locations of the galaxy 
are described in Sect~6. Concluding remarks  follow in Sect~7.

%##########################################################
%################## DADOS E METODOS #######################
%##########################################################

\section{Observations and data analysis}
\label{cap:metodos}

   \subsection{Data}
   
The data used in this work were obtained   from the Multi Unit Spectroscopic Explorer (MUSE), which is an integrated 
field spectrograph installed in the Yepun telescope at the Very Large Telescope (VLT),  Chile.
MUSE covers a field of view of 1'\, $\times$ 1' 
%arcmin  
and  detects spectra in the optical region with a 
projected spaxel size of 0.2" $\times$ 0.2".
%arcsec. 
The average seeing of the observations was $\sim$0.78''.

The data were collected on March 11, 2015 and are available on the ESO website already calibrated in flux and wavelength.
The final reduced data cube has 317 by 319 spaxels and a spectral resolution from 1750 to 4650 at the wavelengths of 3750
and 9300 ~\AA, respectively. In order to enhance the signal-to-noise ratio (SNR), in particular in regions of weak 
emission, a binning was performed so that each new spaxel has a final projected size of $\sim0.6" \times 0.6$". 
Most of the analysis carried out in this work makes use of this rebinned cube with a size of 105 by 106 spaxels, 
resulting in an approximately 10.000 spaxels. When the original cube is employed, it is explicitly mentioned in the text.
    
\subsection{Choice of the Circinus galaxy centre}

In order to determine the precise location of the AGN, we used as a proxy the flux maps obtained for the lines 
of [\ion{Fe}{x}]\,$\lambda$6375, [\ion{Fe}{xi}]\,$\lambda$7892 and [\ion{Ar}{xi}]\,$\lambda$6917 together with the 
continuum emission at 7000~\AA, integrated within a window of 10~\AA\ size. The choice of the above set of lines is 
 due to their very high ionisation potential (IP $>$ 200~eV) and by the assumption that they peak in the innermost 
region of the NLR, very close to the AGN. As each spaxel projects a square region of $\sim12$~pc of side length, 
it is very unlikely that these lines peak outside the spaxel where the AGN is presumably located.

\begin{figure}
\includegraphics[width=\columnwidth]{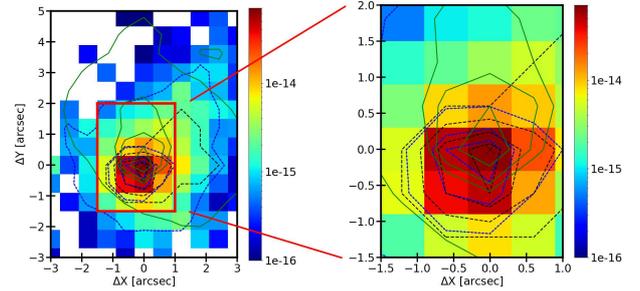}
\caption{Left: map of [Fe\,{\sc x}]~$\lambda$6375 flux (in  erg\,s$^{-1}$ cm$^{-2}$~Spaxel$^{-1}$) overlaid upon
the contours at 7000~\AA\ (green), [Fe\,{\sc xi}]~$\lambda$7892 (blue) and [Ar\,{\sc xi}]~$\lambda$6917 (black). 
Right: zoom to the innermost central region of Circinus. The symbol `X' represents the position of the AGN.}
\label{fig:centros}
\end{figure}

In Fig. \ref{fig:centros}, the flux distribution of [\ion{Fe}{x}] is shown, overlaid to the  contours of the 
continuum at 7000~\AA\ (green), [\ion{Fe}{xi}] (blue) and [\ion{Ar}{xi}] (black). It can be seen that they all peak 
in the same spaxel. The symbol `X' indicates the spaxel that best represents the location adopted for the AGN in 
the Circinus galaxy. As all our maps show relative positions, in all of  them  hereafter the AGN is assumed to be 
located at the position $\Delta$~X = 0, $\Delta$~Y = 0. 

\subsection{Stellar continuum subtraction}    
    
\begin{figure*}
\includegraphics[width=\textwidth]{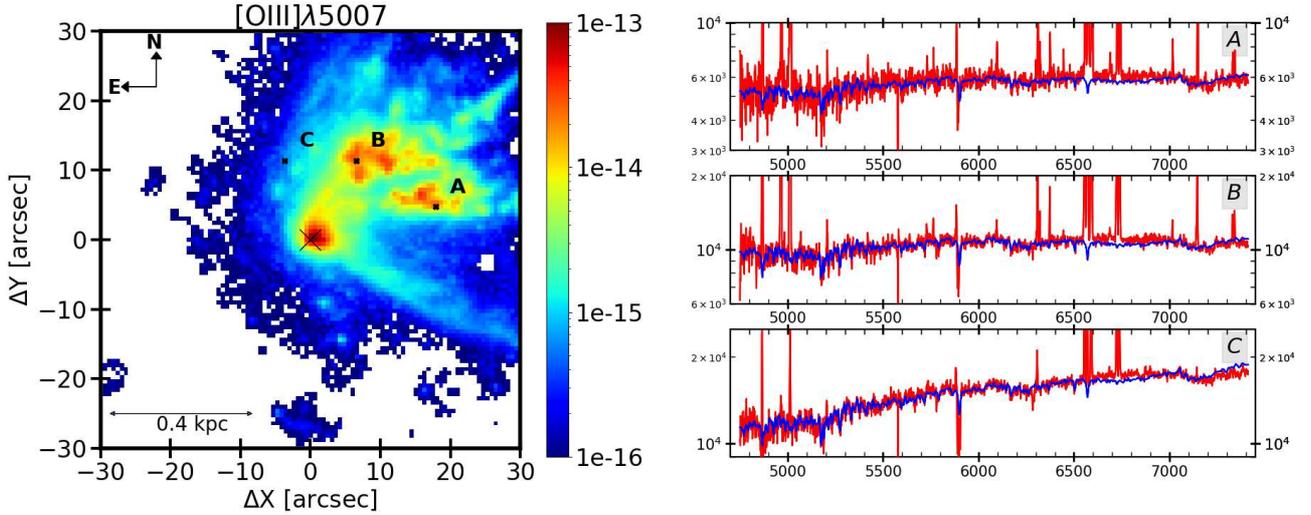}
\caption{Left: map of the emission line flux  for [O\,{\sc iii}]~$\lambda5007$ (in erg~s$^{-1}$~cm$^{-2}$~Spaxel$^{-1}$).
Right:  examples of the continuum fit  by {\sc starlight} at the three selected positions 
indicated in the map of
[\ion{O}{iii}]$\lambda5007$ flux  (in 10$^{-20}$ erg~s$^{-1}$~cm$^{-2}$~\AA$^{-1}$). A and B are in the extended 
North-West region of the ionisation cone. C is in a region dominated by the galaxy disc. The observed spectrum is in red 
and the best fitting stellar population models are in blue. The black cross in the left panel marks the position of the AGN. White areas correspond to spaxels that were masked because the S/N $<$3.}
    \label{fig:espectros}
\end{figure*}

After binning the cube, we subtracted the stellar continuum. This procedure is necessary to fully recover the fluxes 
of the H\,{\sc i} lines affected by the underlying stellar population. Moreover, some weak lines may also be strongly 
diluted by the stellar continuum. To this purpose, the stellar population synthesis code {\sc starlight} \citep{cid_2005} 
was employed, together with the set of stellar populations of E-MILES \citep{vaz_2016}.
{\sc starlight} fits the continuum observed at each spaxel through the whole field-of-view of the data cube.  
Fig.~\ref{fig:espectros} presents examples of this procedure at three different positions. Two points in the North-West 
region (marked as A and B) and one to the North (marked as C) of the AGN. The fitting of the stellar continuum was made in the wavelength range 3750 $-$ 7400~\AA\ to save computational time and to avoid regions of poor sky background subtraction that exist redwards of 7400~\AA. Tests carried out by us showed that the overall fit can be affected by these residuals, while limiting the fit up to that wavelength improved the results.
Redwards of 7400~\AA, we use the original spectra without
stellar light subtraction because that region shows very few lines that are of relevance to this work and are not even
affected by the stellar component. That is the case of [S\,{\sc iii}]~$\lambda$9068, located at the red edge of the spectra.

Then, we subtracted the stellar population 
for the entire cube in the spectral region  3750 - 7400 \AA. Note that because the resolution of the E-MILES and MUSE spectra are different (1~\AA\ and 1.25~\AA, respectively), we rebinnined the later to that of the former. Afterwards, we recovered the original binning of the MUSE data.

Afterwards, the extinction correction due to  the Galaxy was applied. In this process, the CCM extinction law 
\citep{ccm_1989} along with an extinction value $A_{\rm v}$ of 2.1$\pm$0.4 mag, as determined by \citet{for_2012}, was employed.

\subsection{Spectral line fit}        

\begin{figure*}
\includegraphics[width=15cm] {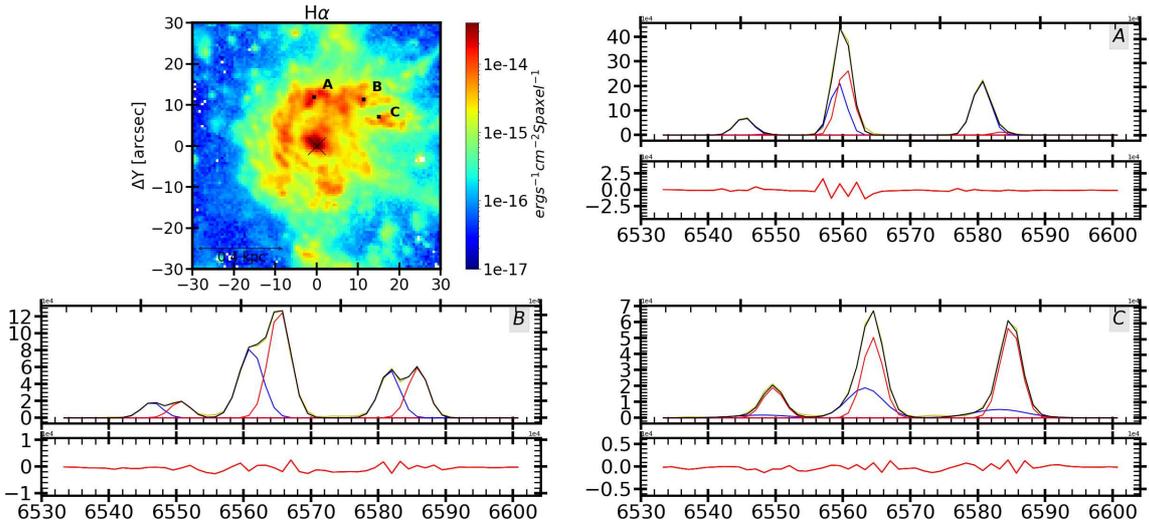}
\caption{Top left: flux  map of  H$\alpha$. The black cross indicates the AGN position. Panels marked as A, B, and C show examples of the Gaussian fits carried out to the H$\alpha$+[\ion{N}{ii}] $\lambda\lambda$ 6549, 6584 lines at some selected positions, identified in the H$\alpha$ map.
In the example fits, the Y-axis scale is in  units of
10$^{-16}$~erg~s$^{-1}$ ~cm$^{-2}$~\AA$^{-1}$ and the X-axis is the wavelength in Angstroms. The blue and red Gaussian components are shown with their respective colour.  The residuals of the fit are shown in the bottom boxes.
The black line  represents the observed spectrum.}
\label{fig:trescomponentes}
\end{figure*}

In order to measure the flux centroid position and the full width at half maximum (FWHM) of the emission lines at each 
spaxel we fit Gaussian functions to individual lines or to sets of blended lines. To this purpose, a set of custom 
scripts written in {\sc python} by our team was employed.  
For each  line one or two Gaussian components were employed to reproduce the observed profile.  Fig. 
\ref{fig:trescomponentes} shows examples of the Gaussian fits carried out to the spectra at the three selected positions 
marked in the H$\alpha$ emission line map (top left panel). It can be seen that two Gaussian component fits to each line 
were necessary to reproduce  H$\alpha$ + [\ion{N}{ii}]~$\lambda\lambda$ 6549, 6583. In this process, some 
constraints were applied. 
For instance, the theoretical  [\ion{N}{ii}] line  flux ratio $\lambda6583/\lambda6548$ was used 
for each component. Moreover, both lines were constrained to have the same width and intrinsic wavelength 
separation. The [\ion{S}{ii}]~$\lambda\lambda$6717, 6731  doublet was constrained to have the same width and a relative theoretical wavelength separation of 14.4~\AA.

It is important to mention that at some positions (i.e., point~A in Figure~\ref{fig:trescomponentes}) if one single Gaussian component is fit, the RMS of the residuals is larger than when two components are employed. In order to assess if two components were indeed necessary, the code examined for a similar solution in adjacent spaxels. We are aware that the Gaussian fitting approach is a mathematical solution and the number of components necessary to reproduce a given profile is strongly related to the S/N and spectral resolution. The physical meaning of two components are always evaluated following the neighbourhood criterion. Moreover, if a component is $< 3\sigma$ of the continuum RMS, that component was discarded.
         
Examples of Gaussian fits carried out to the [\ion{Fe}{vii}]~$\lambda$6087 line in Circinus can be found in Figure~1 of \citet{ardila_2020}.

\begin{figure}
	\includegraphics[width=0.45\textwidth]{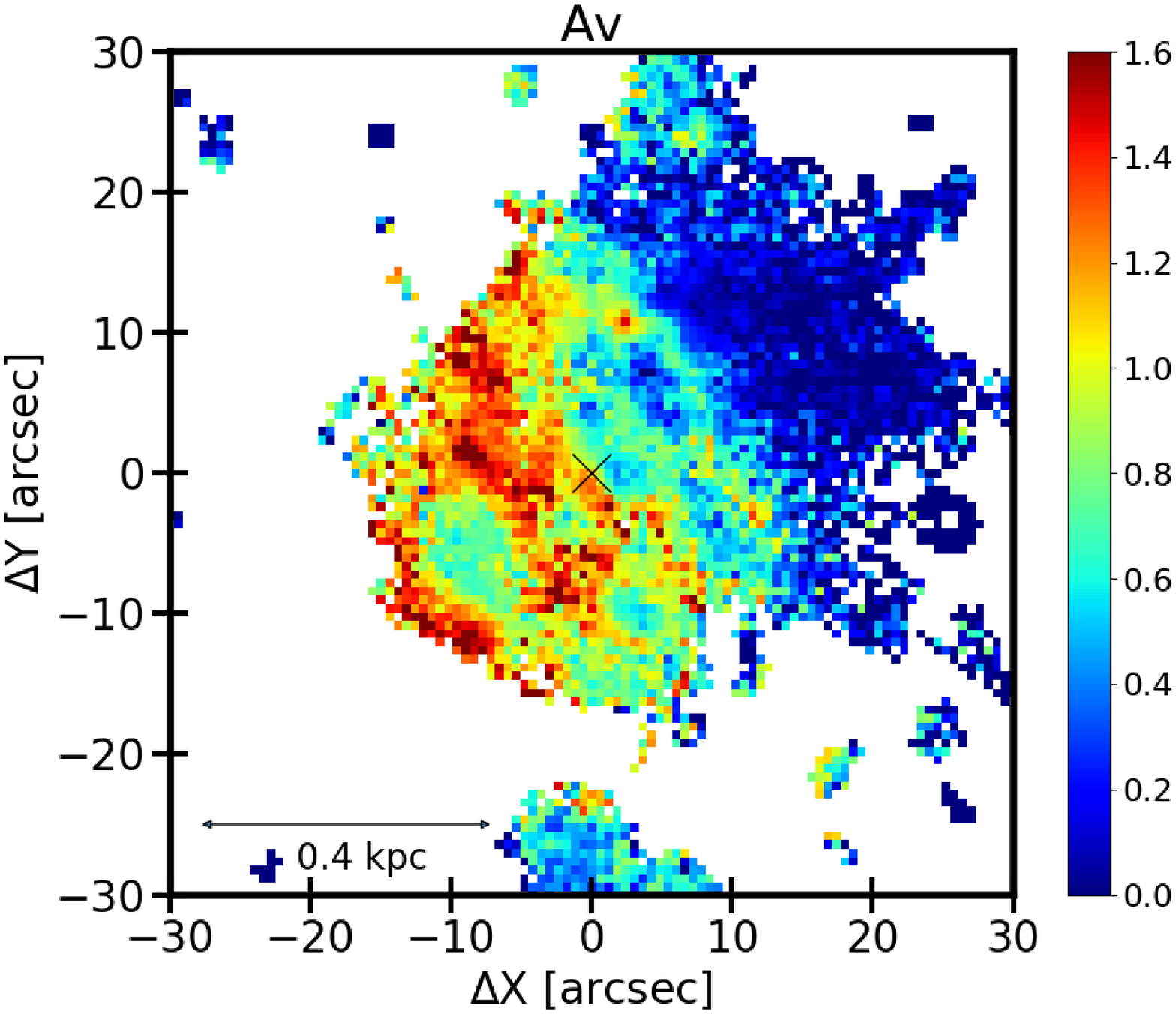}
\caption{Intrinsic extinction map derived for Circinus. The symbol 'X' indicates the AGN position as determined in 
\ref{cap:metodos}. White areas correspond to masked regions where the S/N $<$ 3.}
    \label{fig:av}
\end{figure}

After measuring the fluxes of the emission lines, we determined
the internal extinction affecting the gas at every spaxel. To this purpose, we used Equation~\ref{eq1}, which makes use 
of the observed H$\alpha$/H$\beta$ emission line flux ratio and the CCM extinction law. An intrinsic value of 3.1 for 
the above ratio was assumed and applied to the entire cube. Although we are aware of the presence of star forming regions in the field, we assumed that the AGN photoionisation dominates in the regions of interest to this work. The resulting extinction map is shown in Figure \ref{fig:av}. In the central region 
($<$~200~pc from the AGN) the reddening is above 1.0 mag in the East and around 0.5 mag in the West. In the outermost 
regions ($>$~200 pc away from the active nucleus) to the North-West, the reddening is rather low, with values between 0.1 
and 0.4, in agreement with \citet{mingozzi_2019} results. 

\begin{equation}
    E(B-V)_{H\alpha/H_\beta} = -2.31 \times log \left(\frac{3.1}{H\alpha/H\beta}\right)
	\label{eq1}
\end{equation}

\begin{figure}
\includegraphics[width=\columnwidth]{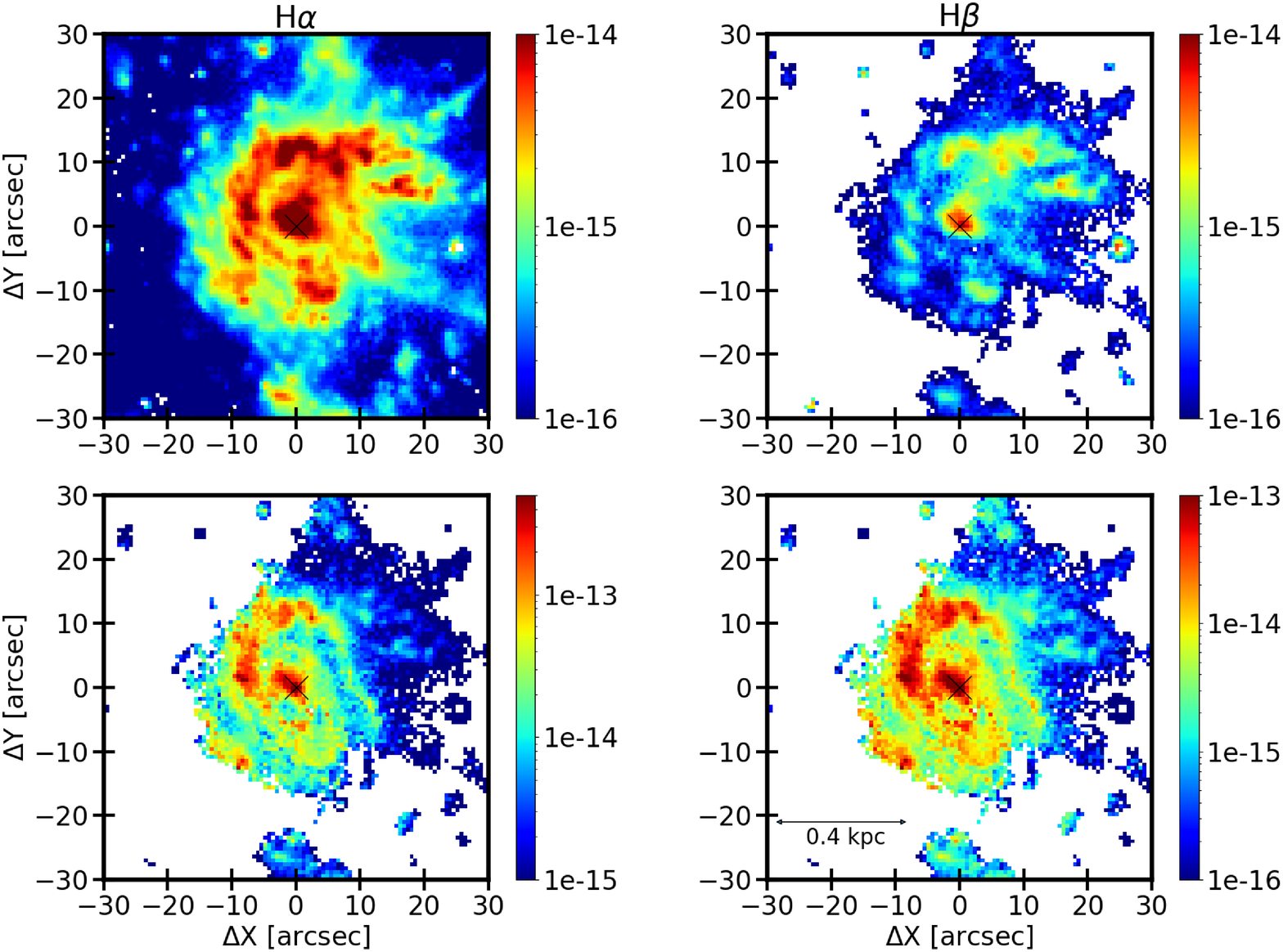}
\caption{Flux maps for H$\alpha$ (left panels) and H$\beta$ (right panels) before (top) and after (bottom) reddening  correction. The colour bar is in units of erg~s$^{-1}$~cm$^{-2}$~ Spaxel$^{-1}$. White areas correspond to mask values
of SNR $<$3. The black cross indicates the AGN position.}
\label{fig:apos_correcao_av}
\end{figure}

Fig. \ref{fig:apos_correcao_av} displays the flux distribution of H$\alpha$ (left panels) and H$\beta$ (right panels) 
before  and after  reddening correction (top and bottom panels, respectively). The  \Ha  emitting  
region is smaller after correction because the H$\beta$ line is detected at 3$\sigma$ in a region smaller than 
that of H$\alpha$.  Only the regions displaying both lines after reddening correction are shown. 

As a last step, the 
spectrum at every spaxel was corrected by  the extinction map of Figure \ref{fig:apos_correcao_av}.
    
%##########################################################
%################## DISTRIBUICAO DO GAS   ########### #
%##########################################################

\subsection{Gas distribution in the Circinus galaxy}

In this section we will analyse the spatial distribution of the different emission lines that were found in the data 
cube obtained by MUSE. After the treatment of the cube (binning, subtraction of the stellar continuum, correction for 
Galactic and intrinsic extinction and measurement of the detected emission lines) as described in 
Section ~\ref{cap:metodos}, flux maps for the most relevant emission lines identified in the different spaxels were 
constructed.  We want to  compare their spatial distribution and confirm the presence of extended emission 
for lines of similar or higher ionisation potential than that of [\ion{Fe}{vii}]. In addition, we estimate 
the physical conditions of  the coronal line region. To this purpose, it is  necessary to find out  suitable diagnostics 
to characterise the density and temperature of the high-ionisation gas. 

\begin{figure}
  \includegraphics[width=0.45\textwidth]{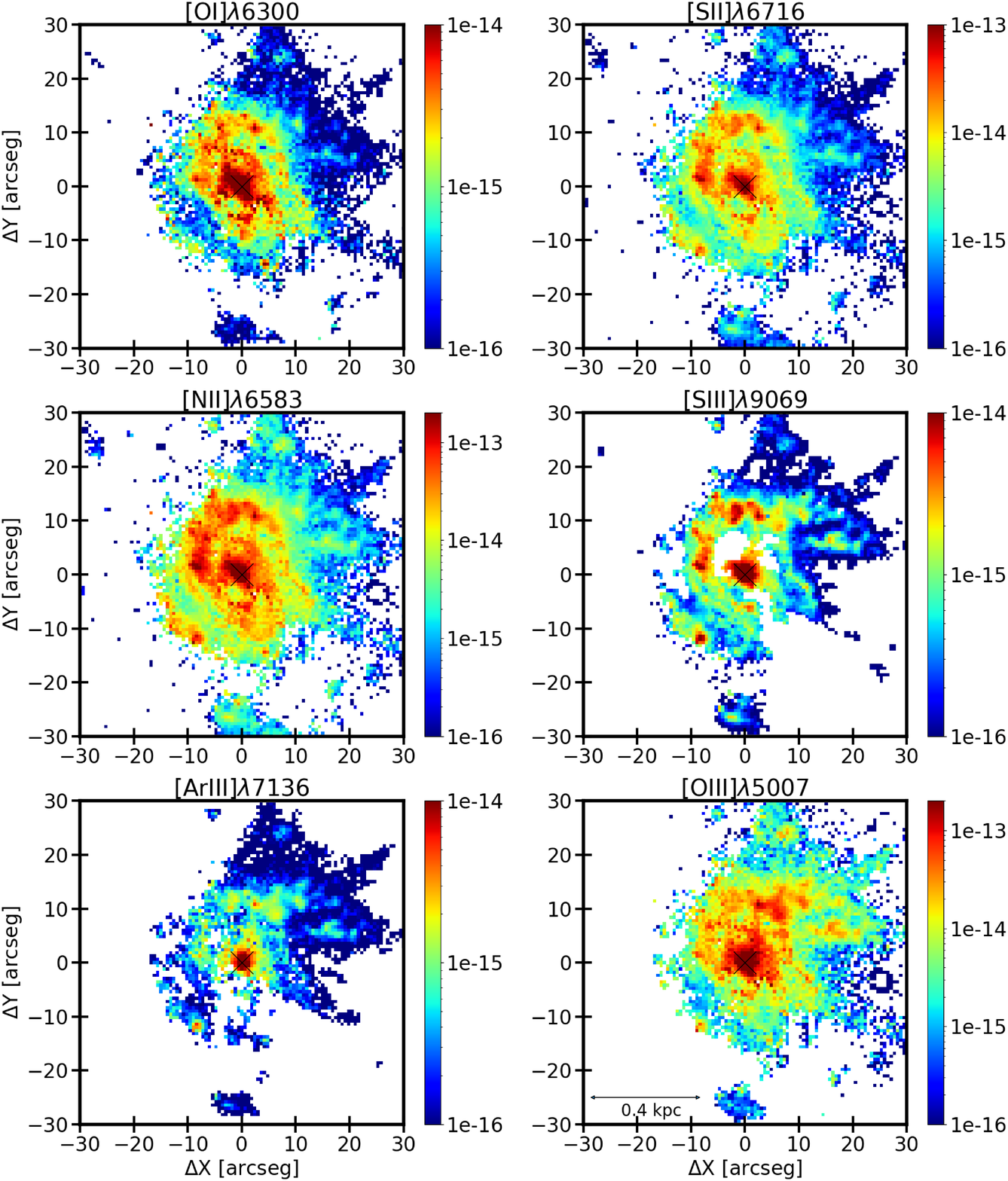}%[width=2.2\textwidth]
    \caption{Emission line flux distribution for [O\,{\sc i}]$\lambda6300$, [S\,{\sc ii}]$\lambda6716$, [N\,{\sc ii}]$\lambda6583$, 
 [S\,{\sc iii}]$\lambda9069$, [Ar\,{\sc iii}]$\lambda7136$ and [O\,{\sc iii}]$\lambda5007$. The colour bar is in units of  
erg~s$^{-1}$~cm$^{-2}$~Spaxel$^{-1}$. North is up and East to the left.}
    \label{fig:fluxos_1}
\end{figure}

Figure~\ref{fig:fluxos_1} presents the flux distribution of  [O\,{\sc i}] $\lambda6300$, 
[S\,{\sc ii}] $\lambda6716$, [N\,{\sc ii}] $\lambda6583$, [S\,{\sc iii}] $\lambda9069$, [Ar\,{\sc iii}] $\lambda7136$ 
and [O\,{\sc iii}] $\lambda5007$. Except for the latter line, all show a distribution similar to H$\alpha$ 
(see Fig.~\ref{fig:apos_correcao_av}). Moreover, they all display a prominent spiral arm that starts  at the East of the 
nucleus towards the North, where it bends and turns to the South. Evidence of a  counter arm is also observed to 
the South and SE of the galaxy centre. It can also be noticed that the higher the gas ionisation, the more conspicuous is
the region where the ion is emitted towards the NW of the AGN. For instance,  [\ion{O}{iii}] and [\ion{Ar}{iii}]  map the 
ionisation cone more suitably than [\ion{O}{i}].

\begin{figure}
\centering
   \includegraphics[width=0.45\textwidth]{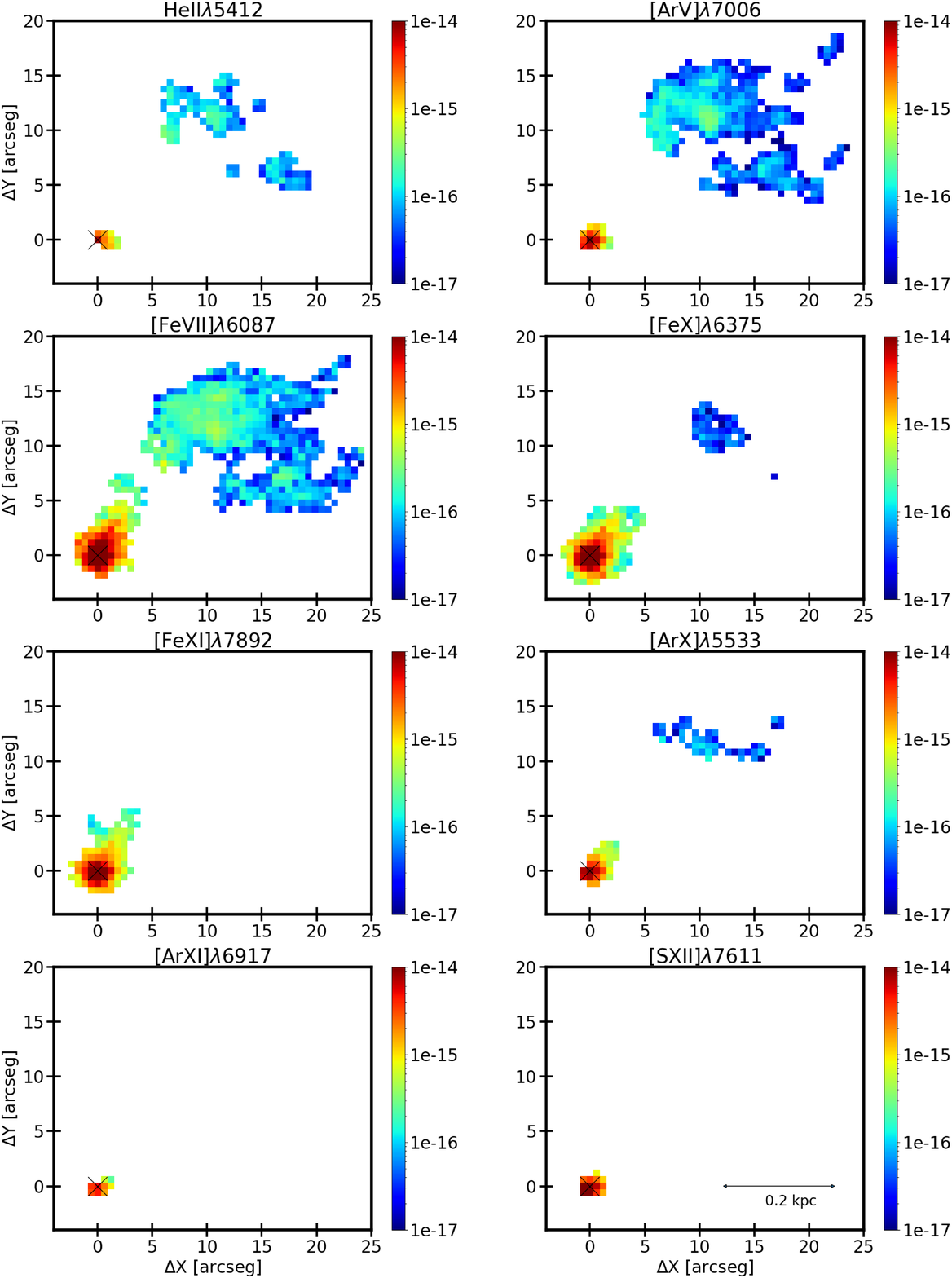}%[width=2.2\textwidth]

    \caption{Flux map distribution for He\,{\sc ii}$\lambda5412$, [Ar\,{\sc v}]$\lambda7006$,  [Fe\,{\sc vii}]$\lambda6087$,  
[Fe\,{\sc x}]$\lambda6375$, [Fe\,{\sc xi}]$\lambda7892$, [Ar\,{\sc x}]$\lambda5533$, [Ar\,{\sc xi}]$\lambda6917$ and  
[S\,{\sc xiii}]$\lambda7611$ in units of  erg~s$^{-1}$~cm$^{-2}$~Spaxel$^{-1}$. On each panel, the cross marks the position of the AGN. Only the NW quadrant of the MUSE detector is shown because the extended, high-ionisation emission is located entirely in the nuclear region and to the NW.}  
    \label{fig:fluxos_2}
\end{figure}

Flux distribution maps for the high ionisation lines were also produced. Fig. \ref{fig:fluxos_2} displays maps for  
He\,{\sc ii}~$\lambda5412$, [Fe\,{\sc vii}]~$\lambda6087$, [Ar\,{\sc v}]~$\lambda7006$, 
[Fe\,{\sc x}]~$\lambda6375$, [Fe\,{\sc xi}]~$\lambda7892$ and [S\,{\sc xii}]~$\lambda7611$. For these lines, the 
emitting gas  is extended only at the North-West and in the nuclear region. Very high ionisation lines, 
such as [Fe\,{\sc xi}] and [S\,{\sc xii}] are only detected in the nuclear region.
[Fe {\sc xiv}]\,$\lambda$5303 with IP~=~361~eV \citep{oliva_1999}
is the  line with the highest IP  already identified in Circinus in the optical/NIR region.
In this paper we report the first detection of the [S\,{\sc xii}]$\lambda$7611 (IP~=~505 eV).
Its emission is not restricted to the unresolved nuclear region but  it is observed   up to a distance
of $\sim40$~pc from the nucleus.
We also report an emission line at 8340~\AA\ in the nuclear spectrum that we tentatively attribute to
[Ar\,{\sc xiii}]~$\lambda$8340 (IP~=~618~eV). However, this identification is not sure as the SNR is $<$ 3.
As already mentioned, \citet{ardila_2020} recently reported on the detection of extended [Fe\,{\sc vii}] emission in
Circinus up to a projected distance of $\sim$700~pc from the AGN. Here, we also found extended
[\ion{Fe}{x}]~$\lambda$6374 emission on scale of hundreds  pc. In the nuclear and circumnuclear region,
the gas emitting this line shows a clearly elongated morphology in the NW-SE direction, nearly coinciding with the
radio jet axis and extending up to 140~pc from the centre.  Beyond it, the line fades below the detection limit.
It reappears
at $\sim$280~pc from the AGN in the same direction as the circumnuclear extended region  forming an extended cloud
of several tens of pc size  that is clearly resolved by MUSE and it is  visible up to $\sim$350~pc from the nucleus.
The region where the extended [\ion{Fe}{x}] emission is observed coincides in position with the brightest  part of
the extended Fe$^{+6}$ gas. To the best of our knowledge, this is the first time that extended Fe$^{+9}$ gas at
such distance scales is reported in a radio-quiet AGN.

\subsection{Distribution of emission line flux ratios}

\begin{figure*}
\centering % POSSO COLOCAR UMA FAIXA VERTICAL PARA DEIXAR AS IMAGENS MAIS LONGES Em vertical 
\includegraphics[width=5.8cm]{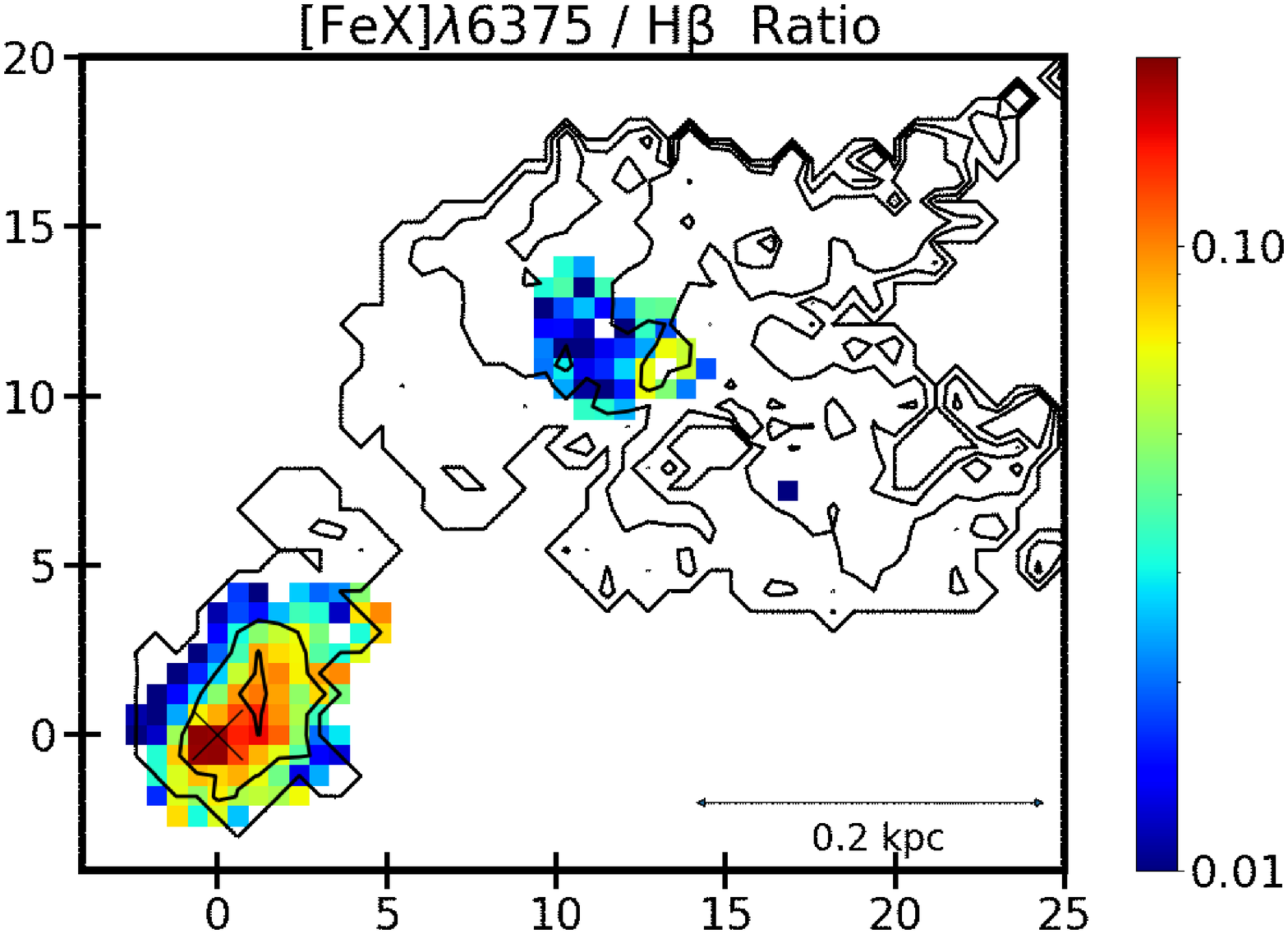}
\includegraphics[width=5.8cm]{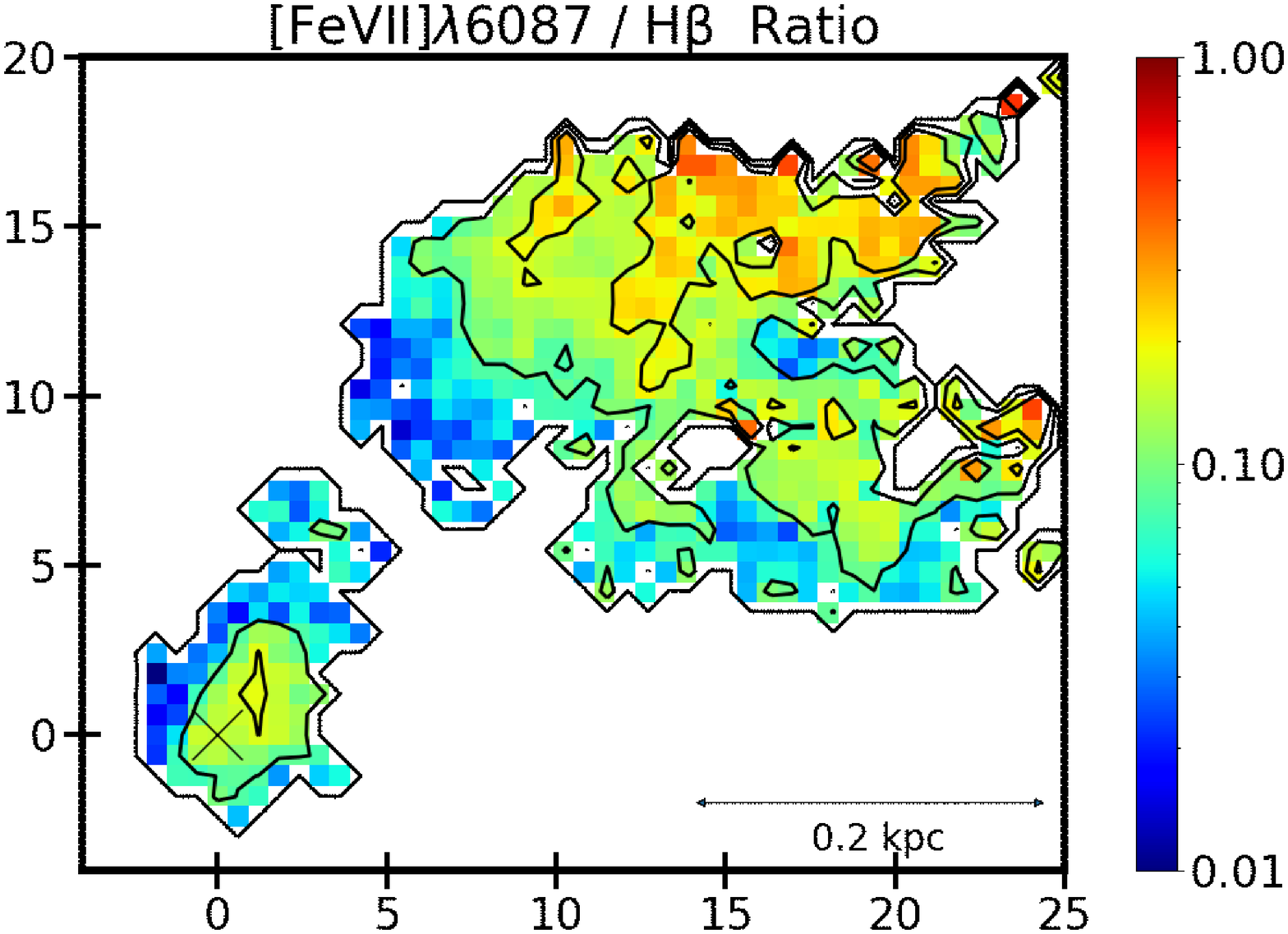} 
\includegraphics[width=5.8cm]{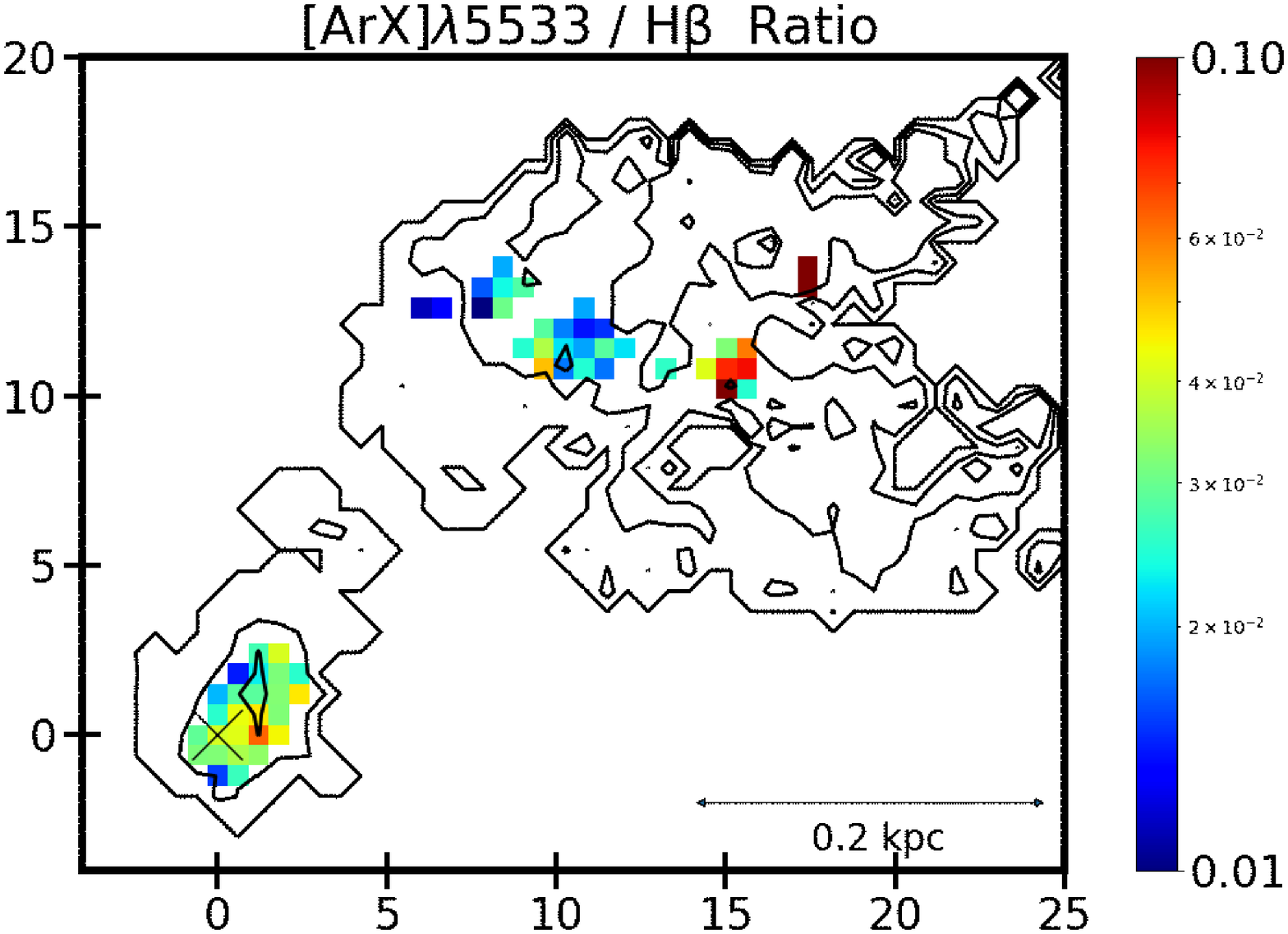}
\includegraphics[width=18cm]{Figuras/fita_branca.eps}
\includegraphics[width=5.8cm]{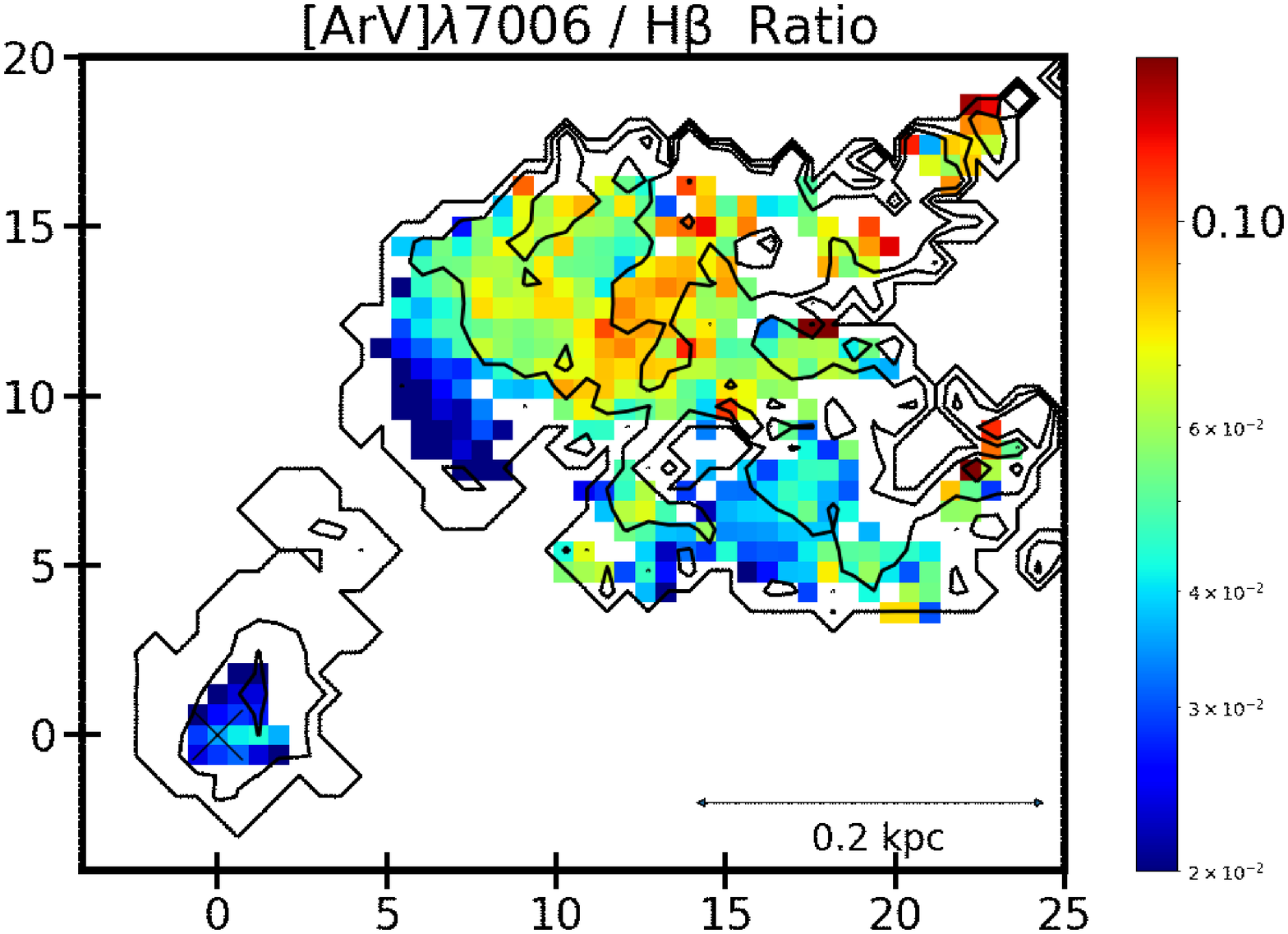} 
\includegraphics[width=5.8cm]{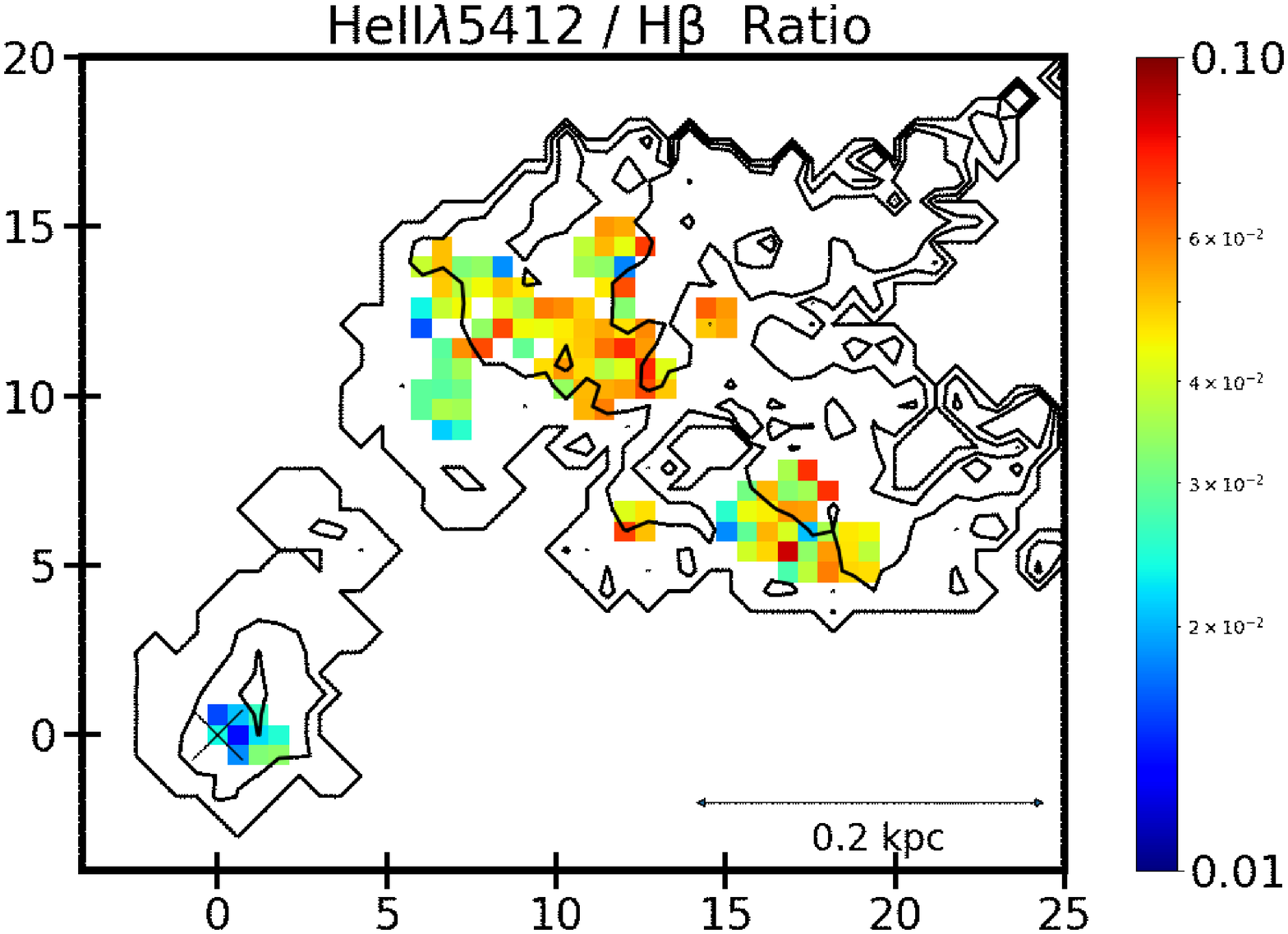}
\includegraphics[width=5.8cm]{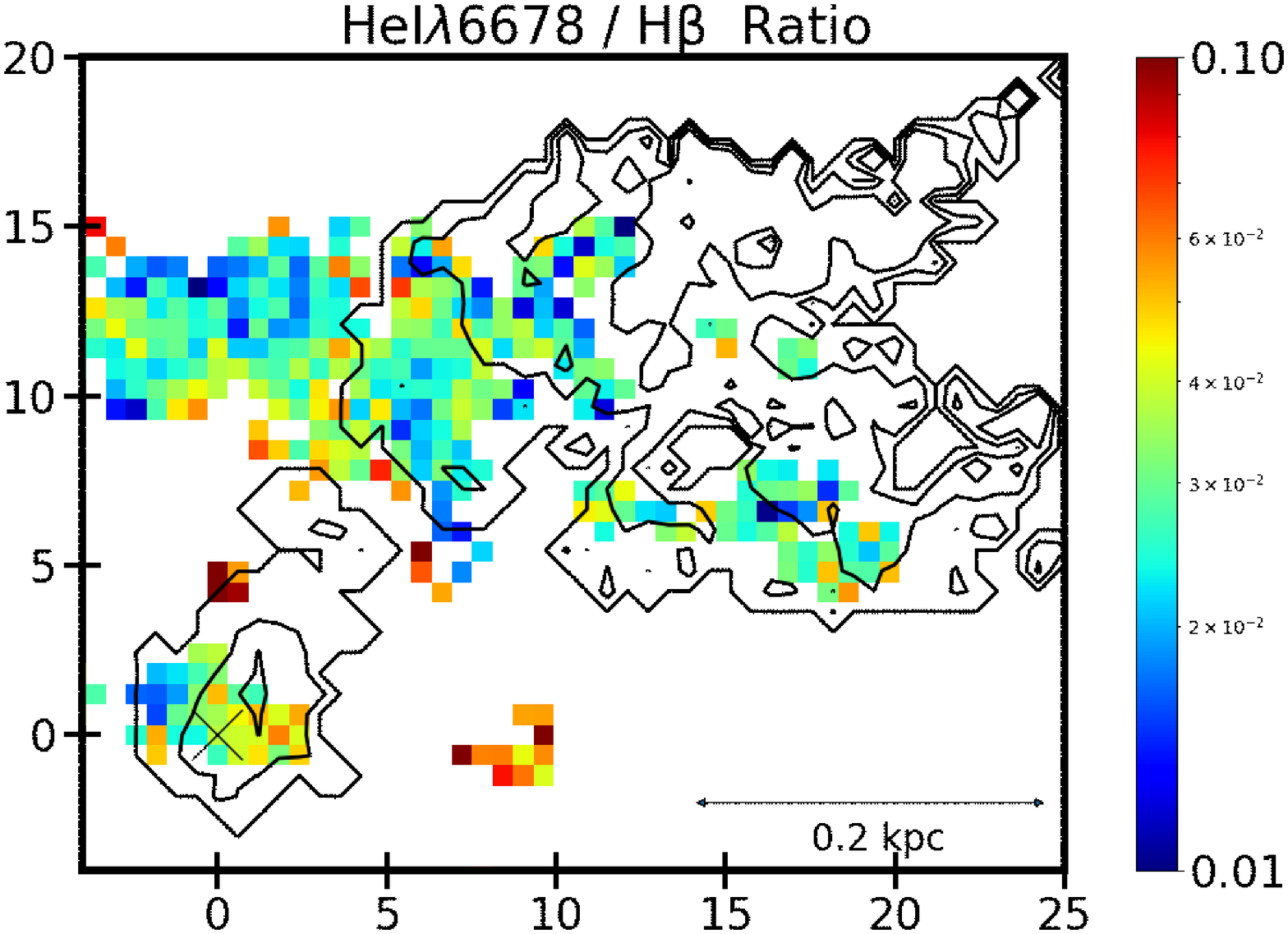}
\includegraphics[width=18cm]{Figuras/fita_branca.eps} 
\includegraphics[width=5.8cm]{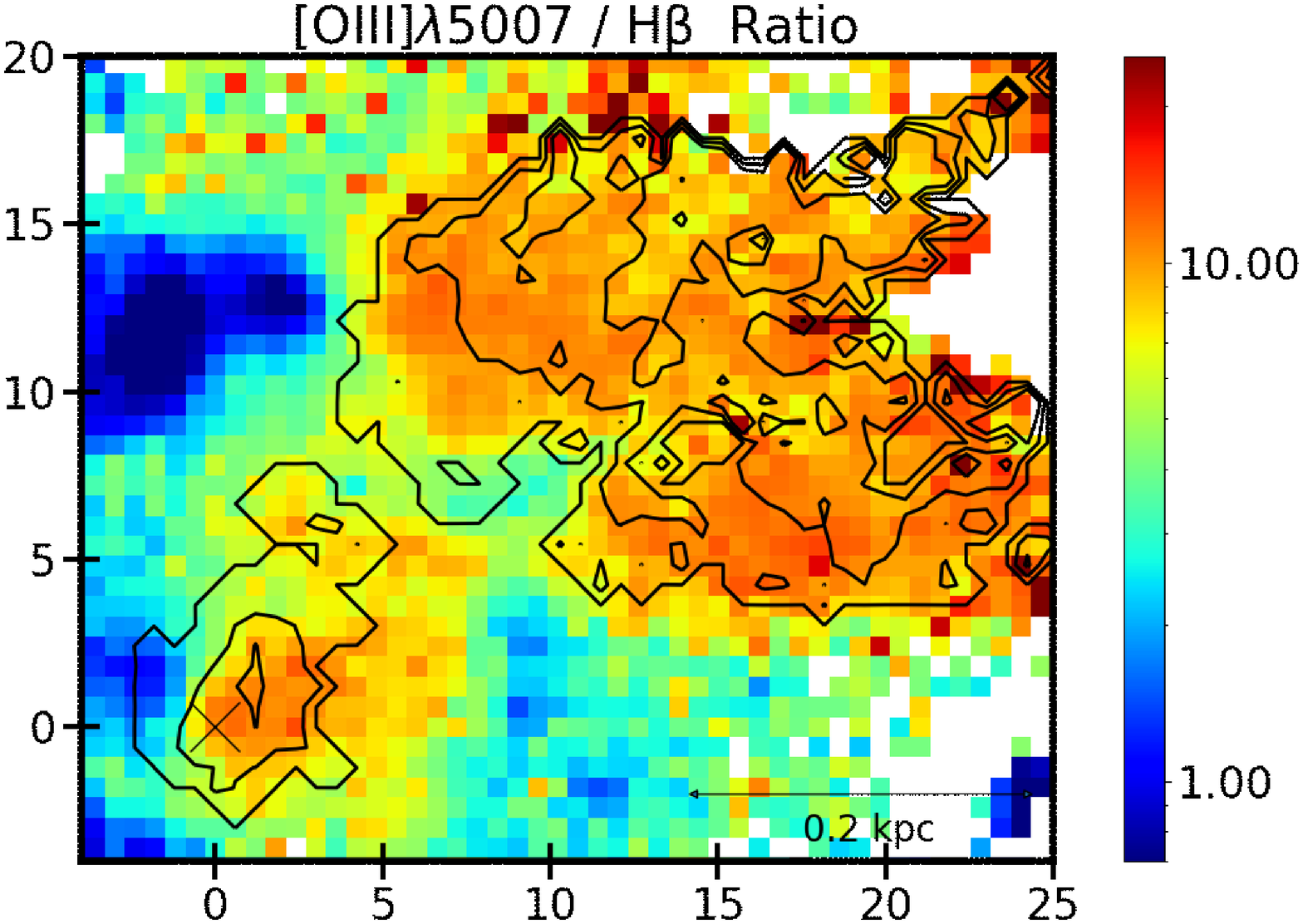}
\includegraphics[width=5.8cm]{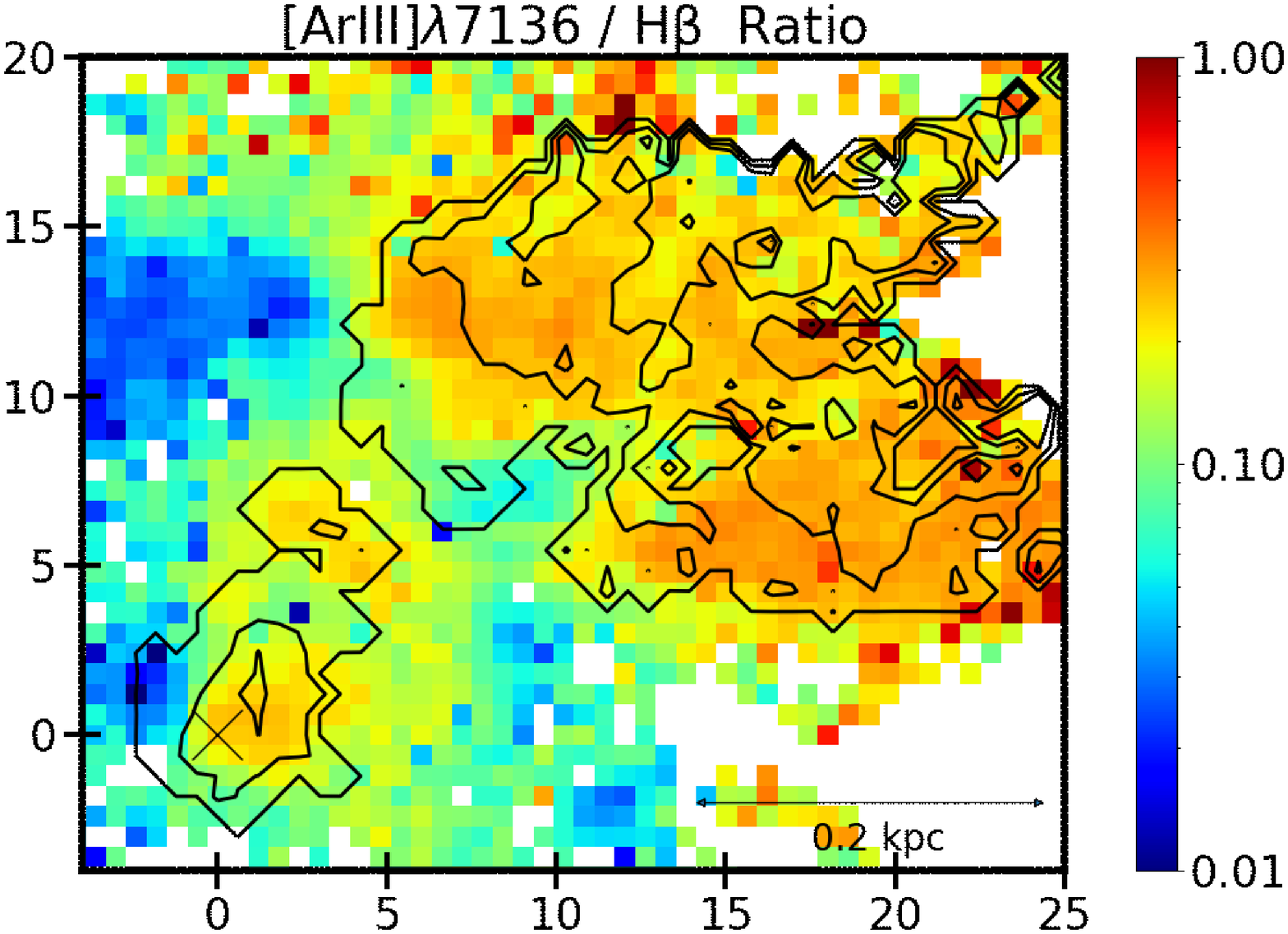}
\includegraphics[width=5.8cm]{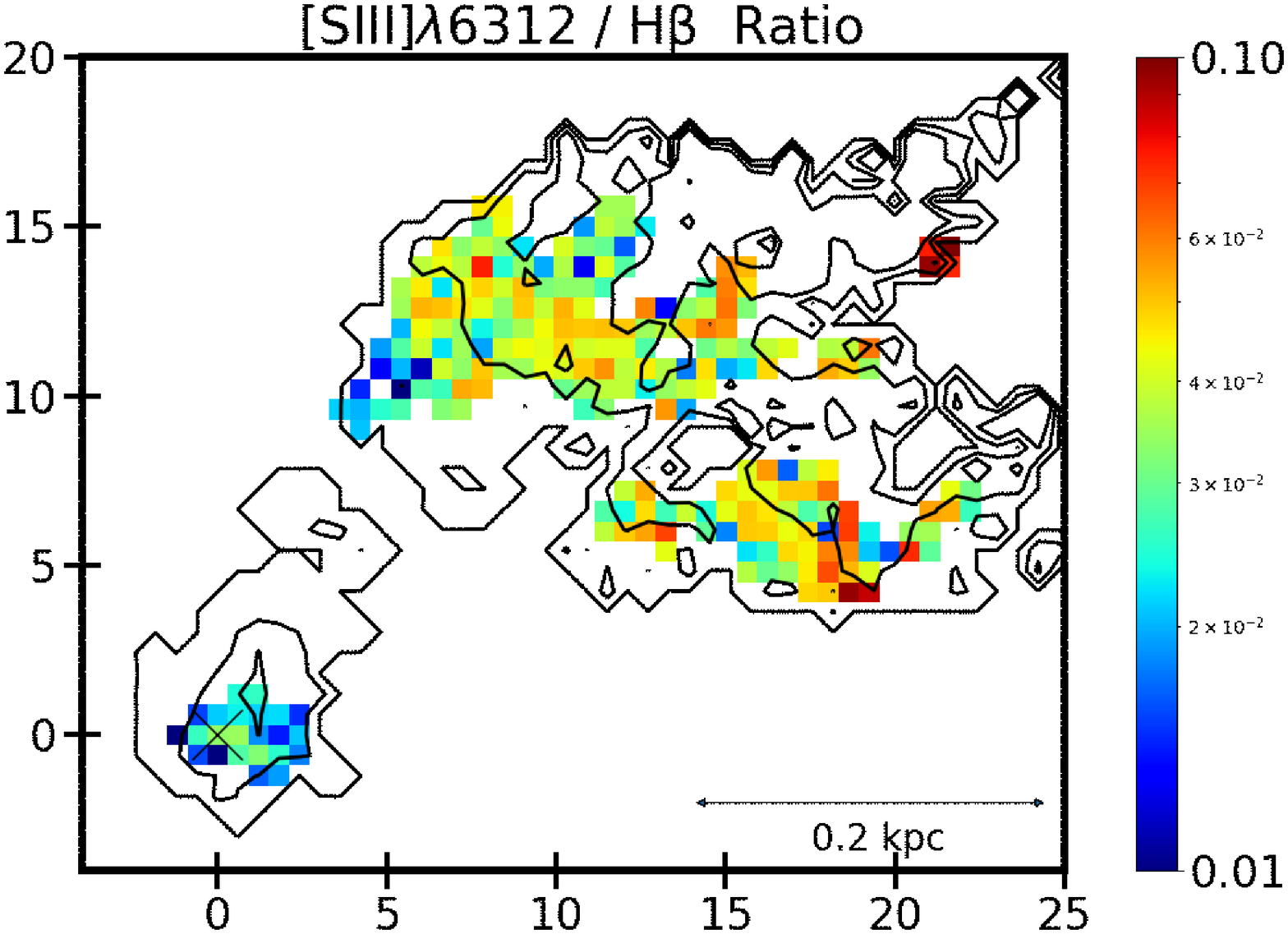}
\includegraphics[width=18cm]{Figuras/fita_branca.eps}
\includegraphics[width=5.8cm]{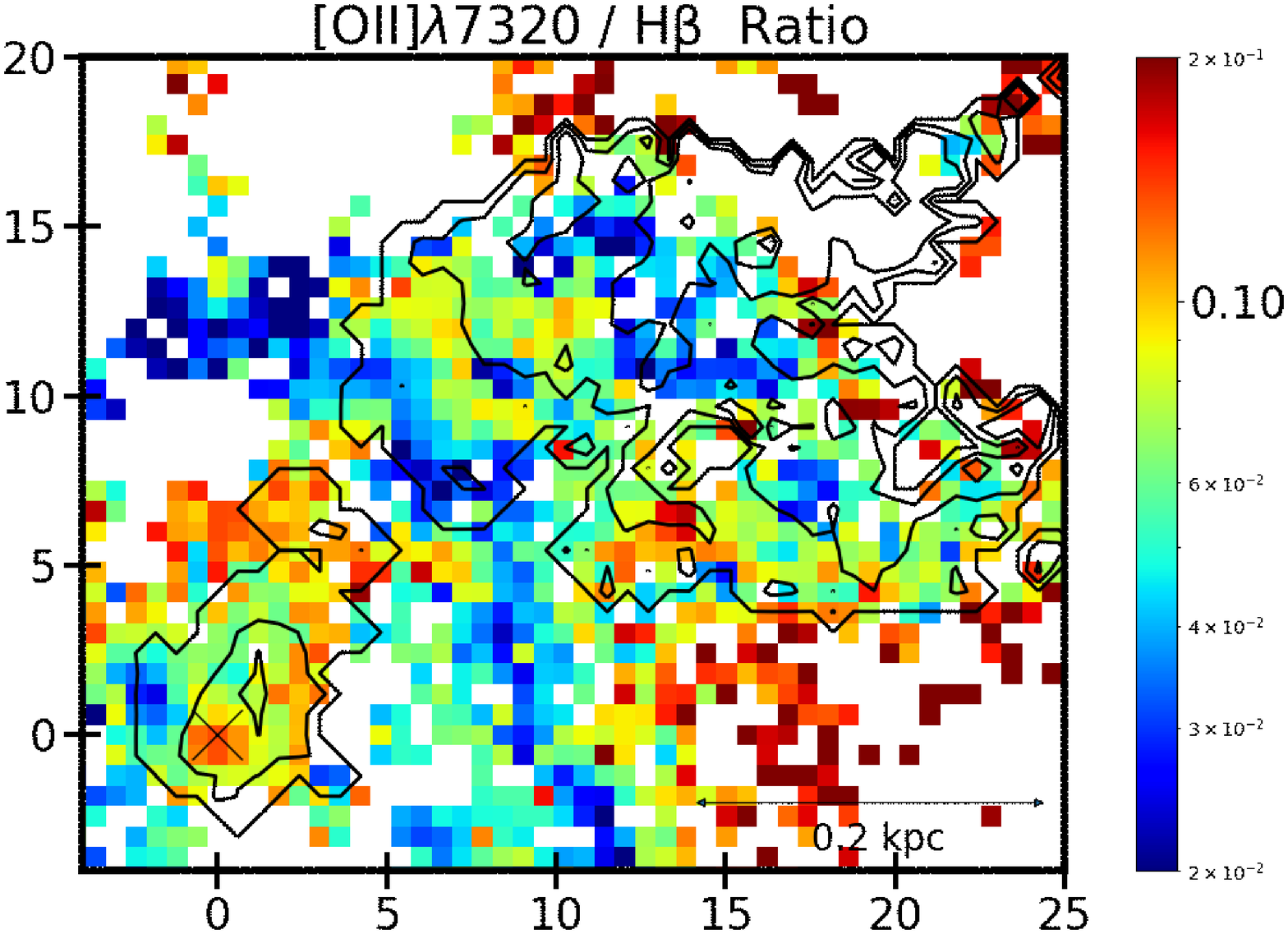}
\includegraphics[width=5.8cm]{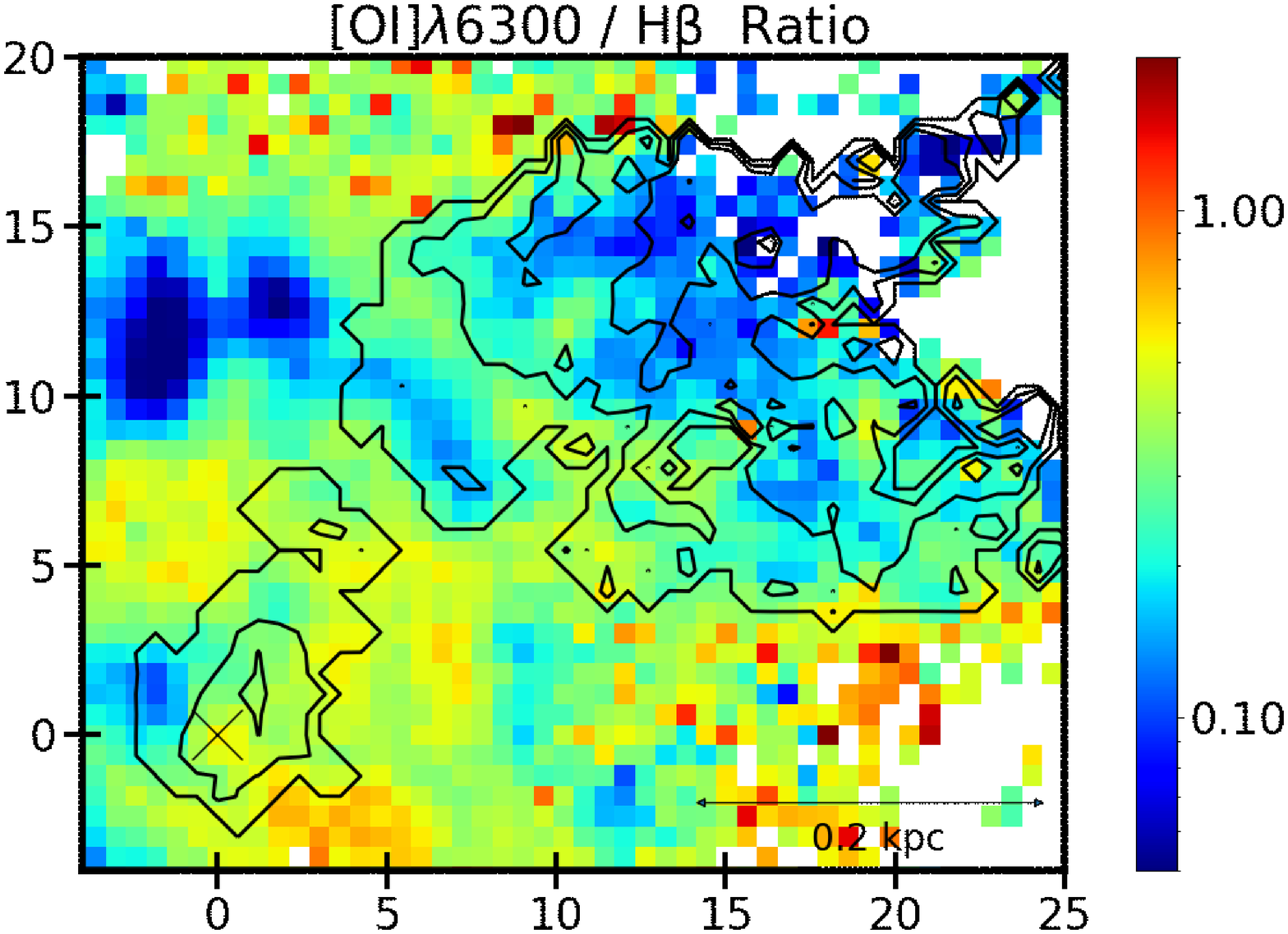}
\includegraphics[width=5.8cm]{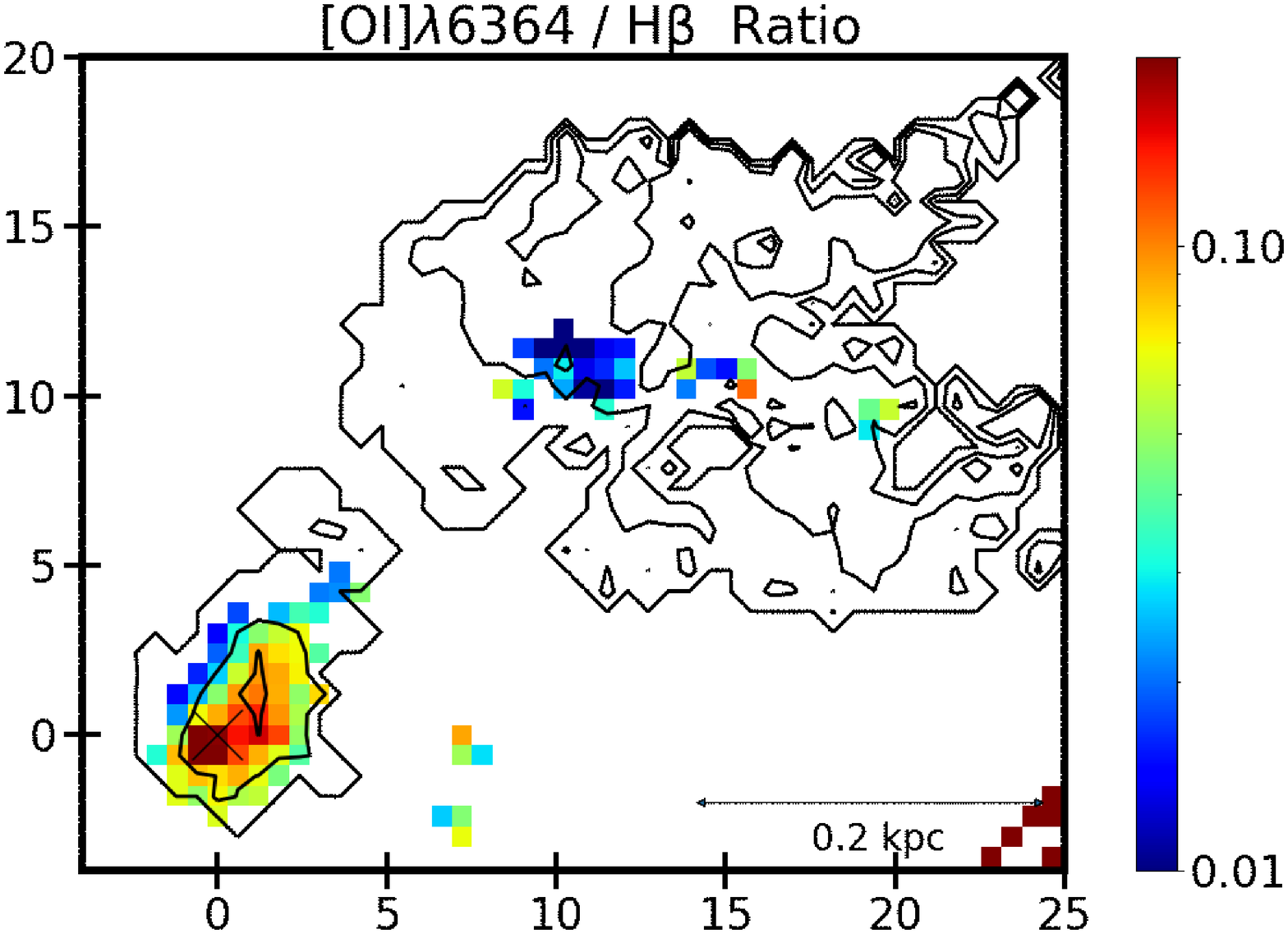}
\includegraphics[width=18cm]{Figuras/fita_branca.eps}
\includegraphics[width=5.8cm]{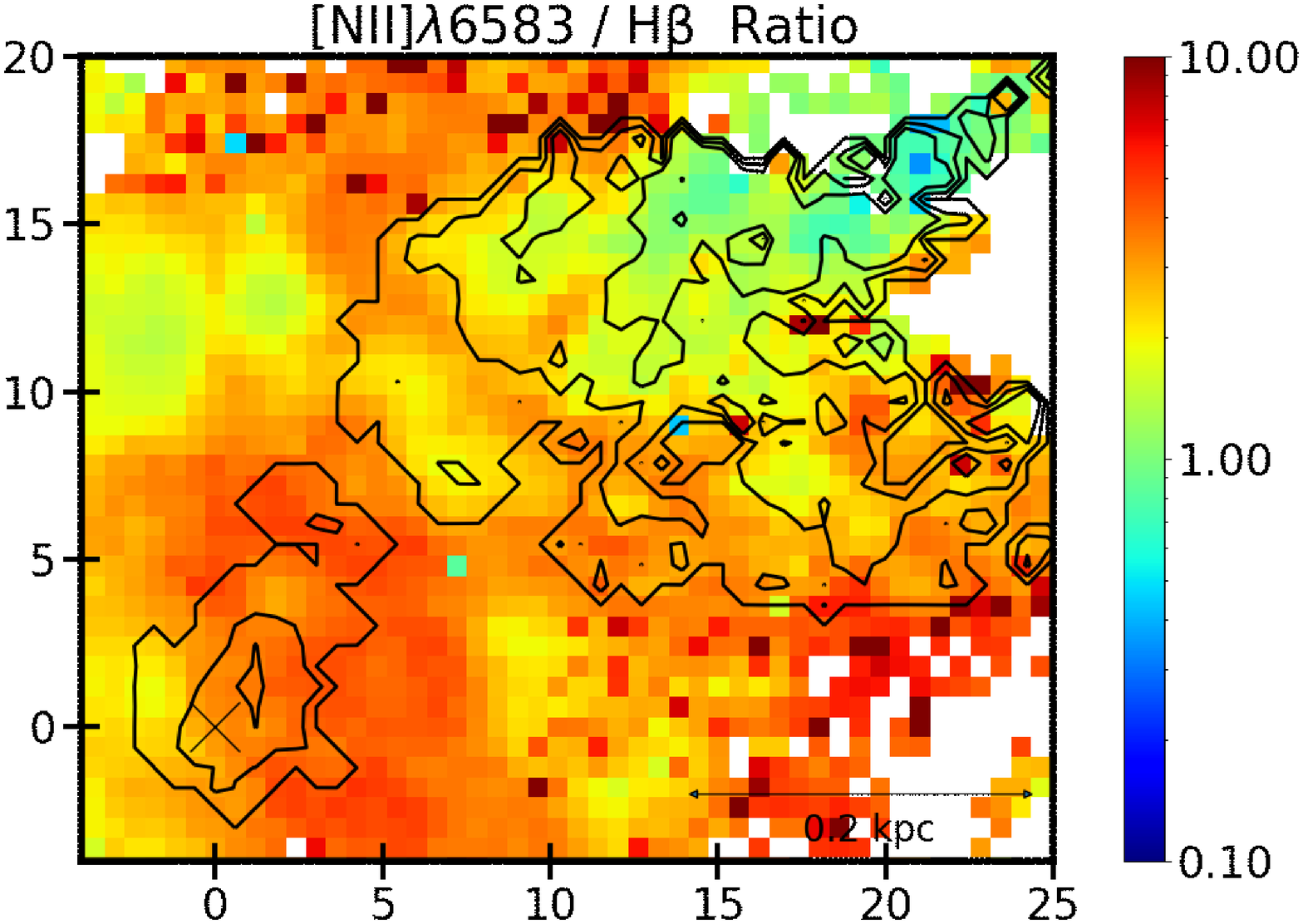}
\includegraphics[width=5.8cm]{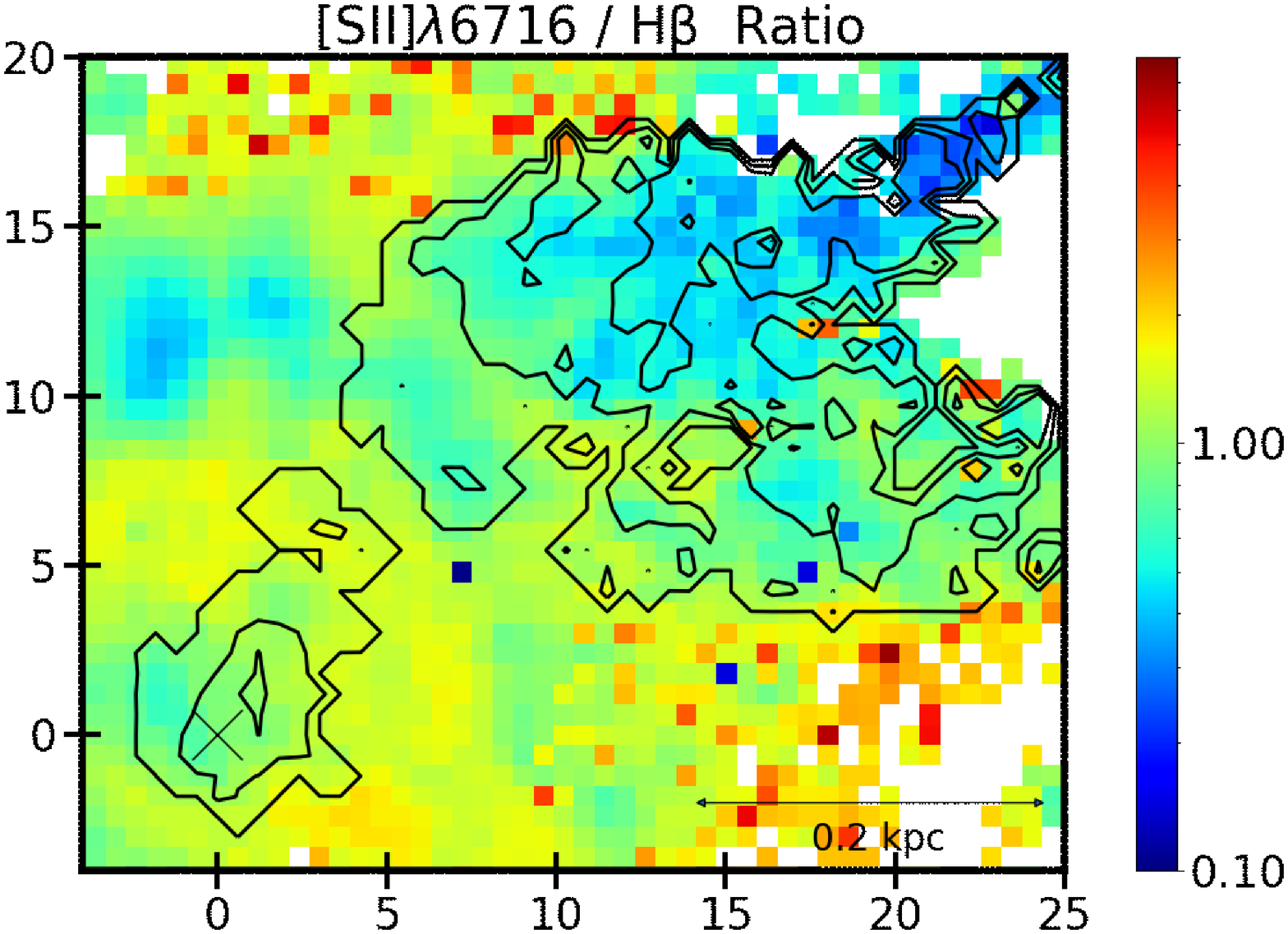}
\includegraphics[width=5.8cm]{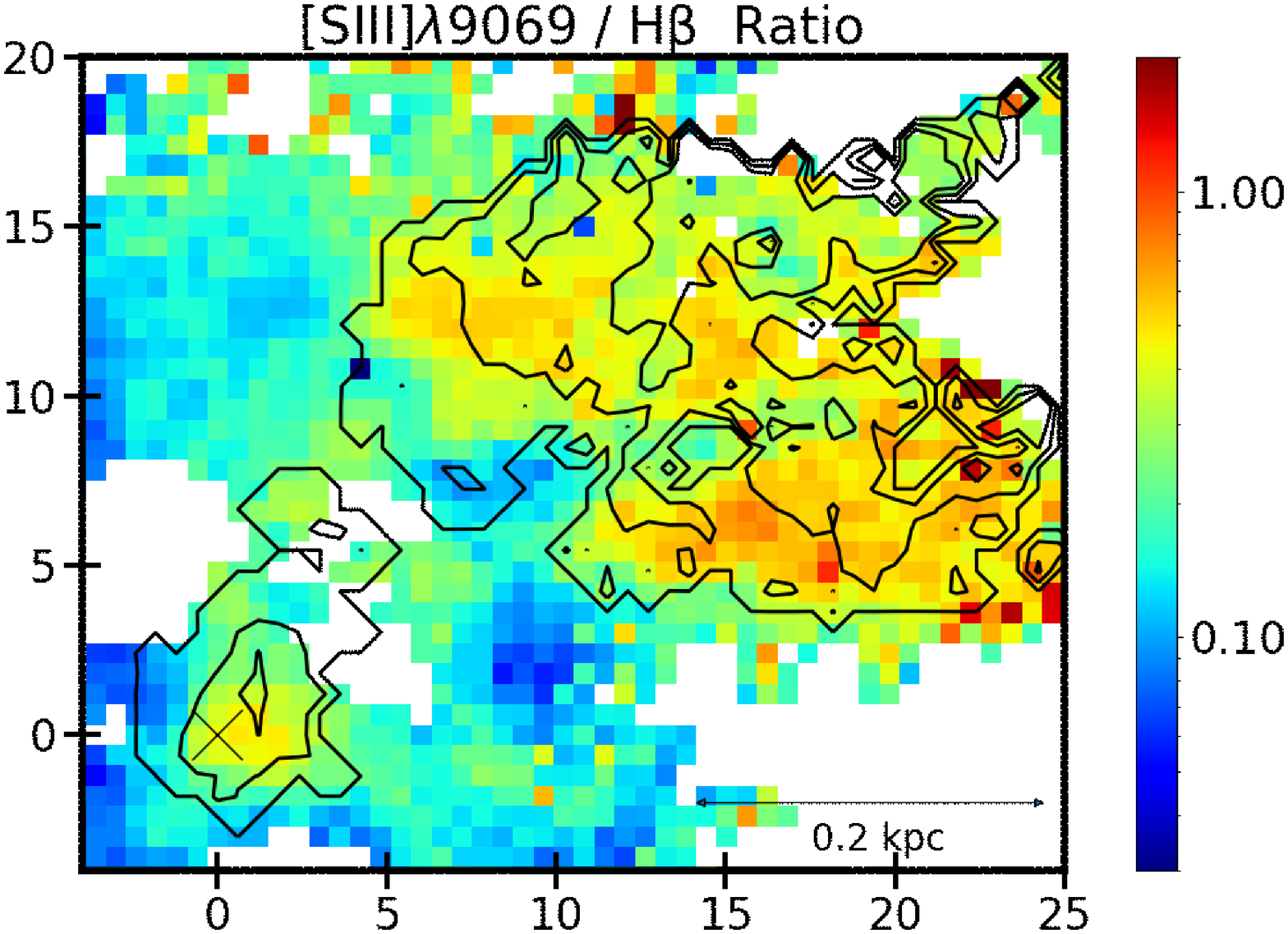}

\caption{Emission line flux ratios for the most prominent ions detected in the spectra of Circinus. In all panels, the black contours represent the ratio of [Fe\,{\sc vii}]/H$\beta$ 
with levels of 0.001, 0.08 and 0.17. The white regions correspond to locations where the lines where not detected.}
\label{fig:ratios_to_hbeta}
\end{figure*}

We use the emission-line fluxes shown in the preceding section to construct line intensity-ratio maps. 
The main goal is to  investigate how the different distributions of the line flux ratios to \Hb 
behave. The results are shown in Fig.~\ref{fig:ratios_to_hbeta}. In all the maps, the lines are reddening corrected. 

Interestingly, Fig.~\ref{fig:ratios_to_hbeta} shows that most of the  high-ionisation line ratios to \Hb
 are enhanced in the extended
 [Fe\,{\sc vii}] region. Moreover, the ratios of these lines increase with  increasing 
distance from the AGN. For example, [Fe\,{\sc vii}]/H$\beta$ has values of $\sim 10^{-2}$ at 240~pc from the AGN while 
at 700~pc it is nearly an order of magnitude larger ($\sim 10^{-1}$). A similar behaviour is observed for
[O\,{\sc iii}] and [Ar\,{\sc v}], where an increase by a factor of $\sim$~5 in the line flux  ratio to H$\beta$ is 
found   outwards.  
Low-ionisation lines are weak relative to H$\beta$ in the region dominated by the extended [Fe\,{\sc vii}]. 
They become stronger outside the region filled with the high-ionisation gas. 

The peculiar extended gas distribution e.g. the increase of the [Fe\,{\sc vii}]/H$\beta$ ratio
 cannot be easily explained by the central source photoionisation alone.

%##########################################################
%################## CODICOES FISICAS DO GAS   ############
%##########################################################
    
\section{Physical conditions of the gas}\label{sec:condicoes_fisicas}

\begin{figure}
 \includegraphics[width=\columnwidth]{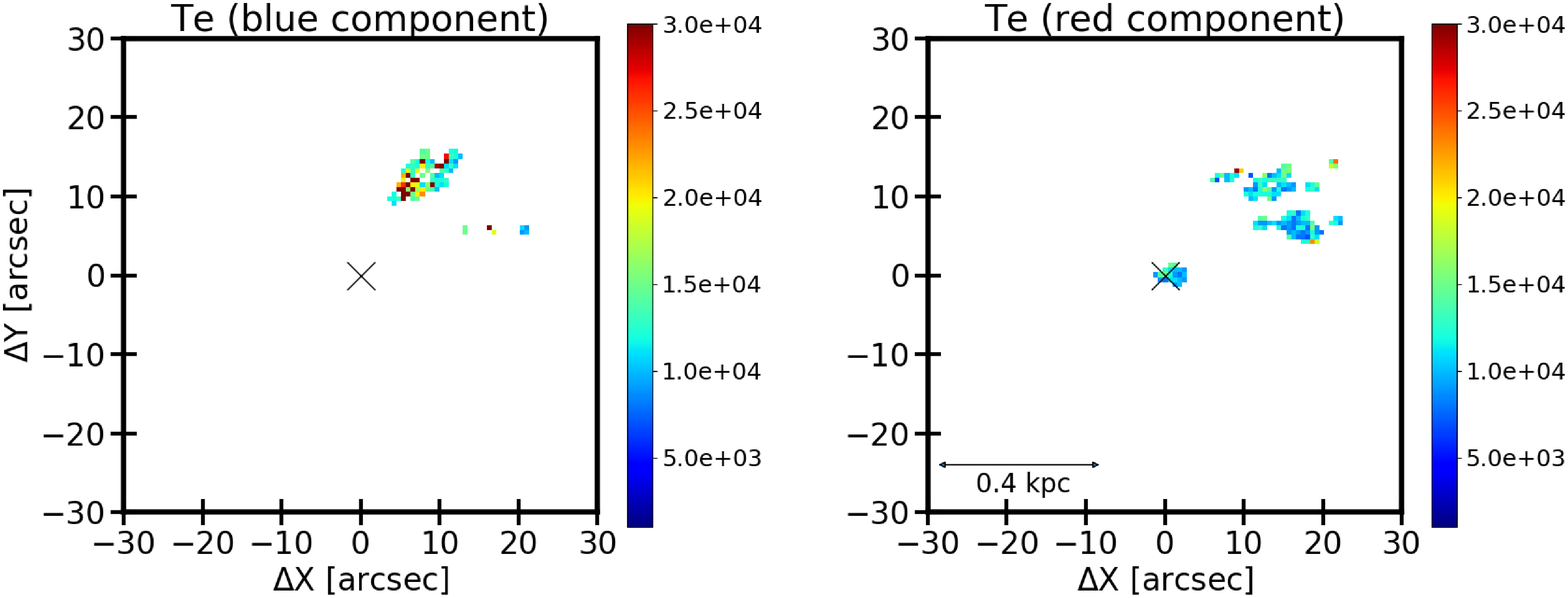} 
\caption{Temperature distribution for the blue and red components (left and right panels, respectively) obtained from the [S\,{\sc iii}] lines. The colour bar is in units of Kelvins.
(see Sect.~\ref{sec:condicoes_fisicas}).}
\label{fig:temperaturas}
\end{figure}

In order to get clues about the processes that power the strong high-ionisation outflow detected in Circinus 
\citep{ardila_2020}, we determine the electron temperature, ~\Te  and density, \ne in the regions  co-spatial to the 
[\ion{Fe}{vii}]~$\lambda$6087 emission. 
\Te in each spaxel was derived  from the  [S\,{\sc iii}] line ratios by means of Equation~\ref{eq2} \citep{osterbrock_2006}. MUSE has a wavelength cutoff at 9200~\AA, so that at the redshift 
of Circinus, the [S\,{\sc iii}]$\lambda$9069~\AA\ is observed. Using the intrinsic ratio determined from quantum 
mechanics, [\ion{S}{iii}]~$\lambda9532 = 2.44 ~\times$ [\ion{S}{iii}] $\lambda9069$ and Eq.~\ref{eq2}, it is possible 
to employ the  S$^{+2}$  ions for temperature diagnostics. Fig.~\ref{fig:temperaturas} shows the results obtained 
for the regions where the lines of [S\,{\sc iii}] have a SNR $>$ 3.  In this calculation, a \ne= 400~cm$^{-3}$ was employed. 
Electron densities were evaluated  from the [\ion{S}{ii}] line ratios (see below). Increasing or decreasing 
the gas density by a factor of 2  has negligible effects on the gas temperature.
The [\ion{S}{iii}]~$\lambda$6312 line is at least a factor of 10 weaker than  
[\ion{S}{iii}]~$\lambda$9069.  \Te  was evaluated  in the spaxels where all three lines have 
SNR $>$3. Thus,  the electron temperature was effectively determined  at just a few particular regions of the MUSE detector. Note that
[\ion{S}{iii}]~$\lambda$6312  was
mostly detected in  the same spaxels where the bulk of [\ion{Fe}{vii}] is observed. Therefore,
on a first approximation, \Te determined above can be  representative of the coronal iron line
emission gas if we consider  that  [S\,{\sc iii}] and [Fe\,{\sc vii}] emitting  clouds are co-spatial.  
Fig.~\ref{fig:temperaturas} shows the temperature distribution derived from the [\ion{S}{iii}] line ratio. 
In the region where [\ion{S}{iii}]~$\lambda$6012 is detected, all the  sulphur lines  display a 
double-peaked profile.
 Thus, we determined $T_{\rm e}$ for the blue and red components separately. The values found for the blue 
component is nearly twice as high as those of the red component. In particular, there is a hot spot where $T_{\rm e}$ 
reaches 3$\times 10^4$~K while in most other regions $T_{\rm e}$ is $\sim$ 1.5$\times$10$^4$~K.  
In some spaxels, Equation~\ref{eq2}  converges to  non-physical values. In these cases, the values were  masked from 
Fig.~\ref{fig:temperaturas}. 
     
\begin{equation}
    [\ion{S}{iii}]\frac{j_{\lambda9532} + j_{\lambda9069}}{j_{\lambda6312}} = 
    \frac{5.44 \times exp (2.28 \times 10^{4}/T)}{1 + 3.5 \times 10^{-4} \times n_{e}/T^{1/2}} 
	\label{eq2}
\end{equation}

 \citet{ardila_2020} proposed a  scenario suitable  to explain the gas morphology seen in 
Fig.~\ref{fig:temperaturas} consisting in  an expanding bubble produced by the passage of a radio jet. 
For both [S\,{\sc iii}] and [Fe\,{\sc vii}] $\lambda6087$ lines, the flux distribution are similar, most likely 
co-spatial. 
This can be noticed by the excellent correspondence between high temperature and high ionisation line represented 
by the [Fe\,{\sc vii}] $\lambda6087$ emission.

We  obtain \ne in each spaxel from the  [S\,{\sc ii}]  $\lambda$6716 / $\lambda$6732 
line ratio along with Equation~\ref{eq:densidade}  assuming \Te = 1.5~$\times~10^4$~K. This value is 
consistent with the $T_{\rm e}$ previously determined in Circinus. Equation~\ref{eq:densidade} was derived by  
\citet{pro_2014} and represents a correction to the one found by \citet{osterbrock_2006}. For the density determination 
we  followed two approaches. First, we used the sum of the blue and red components because at some locations the lines 
were double-peaked while in others a single component was found. Second, we determined the density for each component 
separately for the gas located in the NW region where the sulphur lines are double-peaked.

 \begin{equation}
\begin{array}{l}
log(n_{e}[cm^{-3}]) = 0.0543  \tan{( - 3.0553 R + 2.8506)} \\
+ 6.98 -  10.6905 R  +  9.9186 R^2  -  3.5442 R^3 \\
\\
\end{array} 
\label{eq:densidade}
\end{equation} 
 where $R$ is the line flux ratio [S\,{\sc ii}]~$\lambda6716/\lambda$6732.
 
\begin{figure} 
\includegraphics[width=\columnwidth]{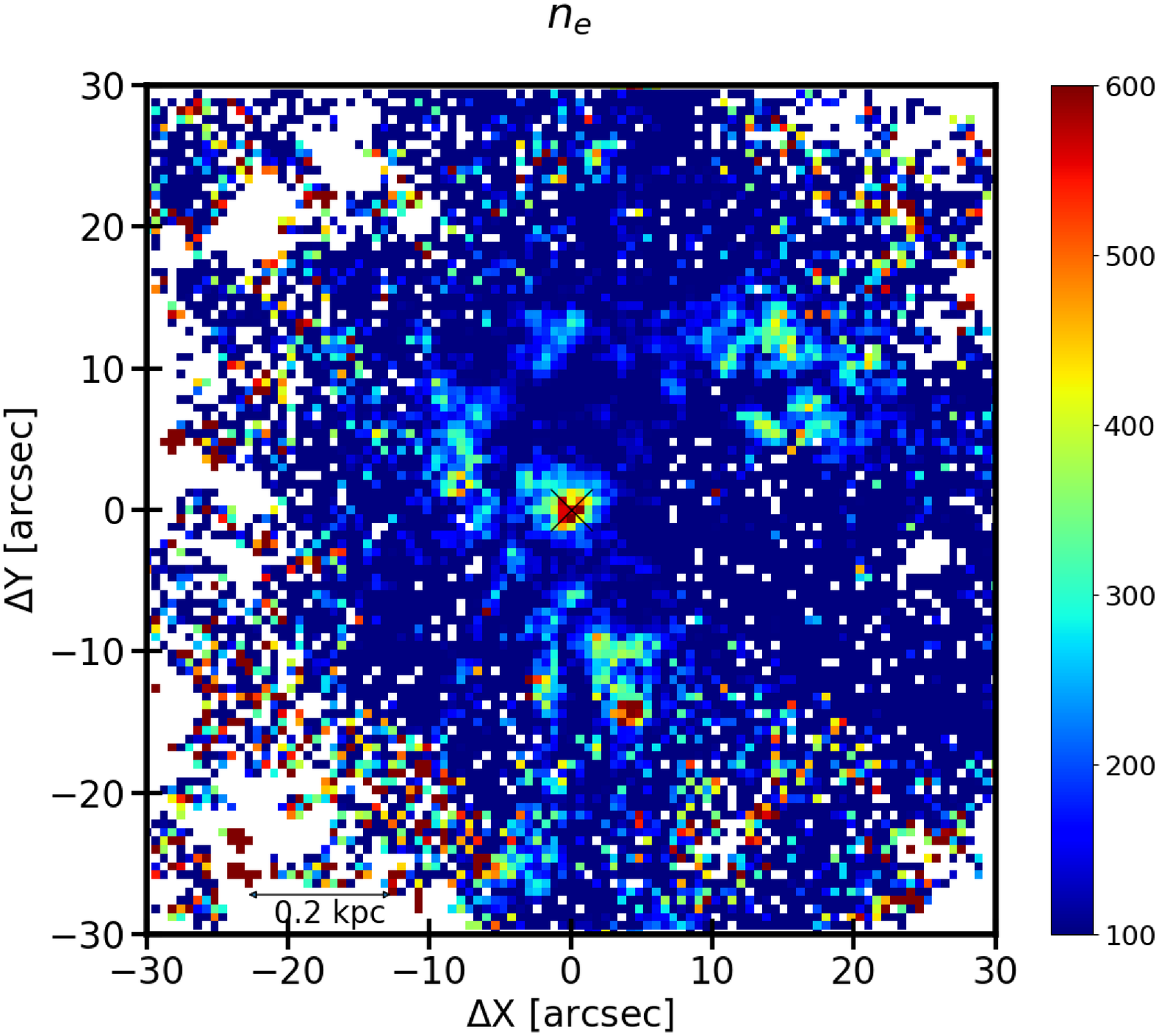}
\includegraphics[width=\columnwidth]{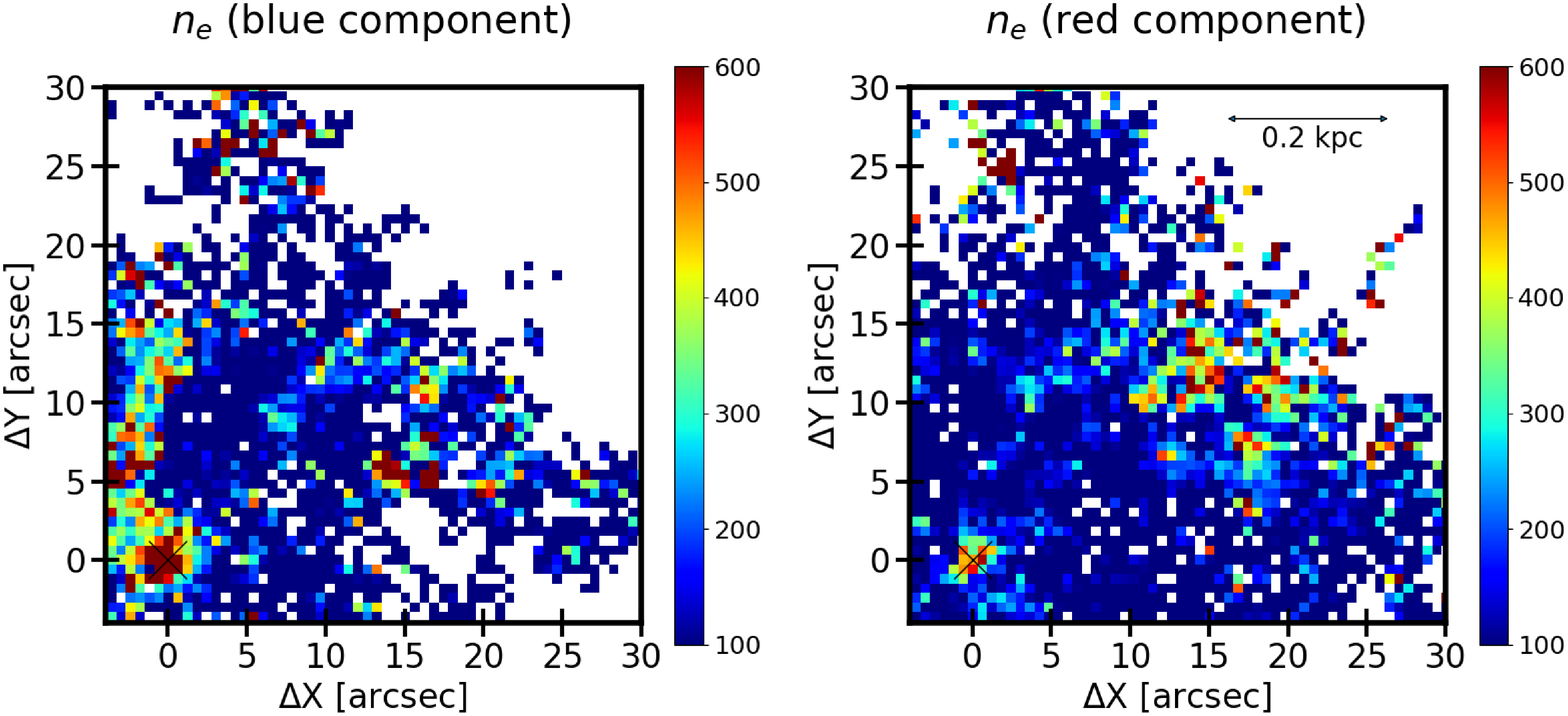}
\caption{Density maps (in \cm3)  obtained from the [\ion{S}{ii}] lines. 
%\textcolor{blue}. 
Top panel: density determined using the sum of the blue and red components of the lines.
Bottom panels:  $n_{\rm e}$ found for the blue (left) and red component (right) fit to each of the two sulphur lines.}
 \label{fig:densidade}
\end{figure}

Fig. \ref{fig:densidade} shows the electron  density in Circinus derived from the sum of the blue and red components of 
the [\ion{S}{ii}] lines (top panel) and the density determined separately for the blue and red components 
(bottom panels). Notice that in this calculation, only sulphur lines with $S/N~>~5$ were employed. The results show two main regions of high density, one at the location of the AGN, and the other 
in the NW side of ionisation cone, the same region where the [\ion{Fe}{vii}]~$\lambda$6087 gas is observed. A gas density of
$n_{\rm e}$ $>$ 400~cm$^{- 3}$ is found in that region. Interestingly, it coincides with the region 
of the highest temperature derived previously, which could be interpreted as another indication of a shocked gas.

%##########################################################
%############# CINEMATICA DO GAS  #######################
%##########################################################

\section{Stellar and Ionised gas kinematics}\label{cap:cinetica}

The study of the gas kinematics in a galaxy is an important tool to separate the contribution of the main stellar disc
component from  other contributions such as a bar, an ionisation cone or a circumnuclear disc. 
It also allows to identify inflowing/outflowing components and their relationship with a possible radio jet. 
In the case of the Circinus galaxy, disentangling the different kinematic components is fundamental because of the 
strong contribution of the host galaxy. To isolate the different components, we will first 
determine the stellar kinematics and afterwards, the gas kinematics.

\subsection{Stellar kinematics}
\label{stellarkin}

We modelled the kinematics of the Circinus host galaxy using a rotation model given by  Equation \ref{eq:rotacao}. 
It was proposed by \citet{bertola_1991} in order to fit the line of sight velocity of a rotation disc.  
To derive the parameters  which best represent the stellar component,  the \ion{Ca}{ii}$\lambda8662$ absorption line 
was employed.

 \begin{equation}
\begin{array}{l}
v_{mod}(R,\Psi ) = v_{sys}  \\
\\
+ \frac{A R \cos{(\Psi - \Psi_{0})} \sin{\theta} \cos^{p}{\theta}   }   {\left \{ {R^{2} [ \sin^{2}({\Psi - \Psi_{0}) + \cos^{2}{\theta} \cos^2{(\Psi - \Psi_{0})}] + c^{2}_{0}\cos^{2}{\theta} }  } \right \}^{p/2}}\\
\\
\end{array} 
\label{eq:rotacao}
\end{equation}

In Equation \ref{eq:rotacao}  $v_{mod}$  is the line of sight velocity, $v_{sys}$ is the systemic velocity of the galaxy,
 $A$ is the amplitude of the velocity, $\Psi_{0}$ is the angular position of the nodes, $i$ is the slope of the disk in 
relation to the sky plane, $c_0$ is the concentration parameter defined as the radius where the curve has 70 percent of 
the maximum rotational velocity and $p$ measures the flatness of a curve of the galaxy. This last parameter can be between 1 and 1.5 for 
external regions of the galaxy. When $p =$ 1 the rotation curve  at large radii is asymptotically flat and for $p =$ 1.5 
it has a finite mass. $R$ is the radial distance to the nucleus and $\Psi$ is the angle where the radius $R$ is measured 
in the sky plane.
 
\begin{figure}
\includegraphics[width=\columnwidth]{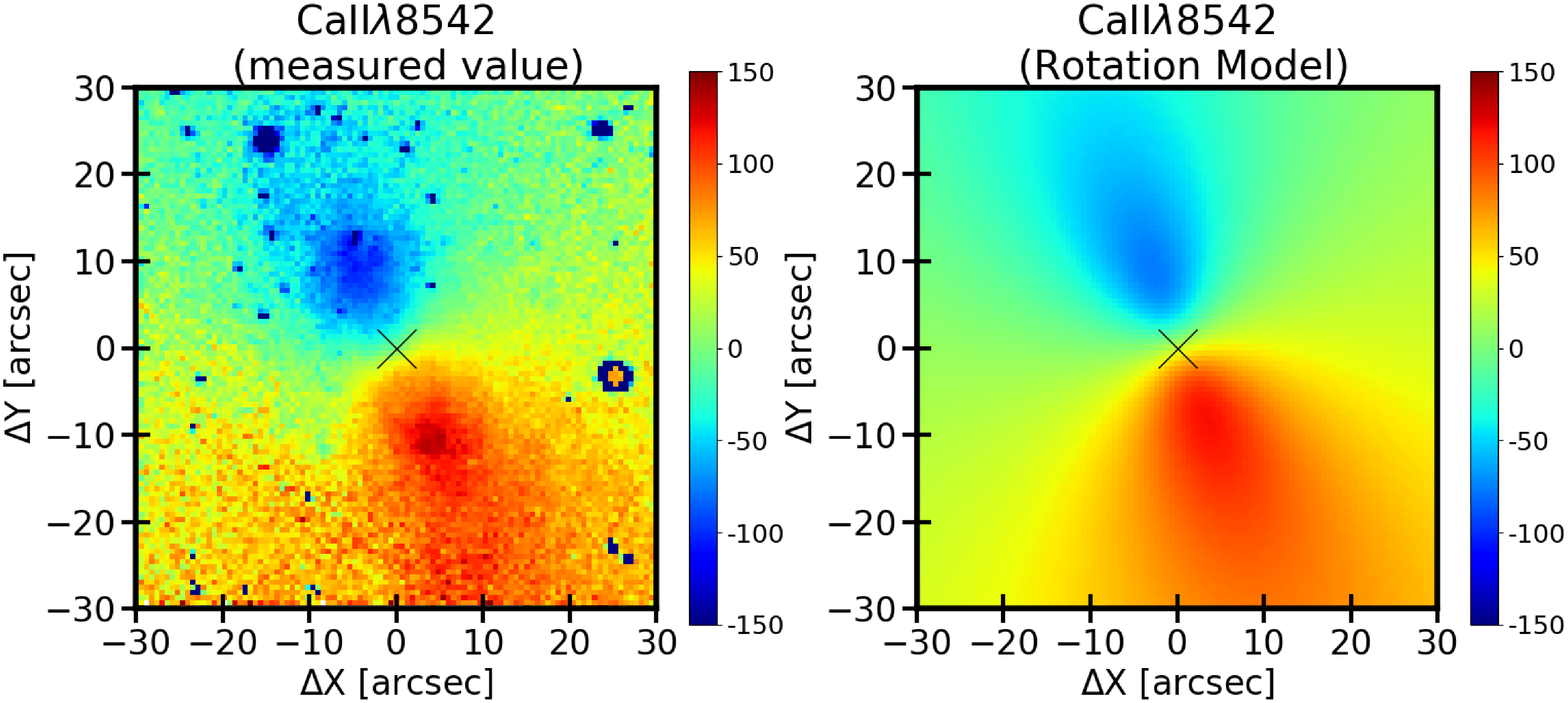}
\includegraphics[width=\columnwidth]{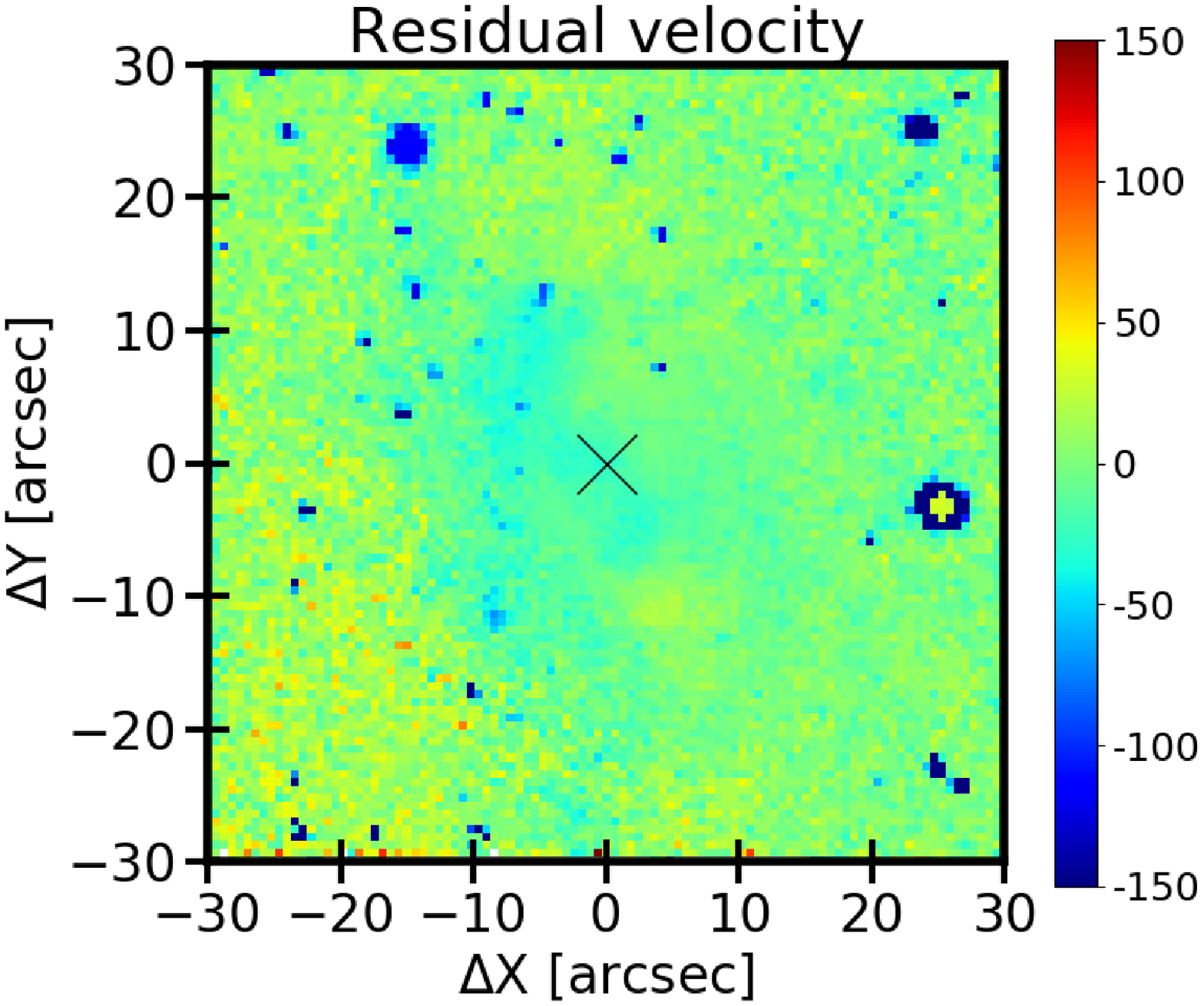}
	\caption{Position-velocity diagram measured for the  \ion{Ca}{ii} line  (top left). The best fit model is in the top right panel. Residuals, after subtraction of the best fit model to the data is in the bottom panel. The colour bar is in units of~km\,s$^{-1}$.}
 \label{fig:velocidades}
\end{figure}

The observed position-velocity map derived for the \ion{Ca}{ii} absorption line is shown
in the top left panel of Fig.~\ref{fig:velocidades}. 
The  top right panel of that figure presents the best model fit  by Equation \ref{eq:rotacao}. In this process,
 the following parameters were obtained: $v_{\rm sys} = 21.2 \pm 0.2$~ km\,s$^{-1}$  (the actual adjusted systemic 
velocity is 455.3~km\,s$^{-1}$  and a systemic velocity correction of 
434.1 ~km\,s$^{-1}$ \citet{meyer_2004} was adopted), $a = 715 \pm 50  $~km\,s$^{-1}$, 
$\Psi_{0} = 289.6^{\circ} \pm 0.2$, $i$ = $50.3^{\circ} \pm 0.5^{\circ}$, $c_{0}= 7.1 '' \pm 0.2$ and $p = 1.61 \pm 0.02$.
 The bottom panel of Fig.~\ref{fig:velocidades} displays the residuals obtained after subtraction from the observed position-velocity diagram the best fit model. In general, the residuals are very close to zero everywhere but at some 
positions (mostly to the SW), where residual velocities  smaller than 50~km\,s$^{-1}$ are found. Overall, our results imply that the stellar 
component is best described in terms of a stellar disc,  that we interpreted  as the galaxy disc. 

\subsection{Emission gas kinematics}

\begin{figure}
\includegraphics[width=\columnwidth]{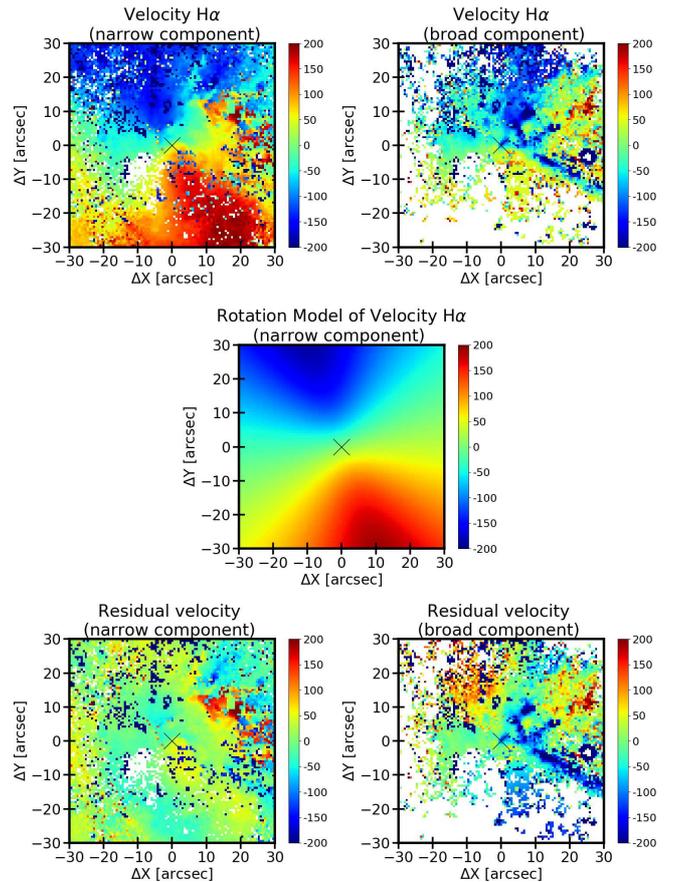}
	\caption{
	 Position-velocity map for the narrow (top left) and broad (top right) components of H$\alpha$. The best model for the narrow component is shown in the central panel. The two bottom panels show the residuals left after subtraction of the best fit model to the observed data for narrow  (left) and broad (right) components. The colour bar is in units of ~km\,s$^{-1}$. }
 \label{fig:velo_alpha}
\end{figure}

The rotation model expressed by Equation~\ref{eq:rotacao} was also applied to the emission-line gas using the narrow component of H$\alpha$. This choice is because we associated it mostly to the rotating gas disc. The second component fit to the H$\alpha$ line is usually broader and likely associated to outflowing gas. 
In Fig. \ref{fig:velo_alpha}, the observed position-velocity (PV) maps derived for the narrow (left) and broad (right) components of H$\alpha$ are shown in the two top panels. The central panel in the second row presents the best fit to the PV map of the H$\alpha$ narrow component. The following parameters were 
obtained:  $v_{sys} = 0.6 \pm 0.7 $~km\,s$^{-1}$  (the actual fit systemic velocity is 434.7~km\,s$^{-1}$), 
$a = 246.2 \pm 4.5  $~km\,s$^{-1}$, $\Psi_{0} = 287.6^{\circ} \pm 0.4$, $i$ = $60.8^{\circ} \pm 1.2^{\circ}$ 
and $c_{0}= 17.3 '' \pm 1.0$. We fixed  $p$ = 1.0. The bottom panels of Fig. \ref{fig:velo_alpha} show the residuals 
obtained after subtracting the best model to the observed PV diagrams of the narrow and broad components. It can be seen that the narrow component of H$\alpha$ suitably describes the ionised gas rotation in the the galaxy disc. Departures from the rotation model are observed in the Northwest region, with velocity values in excess of up to 200~km\,s$^{-1}$. We attribute this discrepancy to the fact that part of the gas in this region is out of the galaxy plane, in the ionisation cone, so that the observed emission is a contribution of both the galaxy disc and outflowing material. As expected, the broad component, which best describes the ionised gas outflow, has residuals of order of 100~\kms\, in absolute value in most locations where it is observed. 
 
\citet{elmouttie_1998}  previously studied the ionised gas in Circinus obtaining a rotation axis of $292^{\circ}$, 
an inclination angle of $40^{\circ} \pm 10^{\circ}$, and a rotation velocity of 350 ~km\,s$^{-1}$. 
It can be seen that the most important parameters agree very well within errors. The difference  between the value of the  inclination angle obtained by  
\citet{elmouttie_1998} data
and ours is mainly due to the size of the field (in their work it is about  30''\, $\times$ 30'' arcsec).  
They also show that the host galaxy is moderately inclined relative to the observer.  Due to its proximity, 
the host should appear in foreground to the gas emission.

\subsection{Channel maps for H$\alpha$ and [O\,{\sc iii}] lines}\label{sec:out2}

\begin{figure}
\includegraphics[width=\columnwidth]{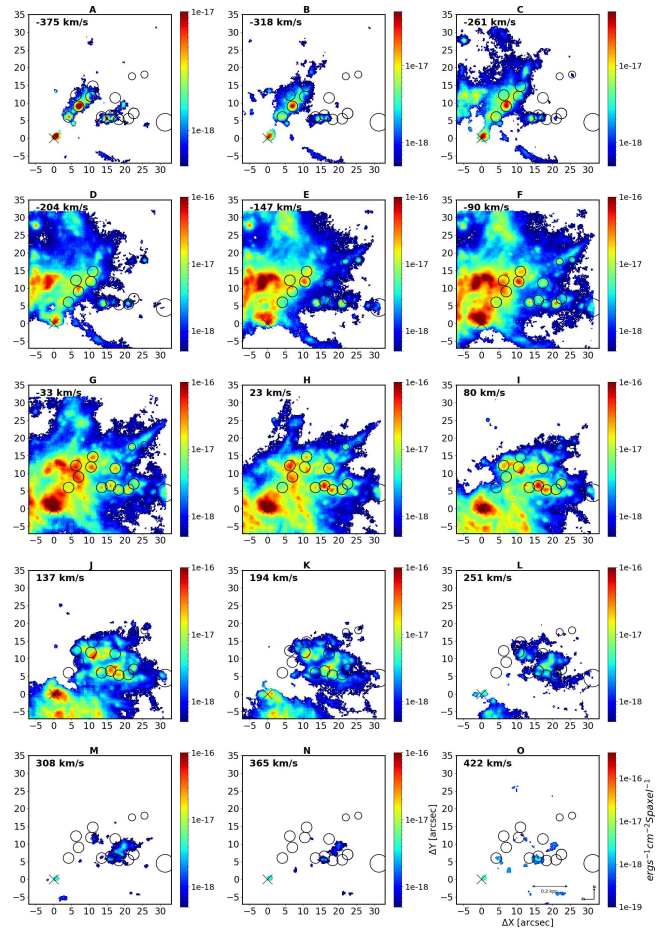}
\caption{Channel maps along the  H$\alpha$ emission-line profile, in order of increasing velocities, shown in the upper left corner of each panel. The black, open circles, mark the position of knots of emission, identified in the maps. The flux is indicated by the colour bar, in units of erg\,s$^{-1}$\,cm$^{-2}$\,spaxel$^{-1}$.}
\label{fig:channel_maps_h}
\end{figure}

\begin{figure}
\includegraphics[width=\columnwidth]{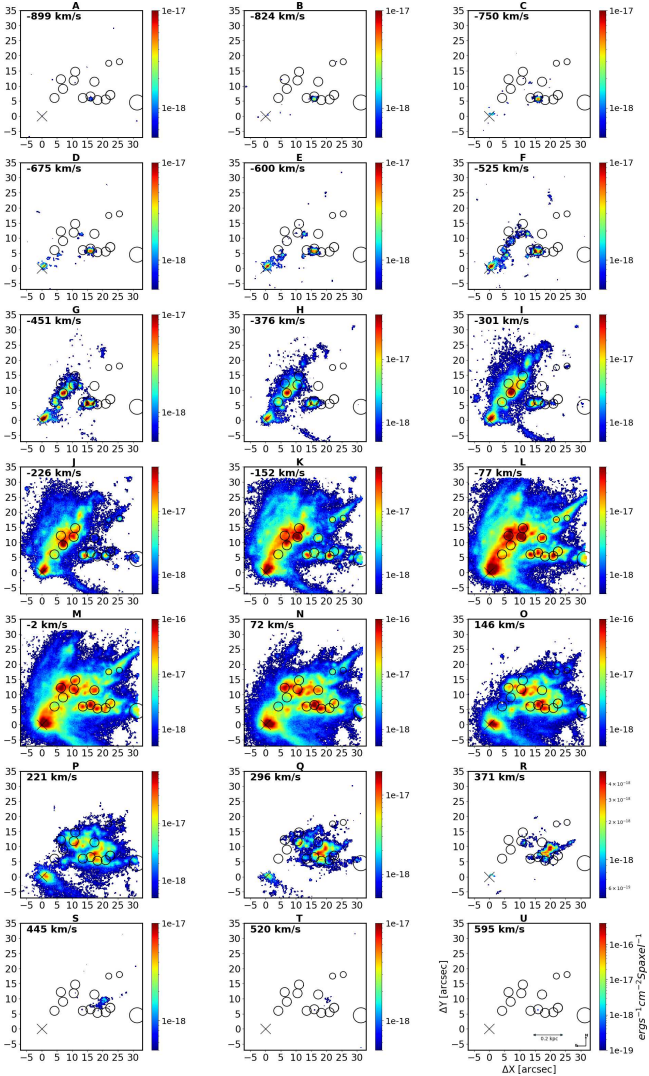}
\caption{Same as Figure~\ref{fig:channel_maps_h} but now for the [O\,{\sc iii}] emission-line profile.}
\label{fig:channel_maps_o3}
\end{figure}

We have also studied the gas kinematics using channel maps extracted  from the H$\alpha$ and [O\,{\sc iii}] emission line 
profiles. To this aim we use the MUSE cube with the original binning in order to preserve the best possible angular 
sampling (spaxel size of 0.2"). The procedures described in Sect.~\ref{cap:metodos} for processing the binned cube 
(i.e. stellar continuum subtraction and extinction correction) were also applied to the original cube.   
Figs.~\ref{fig:channel_maps_h} and~\ref{fig:channel_maps_o3} show channel maps for the H$\alpha$ and [O\,{\sc iii}], 
respectively, both in bins of  $\Delta \lambda$ = 1.25 \AA. Regions where the measured flux is less than 
$1.5 \times{\sigma}_{sky}$  have been removed from the maps and appear in white.  ${\sigma}_{sky}$ is the 
standard deviation of the sky noise.

We found that for H$\alpha$ the most extended emission features within the NW ionisation cone display velocities 
between -375 km\,s$^{-1}$ and 422 km\,s$^{-1}$. Moreover, filaments and several emission knots associated to 
them are evident in the channel maps. The most conspicuous filament is  seen in the velocity 
channels -204~km\,s$^{-1}$, -147~km\,s$^{-1}$ and -90~km\,s$^{-1}$. It runs from the AGN to the SW and it is usually 
identified with the Southern edge of the ionisation cone.  There is also a  filament in the velocity 
channels -33~km\,s$^{-1}$ and 23~km\,s$^{-1}$,  which seems to arise in the galaxy nucleus and it is projected towards the 
NW at a PA of $\sim305^{\rm o}$. 
We have clearly identified 14 knots of  H$\alpha$ emission in the channel maps, marked with black circles in both figures.
 These knots are better  seen in the velocity channels -90~km\,s$^{-1}$, -33~km\,s$^{-1}$, 23~km\,s$^{-1}$ and 
80~km\,s$^{-1}$, although some of them appear in the channel maps corresponding to velocities of -375~km\,s$^{-1}$ 
and 422~km\,s$^{-1}$.
The same knots are also detected in [O\,{\sc iii}], as can be seen in Fig. \ref{fig:channel_maps_o3}. Moreover, they 
show up with larger projected blueshifted velocities, being detected at -675~km\,s$^{-1}$ relative to the systemic 
velocity. The connection with the filaments  is even more evident here, mainly at the centre and at the edges of the cone.
 The filamentary structure in the ionisation cone already noticed  in H$\alpha$ is enhanced in [O\,{\sc iii}], with 
at least six filaments of emission detected in the channel maps, in particular  for  velocities corresponding 
to -2~km\,s$^{-1}$ and 71~km\,s$^{-1}$.  
Some of the filaments and knots we found in Circinus were  already  presented by \citet{veilleux_1997}. These authors 
report that one of these filaments, the one that runs from the nucleus to the West, extends up to 40'' ($\sim$~800 pc). 
Fig.~\ref{fig:channel_maps_o3}  shows that  this same filament is detected here, up to the West edge of the MUSE 
detector.  Moreover, the filament  closely aligned  with the radio jet axis is  observed up to a distance 
of $\sim$1~kpc from the AGN. 

The excellent sensitivity achieved with MUSE also allows us to detect a larger number of knots, called ``bullets'' 
by \citet{veilleux_1997}. They identified four main bullets at $\sim$11", 18", 30" and 37" from the AGN with a PA 
$\sim$270$^{\rm o}$ from the nucleus. In addition, we identified other systems of bullets at distances 
smaller than 10" from the AGN.  
\citet{veilleux_1997} argue that the motions observed across the ionisation cone are highly supersonic, therefore
 high-velocity (V$_{\rm s}>100$~km\,s$^{-1}$) shocks are likely to contribute to the ionisation of the line-emitting 
gas. Additional support for this scenario comes from the detection  of high-ionisation emission on large scales
 filling up the region where most of the bullets are located.  The velocity field of the knots and their strong 
association with the filaments suggest that they represent material either expelled  from the nucleus or dragged by 
a wide-angle wind aligned to the polar axis of the galaxy.  
The current radio data available for Circinus support that the
wide-angle wind is jet-driven. Indeed, the northwest ionisation cone appears to have a radio counterpart at both 13 
and 20~cm \citep[figs. 5 and 6 of][]{el_1998b}. This conclusion is drawn from the comparison between  
Figs.~\ref{fig:channel_maps_h}  and~\ref{fig:channel_maps_o3} and their radio maps.

%##########################################################
%################# MAPA DE MOMENTOS #######################
%##########################################################

\subsection{Moment maps for the low and medium ionisation lines}

\begin{figure*}
 \includegraphics[width=0.70\textwidth]{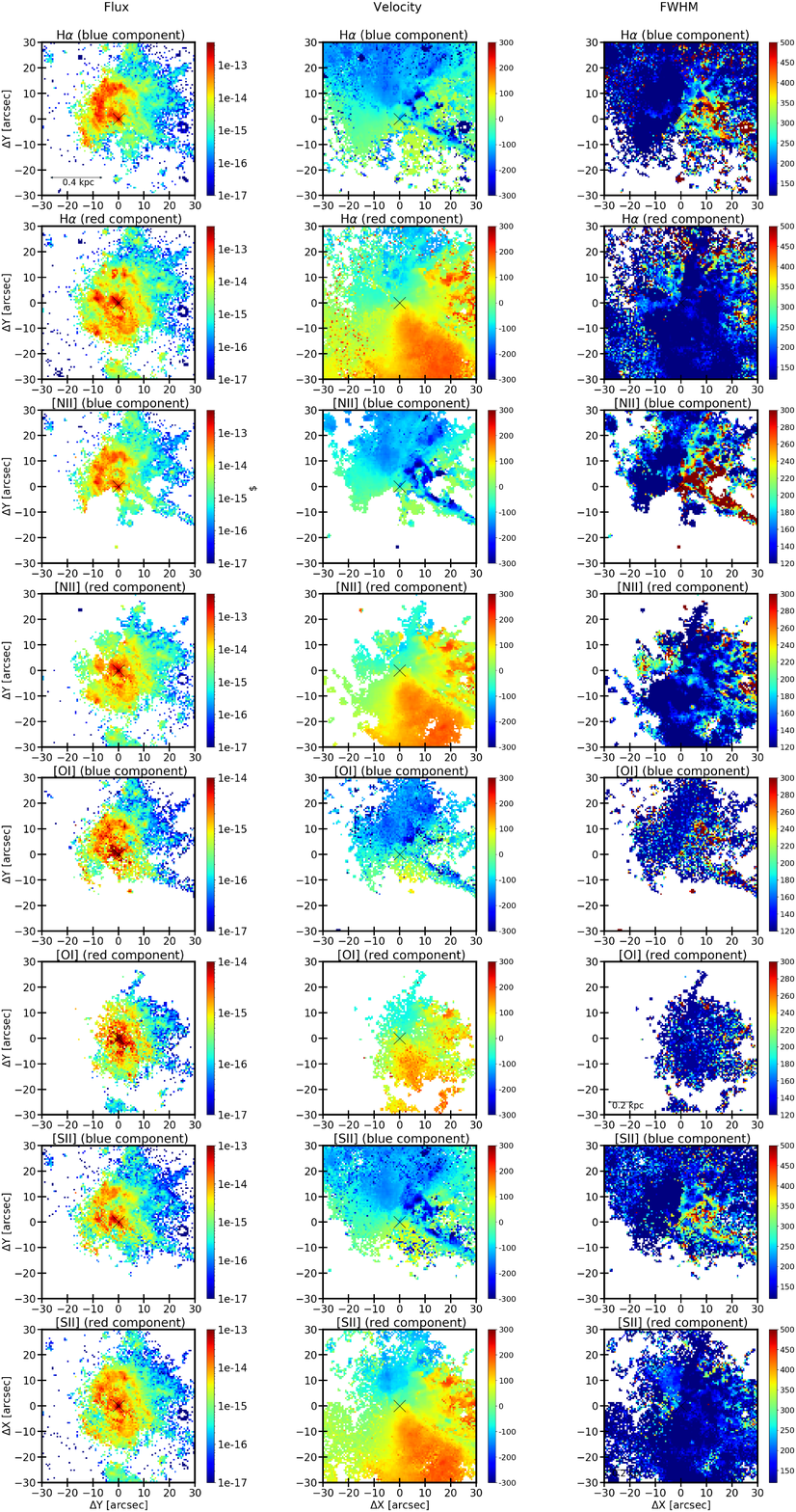}%[width=2.2\textwidth]
    \caption{Moment maps for the low ionisation lines H$\alpha$, [O\,{\sc i}]$\lambda6300$, [S\,{\sc ii}]$\lambda6716$ 
and  [N\,{\sc ii}]$\lambda6583$. Flux scale is in units of  erg\,s$^{-1}$\,cm$^{-2}$\,Spaxel$^{-1}$. Velocity and FWHM are in units of km\,s$^{-1}$.}
    \label{fig:momentos_baixa}
\end{figure*}

In order to further characterise the kinematics of the ionised gas in Circinus, we constructed moment maps (flux, velocity, FWHM)
 for the most prominent emission lines detected in the galaxy. Fig. \ref{fig:momentos_baixa} displays the results for  
H$\alpha $, [O\,{\sc i}] $\lambda6300$, [S\,{\sc ii}] $\lambda6716$ and [N\,{\sc ii}] $\lambda6583$. At some positions (i.e. within the ionisation cone), 
each of the above lines were fit using two Gaussian components  while at others, one component was enough to fully represent the observed profile. For the sake of clarity, we present maps for the two components, separately. When only 
one component is observed, the line appears either as a blue or as a red component. 

\begin{figure*}
   \includegraphics[width=0.67\textwidth]{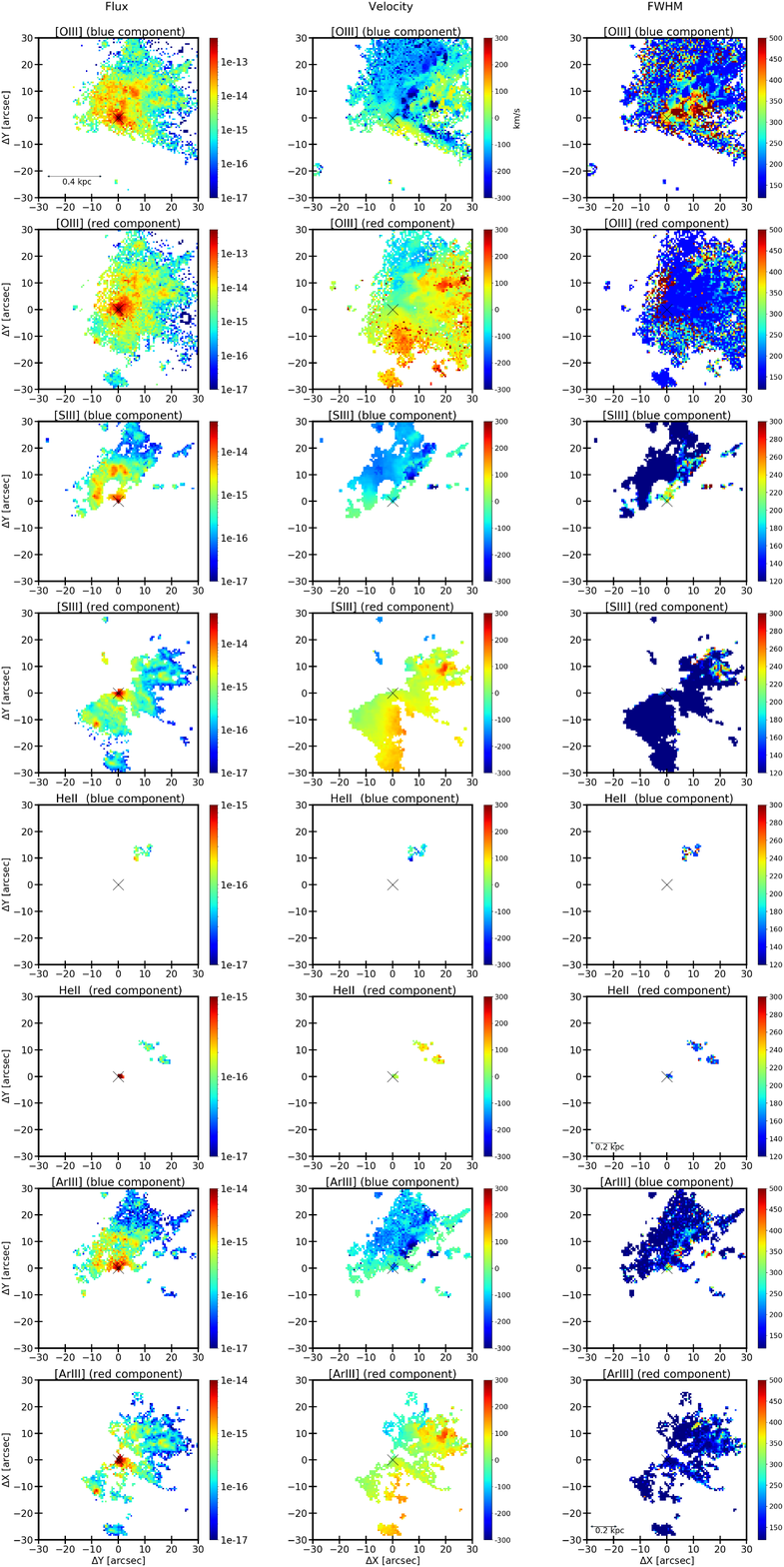}%[width=2.2\textwidth]
    \caption{Moment maps for mid-ionisation lines: [O\,{\sc iii}]$\lambda5007$,  [S\,{\sc iii}]$\lambda9069$,  
[HeII]$\lambda5412$ and  [Ar\,{\sc iii}]$\lambda7136$. Flux scale is in units of  erg\,s$^{-1}$\,cm$^{-2}$\,Spaxel$^{-1}$. Velocity and FWHM are in units of km\,s$^{-1}$. }
    \label{fig:momentos_media}
\end{figure*}

The flux maps displayed in the first column of Fig. \ref{fig:momentos_baixa} show that the low-ionisation lines are 
particularly enhanced in the nuclear and circumnuclear region of the galaxy (central 10''~$\times$ 10'' around the AGN).
 They nicely trace the prominent ring of star formation that characterises Circinus, previously reported and studied by 
other authors \citep{marconi_1994,curran_1998}. Moreover, the low-ionisation gas is also clearly extended towards the NW,
 within the ionisation cone.  It fills all the FoV covered by the detector
 both towards the North and West sides of the galaxy. 
 Some extended emission is also detected  towards the South but it is not possible to state so far  whether it is located
 in the counter-cone or in the galaxy disc. Anyway, the emission line flux in the cone and Southern part is weak, 
nearly two to three orders of magnitude fainter than in the central regions of the galaxy.  
From the position-velocity diagrams in Fig.~\ref{fig:momentos_baixa} (central panels) we found that in most parts of 
the galaxy, the low-ionisation gas is in rotation, similar to what is observed in H$\alpha$,  except in the 
NW side, where the gas is largely  out of the galaxy plane. This region spatially coincides with the high-ionisation 
component of the outflow.  A long radial filament of gas,  with velocities between -200~km\,s$^{-1}$ 
and -100~km\,s$^{-1}$, is also evident towards the SW, mainly from  the blue component of the lines. 
We associate this feature to the 
SW edge of the ionisation cone, already noticed by \citet{mingozzi_2019}. This filament is also characterised by a 
larger velocity dispersion, particularly seen in the [N\,{\sc ii}] and [S\,{\sc ii}] lines. The FWHM  reaches values 
larger than 300~km\,s$^{-1}$, similar to the widths  found in the central part of the ionisation cone. 
Fig. \ref{fig:momentos_media} shows moment maps for the blue and red components of the mid-ionisation lines 
 [O\,{\sc iii}] $\lambda5007$, He\,{\sc ii} $\lambda5412$, [Ar\,{\sc iii}] $\lambda7136$ and [S\,{\sc iii}] $\lambda9069$.
 As for the low-ionisation emission discussed above, the prominent starburst ring is also enhanced by these lines  
 except He\,{\sc ii}. 
This latter emission is only detected at the AGN position and in the highest ionised portion of the NW cone.    
 
There is also some  evidence  that part of the gas emitting  [O\,{\sc iii}], [Ar\,{\sc iii}] and 
[S\,{\sc iii}] is located in  the galaxy disc. This result is  gathered  from the detection of blue- and red-shifted 
emission following the rotation pattern found in H$\alpha$. Moreover, towards the Northwest, the outflow component is 
evident and highlighted by the mid-ionisation emission. 
A wide range of velocities is also detected by the mid-ionisation gas (see central panel of Figure \ref{fig:velocidades})
which displays a larger width to the Northwest (FWHM~$>$~ 300~km\,s$^{-1}$) and gas velocities  which strongly deviate
from rotation, confirming that most of the gas in the ionisation cone is out of the galaxy plane.

\subsection{Moment maps for high ionisation lines}

\begin{figure*}
\includegraphics[width=0.85\textwidth]{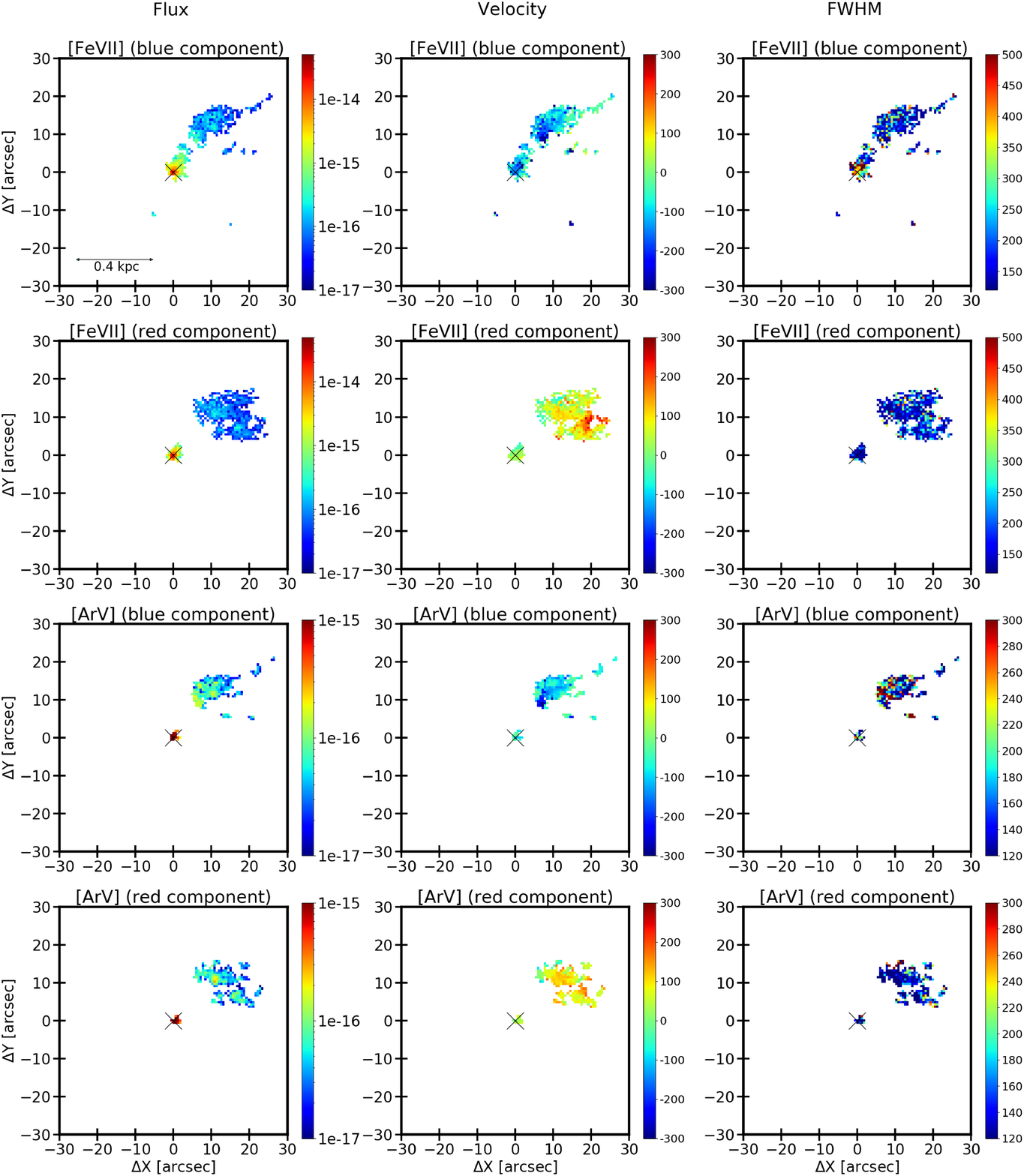}%[width=2.2\textwidth]
 \caption{Moment maps for the high ionisation lines [Fe\,{\sc vii}]$\lambda6087$ and 
[Ar\,{\sc v}]$\lambda7006$. Flux scale is in units of  erg\,s$^{-1}$\,cm$^{-2}$\,Spaxel$^{-1}$. Velocity and FWHM are in units of km\,s$^{-1}$.}
\label{fig:momentos_alta1}
\end{figure*}

\begin{figure*}
 \includegraphics[width=0.85\textwidth]{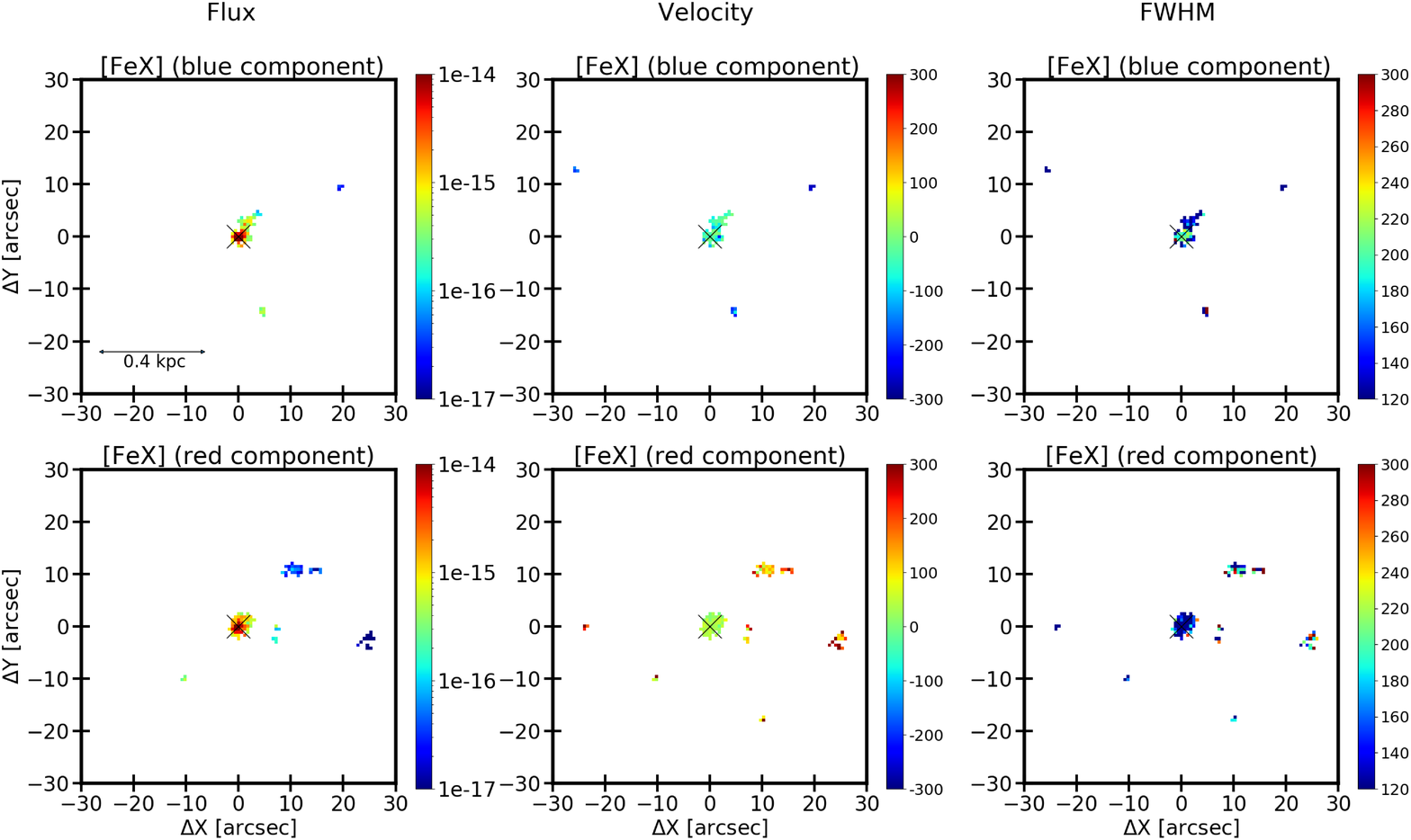}%[width=2.2\textwidth]
    \caption{Moment maps for the high ionisation line  [Fe\,{\sc x}]$\lambda6375$. Flux scale is in units of  erg\,s$^{-1}$\,cm$^{-2}$\,Spaxel$^{-1}$. Velocity and FWHM are in units of km\,s$^{-1}$. }
    \label{fig:momentos_alta2}
\end{figure*}

We also study the kinematics of the most important high ionisation lines detected in Circinus in the wavelength range 
covered by the MUSE data (3750 to 9300~\AA). In Figure \ref{fig:momentos_alta1} we present the moment maps of the blue 
and red components of [Fe\,{\sc vii}]~$\lambda6087$ and [Ar\,{\sc v}]~$\lambda7006$ and in 
Figure~\ref{fig:momentos_alta2} the moments for [Fe\,{\sc x}]~$\lambda6375$. 
It is evident from Figures~\ref{fig:momentos_alta1} and~\ref{fig:momentos_alta2} that, in contrast to the low- and 
mid-ionisation lines, coronal lines although extended are observed only in two regions:  in the nucleus and in the 
highest ionised portion of the NW cone. Indeed, no evidence of such lines are found in the cone edges, as seen in e.g. 
H$\alpha$ or [O\,{\sc iii}]. Notice that [Fe\,{\sc x}] is also detected in the cone but at two very specific locations. 
In none of the three coronal lines considered here evidence of gas rotation is observed, confirming previous results 
derived from the low- and mid-ionisation lines, that the bulk of the ionised gas in the cone is out of the 
galaxy plane. The detection of split line profiles at  most positions in the central part of the ionisation cone further 
supports this scenario. The data suggest an expanding bubble of gas with the approaching and receding maximum 
velocities reaching nearly -300~km\,s$^{-1}$ and 200~km\,s$^{-1}$, respectively.

The [Ar\,{\sc v}] emission (Figure \ref{fig:momentos_alta1}) shows a flux distribution very similar to that of 
[Fe\,{\sc vii}] but restricted to a more compact region. The [Fe\,{\sc x}] line (see Figure \ref{fig:momentos_alta2}) 
was detected in the nucleus and up to $\sim$40~pc towards the NW. Moreover, to the best of our knowledge, we detected, 
for the first time in the literature, [Fe\,{\sc x}] emission at a distance of 340~pc from the AGN. The emission is 
produced by a cloud with a diameter of ~60~pc, best traced by the red component of the [Fe\,{\sc x}] line profile. 
It agrees in velocity, FWHM and spatial position with the most intense emission detected in [Fe\,{\sc vii}].

\subsection{A 3D kinematic model for Circinus}

\begin{figure}
\includegraphics[width=\columnwidth]{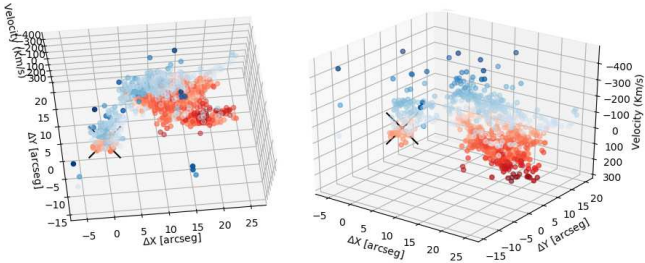}
\caption{Velocity maps of  the [Fe\,{\sc vii}]$\lambda$6087 line  (left) on the XY plane (similar to Figure 
\ref{fig:momentos_alta1}) and  (right) by a different angle of view. The centre of the galaxy is represented in black.}
\label{fig:f7_3d}
\end{figure}

\begin{figure}
\includegraphics[width=\columnwidth]{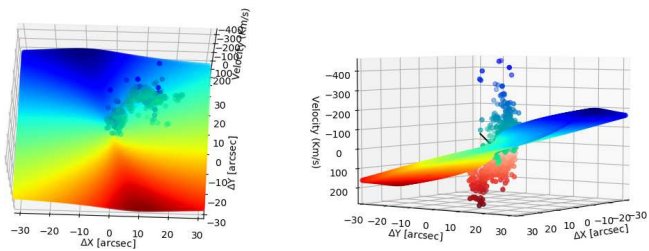}
\caption{
Rotation model in a 3d map already presented in  section \ref{cap:cinetica}.  Left: the rotation model for the 
H$\alpha$ line.  The blue dots indicate the velocity for the [Fe\,{\sc vii}]$\lambda$6087 component.
Right:  rotation model together with the blue and red speed components of [Fe\,{\sc vii}].}
\label{fig:fe7_e_rotacao}
\end{figure}

Two main results  about the kinematics of the coronal gas were derived in previous sections. First, 
the coronal gas is out of the galaxy plane with most of that emission taking place in the central region of the 
ionisation cone. Second, the gas is clearly in expansion with two main components. The northern portion is outflowing 
towards the observer while the bulk of southern component is receding from us. 
3D projection modelling of this scenario is presented in Figure \ref{fig:f7_3d} for the velocity components of 
[\ion{Fe}{vii}]. The left panel depicts the $\Delta$X and $\Delta$Y axes which define the two spatial dimensions. 
The $z$-axis shows the gas velocity component.  Negative values are coloured in blue while positive ones are in red. 
Notice that this model resembles the current gas morphology observed, for instance, in Figure~\ref{fig:momentos_alta1}. 
The panel to the right is equivalent to the left panel after applying a clockwise rotation of $\sim$90$^{\rm o}$ to the 
XY plane and untilting the velocity axis. It can be seen that the gas velocity field suggests an expanding cocoon. 
We speculate that this feature could be formed by the passage of a radio jet which, since then, is expanding freely out 
of the galaxy plane. 
The left panel of Figure \ref{fig:fe7_e_rotacao} shows the rotation model fit to the galaxy disc overlaid upon the 
velocity components of [\ion{Fe}{vii}]. The right panel shows the rotation model derived for H$\alpha$. 
The blue dots indicate negative velocities while the red dots positive velocities. It can be clearly seen that 
the [\ion{Fe}{vii}] gas is detached from the rotation of the galaxy disc. 
 
\subsection{Velocity-velocity dispersion diagram}

\begin{figure}
\includegraphics[width=\columnwidth]{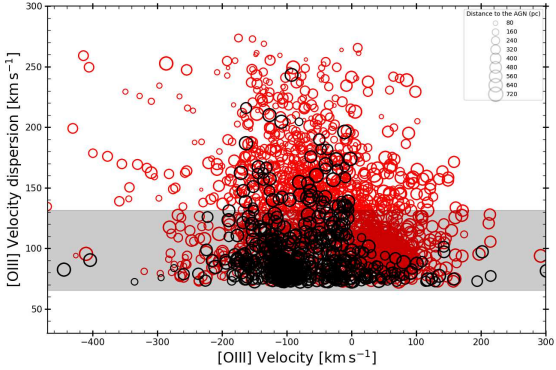}
\caption{VVD diagram of the narrow (black) and broad (red) components of [\ion{O}{iii}] measured in individual spaxels of the ionisation cone of Circinus. The symbol size indicates the radial distance from the AGN (see the box at the upper right corner of the panel). The horizontal shaded area indicates the $\sigma_*$ values measured from the CaT absorption lines (see Sect.~\ref{stellarkin}).}
\label{vvd}
\end{figure}

The gas kinematics can also be studied by means of the so-called velocity versus velocity dispersion (VVD) diagram \citep[i.e.,][]{kar+16,woo+16}. It allows us to investigate if the observed velocity shifts
and velocity dispersion are related. Figure~\ref{vvd} presents
the VVD diagram for the narrow (black circles) and broad (red circles) components
of the [\ion{O}{iii}] line   measured in the ionisation cone of Circinus. The grey horizontal area indicates the stellar velocity dispersion, $\sigma_*$, measured along the galaxy disc by means of the CaT lines. The V$=$0 \kms\ represents the systemic velocity of the galaxy.

Several interesting results can be observed in Figure~\ref{vvd}. First, there is a clear difference between the kinematics of the broad and narrow components. 
The former shows larger velocity dispersion, with values of up to 260~\kms\ and larger velocity offsets, with blueshifts reaching $-$400~\kms. In contrast, in the latter the velocity dispersion is restricted to the $\sigma_*$ values of the galaxy.  Second, the VVD diagram of the narrow component shows two distinct distributions: one clustered at $-$40~\kms\ and another at $-$120~\kms. As all the points shown refer to the NW portion of the ionisation cone, we discard that the separation is due to the galaxy rotation. Moreover, the two clustered velocities are not distributed evenly along V=0~\kms\ but systematically shifted to the blue. We interpret this result in terms of a double-peak structure observed in the [\ion{O}{iii}] lines in the region that is coincident with the [\ion{Fe}{vii}] emission. We also discard that the low velocity dispersion is due to the fact that the gas is along the galaxy disc. Indeed, it is a physically connected to the region that goes along the radio-jet, which is nearly perpendicular to the galaxy disc.

We also observe a negative radial trend for the velocity and the velocity dispersion of the broad component. As pointed out by \citet{kar+16},  the negative radial trend of the broad [\ion{O}{iii}] component indicates
non-gravitational effects such as outflows, whose velocity
decreases radially. 
Still, the prevalence of negative velocities in the broad component indicates that we preferentially see the approaching component of the outflow. However, at some positions, we clearly see both the approaching and receding components of the outflow. 

It is also interesting to note that the bulk of the outflow is characterised by relative small values of velocity dispersion ($\sigma~\sim~200$~\kms) with moderate shifts, of up to $-$200~\kms.   Outflowing gas with the above characteristics is also emitting [\ion{Fe}{vii}], with similar line shifts and velocity dispersion. More extreme [\ion{O}{iii}] outflowing gas ($\sigma~>~250$~\kms\ and V$<-200$~\kms) has no counterpart in the coronal gas. This is probably due to S/N issues, as the [\ion{Fe}{vii}] line is about an order of magnitude weaker relative to [\ion{O}{iii}].

%This is likely related to the fact that low-luminosity AGNs tend to have small velocity shifts and dispersion \citep{woo+16}. However, the 

We interpret the above results in terms of the launching point of the outflow: the highest ionised portion of the outflow in Circinus is not launched at the AGN position but locally, likely inflated by the passage of the radio-jet. Another alternative is that it is produced by a stellar wind. However, the global star-formation rate in Circinus seems to be moderate \citep[3 to 8~M$\odot$~yr$^{-1}$,][]{for_2012} to produce stellar outflows extending over scales of hundred of parsecs. 

In summary, the data presented in this section point out to the presence of an ionised outflow in Circinus, with the NW side out of the galaxy plane and dominated by the approaching component. Part of the outflow is likely originated by a wind launched by the AGN. However, the highest-ionised portion of the outflow is most likely produced by a bubble of ionised gas inflated by the passage of the radio-jet.

\section{Determination of the mass outflow rate}\label{sec:outflow}

Determinations of the mass and outflow rate are traditionally carried out using the [O\,{\sc iii}] line. However, in the case of Circinus, it is 
difficult  to separate the contribution of [O\,{\sc iii}] produced by the disc of the galaxy  from that of the outflow. This difficulty is due to the fact that the galaxy disc appears in background relative the outflowing gas.
\citet {mingozzi_2019} used a velocity criterion of [O\,{\sc iii}], where solely components with values smaller than -150 km\,s$^{-1}$ and larger than 150~km\,s$^{-1}$ (relative  to the systemic velocity of the galaxy) would contribute to the outflow. However, \citet{ardila_2020}  presented the velocity map of the 
[Fe\,{\sc vii}]$\lambda$6087 gas and found that the outflow velocity determined from that line would have values 
-150 km\,s$^{-1} <$ v $< 150$ km\,s$^{-1}$ in most of the region where it is located. Assuming that [Fe\,{\sc vii}]$\lambda$6087 entirely represents the outflowing gas, [O\,{\sc iii}] 
and [Fe\,{\sc vii}] emissions are expected to agree in velocity on each spaxel if the only contribution is the outflow. Therefore, the [O\,{\sc iii}] emission attributed 
solely to the disc component by \citet{mingozzi_2019} should also contribute to the outflow.
Thus, for the calculation of the outflow mass in Circinus, we assume that the contribution of [O\,{\sc iii}] in the 
region co-spatial with [Fe\,{\sc vii}] is dominated by the outflow component.

\subsection{Calculation of outflow in a model for  a biconic structure}\label{outf_1}

We first calculated the outflow rate and the mass of the outflow for the high ionisation region traced by the 
[Fe\,{\sc vii}] line. To this purpose, a mask was initially generated in the region where the high ionisation gas is 
found  allowing to calculate the integrated flux of the [O\,{\sc iii}] line in that region.
For the results obtained here we use the calculations presented by \citet{diaz_2012} given by Equation \ref{eq:massa_outflow}.

\begin{equation}
\begin{array}{l}
M^{out}_{ion}  =  5.3 \times 10^{7}  \frac{ L_{44}([OIII]) }{ n_{e} 10^{ [O/H] } }  (M{\odot}) \\
\\
\end{array} 
\label{eq:massa_outflow}
\end{equation}

where $ L_{44}([OIII])$ is the luminosity of the [\ion{O}{iii}] line which contributes to the outflow
in units of $10^{44}$\,erg\,s$^{-1}$, $n_{\rm e}$ is the gas density in units of $10^{3}$\,cm$^{-3}$, $\nu_{3}$  is the outflow speed in units of $10^{3}$\,km\,s$^{-1}$ and 
$R_{\rm Kpc}$ is the radius, in kpc, of the biconical region  which participates in the outflow.
Thus, from the mask of the [Fe\,{\sc vii}] region we found a [O\,{\sc iii}] flux for a blue and red components of 
$2.7 \times 10^{-12}$ erg\,s$^{-1}$\,cm$^{-2}$\,Spaxel$^{-1}$  and 
$2.4 \times 10^{-12}$ erg\,s$^{-1}$\,cm$^{-2}$ Spaxel$^{-1}$, respectively. The corresponding gas luminosity is 
$1.5 \times 10^{6}~$L$\odot$ (i.e. $5.7\times 10^{39}$~erg\,s$^{-1}$) and $1.3 \times 10^{6}~$L$\odot$ 
(i.e. $5.1\times 10^{39}$~erg\,s$^{-1}$). We thus find an outflow rate for the blue and red components of 0.12 M$\odot$\,yr$^{-1}$ and 
0.10 M$\odot$\,yr$^{-1}$. The mass of the gas involved in the outflow is $10^{4}$ M$\odot $ and 
$9 \times 10^{3}$ M$\odot$, for the blue and red components, respectively.

\subsection{Calculation of outflow rate using the $H{\beta}$ emission}\label{outf_2}

We also estimated the mass and outflow rate  using the flux of H$\beta$  following a method presented by 
\citet{santoro_2018}. The ionised gas mass can be estimated by the Equation \ref{eq:massa2}.

\begin{equation}
\begin{array}{l}
M_{gas}  =  \frac{ L\left ( H_{\beta} \right ) m_{p} }{ n_{e} {\alpha}^{eff}_{H \beta} h \nu_{H_{\beta}}  }
\\
\\
\end{array} 
\label{eq:massa2}
\end{equation} 

Where $L(H\beta)$ represents the luminosity of the H$\beta$ line, $m_{p}$ is the mass of the proton ($1.673 \times 10^{-27}$\,kg), $n_{\rm e}$ is the gas electron density  
($\sim 300$\,cm$^{-3}$), ${\alpha}^{eff}_{H\beta}$ is the effective recombination coefficient for H$\beta$ 
(we use $3.03 \times 10^{-14}$\,cm$^{3}$\,s$^{-1}$, for case B, \citep{osterbrock_2006}),
$h\nu_{H_{\beta}}$ is the frequency of the H$\beta$ line (4861.3 \AA) and  $h$ is Planck constant 
($6,626 \times 10^{-34}$ J.s).
The outflow rate was estimated using  Equation \ref{eq:tx2}. The calculations herewith are presented in detail  by
\citet{rose_2018}.

\begin{equation}
\begin{array}{l}
\dot{M}_{gas}  =  \frac{ L\left ( H_{\beta} \right ) m_{p} \nu_{out} }{ n_{e} {\alpha}^{eff}_{H \beta} h \nu_{H_{\beta}} r }
\\
\\
\end{array} 
\label{eq:tx2}
\end{equation} 

In Equation \ref{eq:tx2} the term $\nu_{out}$ represents the expansion velocity of the outflow. Here, we assume a value of $\sim 750$ km\,s$^{-1}$), derived from the channel maps of [\ion{O}{iii}].
The term $r$ is the gas shell radius of the expanding outflow (200~parsec).
We found a H$\beta$ flux for the blue and red components of $2.9 \times 10^{-13}$~erg\,s$^{-1}$\,cm$^{-2}$ Spaxel$^{-1}
$  and $2.4 \times 10^{-13}$\,erg\,s$^{-1}$\,cm$^{-2}$\, Spaxel$^{-1}$ respectively. In terms of luminosity, these values
translate into $1.6 \times 10^{5}$\,L$\odot$ (i.e. $6.1\times 10^{38}$~ergs\,s$^{-1}$) and $1.4 \times 10^{5}~$L$\odot$ 
(i.e. $5.1\times 10^{38}$~ergs\,s$^{-1}$). Thus, we find an outflow rate for the blue and red components of 
0.05 M$\odot$\,yr$^{-1}$  and 0.04 M$\odot$\,yr$^{-1}$  and a mass of gas of $1.4 \times 10^{4}$ M$\odot$ and 
$1.2 \times 10^{4}$ M$\odot$, respectively.

%##########################################################
%################## EMISSAO DE RAIO-X   ############
%##########################################################

%##########################################################
%################## MODELOS COM SUMA  ############
%##########################################################

\section{Modelling the spectral line ratios}

In the preceding sections, we have accumulated compelling observational evidence of the presence of a radio jet interacting with
the ionisation cone ISM in Circinus. Although the jet has not been imaged at high-resolution, the detection of extended 
highly-ionised gas as well as the peculiar kinematics exhibited by [O\,{\sc iii}]~$\lambda$5007 suggest the presence of 
localised jet-driven shocks across the region where the extended, high-ionisation emission is detected.

\begin{figure*}
\centering
\includegraphics[width=13.cm]{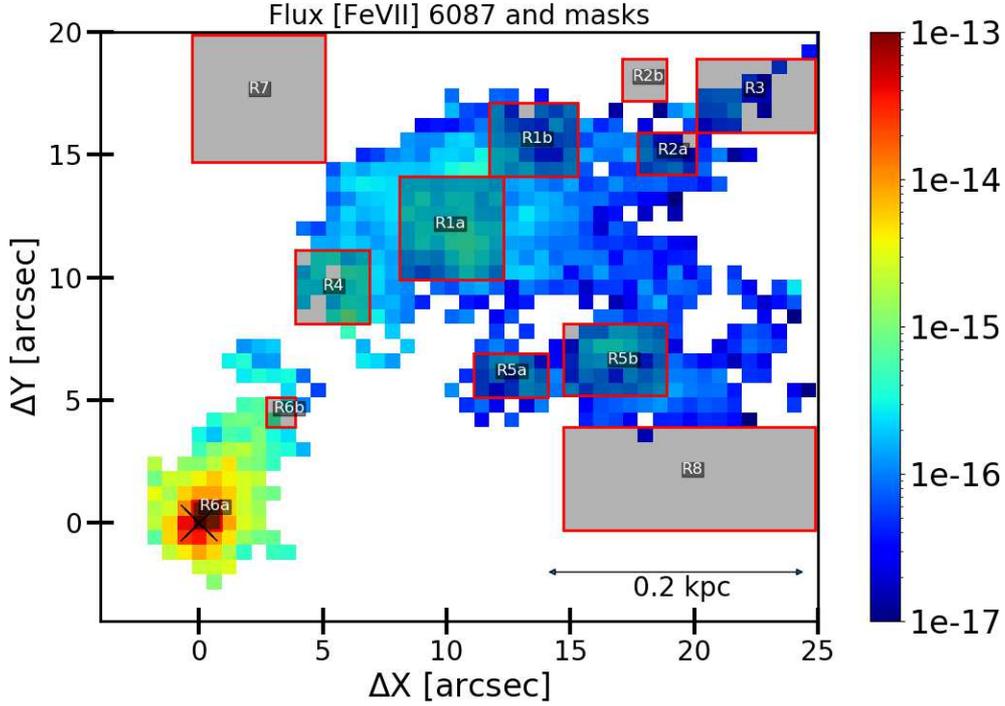}
\caption{In colour scale we show the [Fe\,{\sc vii}]~$\lambda$6087 emission.
%%while the grey contours show the [O\,{\sc iii}]~$\lambda$5007 emission. 
Twelve regions, bounded in red, were selected as representative of the ionisation cone. The colour bar is in  
erg\,cm$^{-2}$\,s$^{-1}$\,spaxel$^{-1}$.}
\label{fig:regions}
\end{figure*}

We investigate if shocks can indeed power much of the coronal line emission observed in the Circinus ionisation 
cone. Note that by no means we are discarding the influence of the radiation from the central source. Its effects, 
however are mostly restricted to the low-ionisation gas. For instance, we have seen in 
Fig.~\ref{fig:ratios_to_hbeta} the increase of the gas ionisation with increasing  distance from the AGN. Moreover, 
the enhancement of the gas ionisation is preferentially observed in the directions where the jet has presumably crossed 
the ionisation cone. 
To study the main source of gas ionisation in
Circinus, we selected 11 different regions located within the ionisation cone. Seven of them include areas where the 
highest ionisation gas, as probed by the [Fe\,{\sc vii}]~$\lambda$6087 line, is observed. We also include regions 
along the edges of the ionisation cone, as traced by  [O\,{\sc iii}] as well as clumps with bright [Fe\,{\sc vii}] 
and [Fe\,{\sc x}] emission. Figure~\ref{fig:regions} shows the 11 selected regions marked with red boxes in addition 
to region R6a, which is centred at the nucleus of Circinus. 

The observed lines and the ionisation potentials (IP) relative to their ions are given in Table~\ref{tab:ions}.
The IPs are ordered with decreasing energies  to give a hint about 
the distribution of the physical parameters throughout the observed regions.

\begin{table*}
\centering
	\caption{Ionisation potentials of different level ions.} \label{tab:ions}
\begin{tabular}{lccccccccccccc} \hline  \hline
\ line          & [Ar\,{\sc x}]&  [Fe\,{\sc x}] & [Fe\,{\sc vii}]&[Ar\,{\sc v}]& [O\,{\sc iii}] & HeII & [Ar\,{\sc iii}]& [O\,{\sc ii}]&[S\,{\sc iii}]&[N\,{\sc ii}]& HeI  & [S\,{\sc ii}]& [O\,{\sc i}] \\ 
\ $\lambda$(\AA)& 5533& 6375   & 6087   & 7006& 5007  & 5412 & 7136   & 7320 & 9069 & 6583 & 6678& 6716 & 6300\\
\  IP (eV)       &422.6& 233.6  &99.0  &59.6 & 35.5   & 54.4& 27.6   & 13.6 & 23.3 & 14.5  & 24.6 & 10.4 & 0 \\ \hline 
\end{tabular}

\end{table*}

Fig.~\ref{fig:regions} shows, at a first glance, some interesting issues.
Four regions, characterised by the  highest ionisation lines, are revealed by [Fe\,{\sc x}].
They are identified as R1a, R5b, R6a and R6b. 
[Fe\,{\sc vii}]/\Hb is relatively strong in the R1a region, increasing towards the North-West (R1b, R2a, and R3).
[Ar\,{\sc iii}]/\Hb,  [O\,{\sc iii}]/\Hb, and [S\,{\sc iii}]/\Hb  have similar intensity-ratio distributions covering 
all  the regions 
named R1, R2, R3, R4, and R5 with smooth differences. This extended emission  is divided in different regions according to
the distribution of the [Fe\,{\sc vii}]/H$\beta$, [O\,{\sc i}]/H$\beta$, and [N\,{\sc ii}]/H$\beta$ line ratios 
(see Fig.~\ref{fig:ratios_to_hbeta}).
In particular, the [N\,{\sc ii}]/H$\beta$ map indicates a kind of "jet-driven" region in R3. This is because it is
characterised by a relatively low line ratio. The same can be said for the [S\,{\sc iii}]/\Hb map, although with lower values.
In the opposite position (South-East) the detached `island' (R6) shows a concentric distribution with
a high [Fe\,{\sc x}]/\Hb and [O\,{\sc iii}]/\Hb core (R6a) decreasing towards  the edge (R6b). 
[O\,{\sc ii}]/\Hb  has a fragmented structure, most likely depending on the relatively low critical density for
collisional deexcitation. A region of  slightly enhanced low ionisation-level lines can be seen 
in the [O\,{\sc i}]/\Hb, [S\,{\sc ii}]/\Hb, and [O\,{\sc ii}]/\Hb maps at the South-West. It is identified as R8 in 
Fig.~\ref{fig:regions}.  On the other hand, [O\,{\sc ii}]/\Hb disappears
in the North-East where [S\,{\sc ii}]/\Hb and [O\,{\sc i}]/\Hb are still relatively high (reported as R7). 
Regions R7 and R8 show the spectroscopic characteristics of  the ISM.

Tables~\ref{tab:fwhmR1a} to~\ref{tab:fwhmR8} list the average FWHM (column~2), the maximum FHWM averaged over all the spaxels in the region (column~3) and the maximum FWHM (column~4) measured 
for the different emission lines in each region defined  by its coordinates (\dx/ $\Delta$Y). 
The scatter in FWHM for each region reveals that the observed values are the results of the contribution of
many clouds with different physical conditions.
Therefore, to reproduce the observations for the averaged line ratios in the different regions,
we employed for the calculation of the spectra the code {\sc suma} \citep{contini_2011}, which accounts for the coupled effect of shock 
and photoionisation by the central source. The code and the input parameters are briefly described in 
Sect.~\ref{sect:calculation}.

\begin{figure}
\centering
\includegraphics[width=5cm]{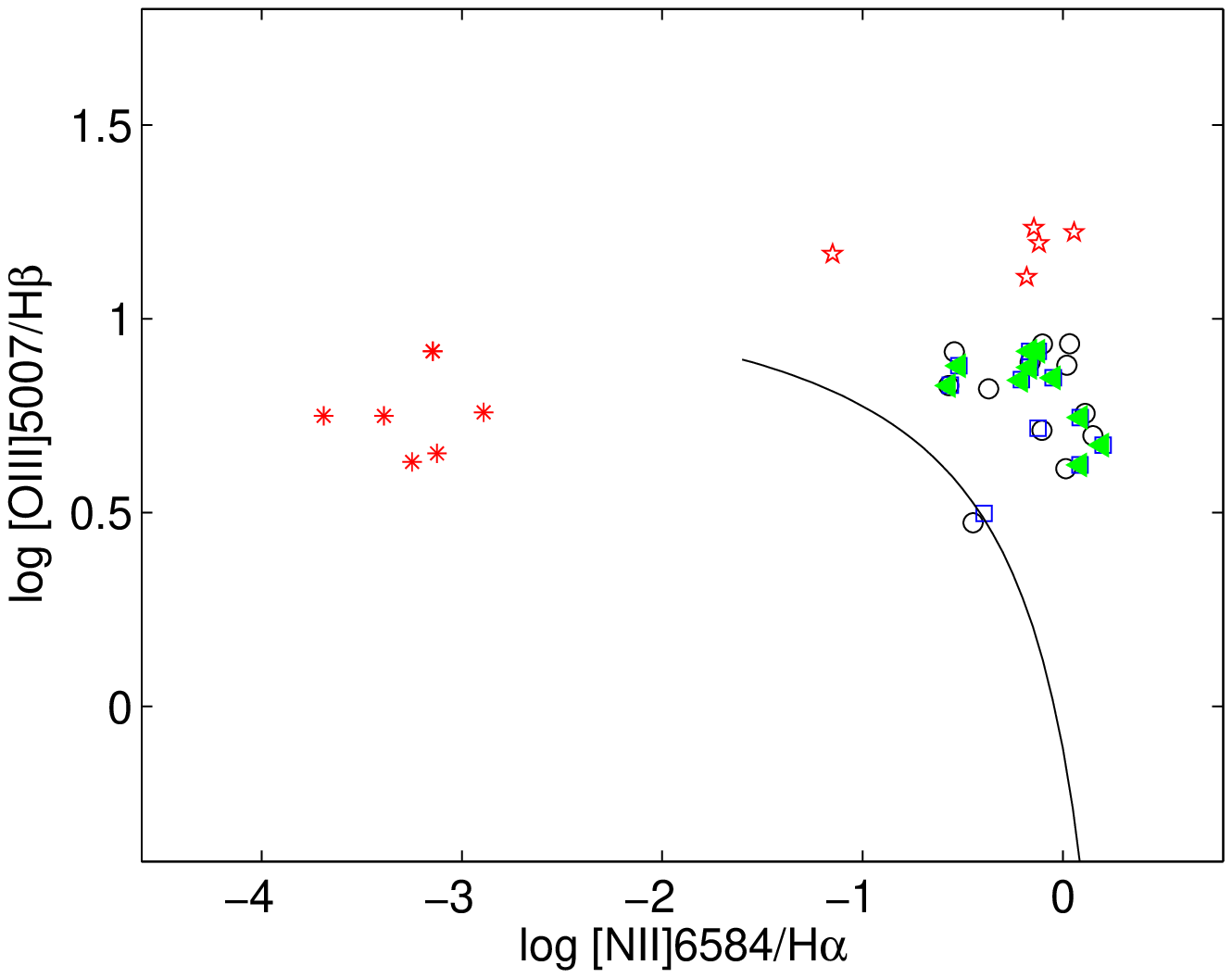}
\includegraphics[width=5cm]{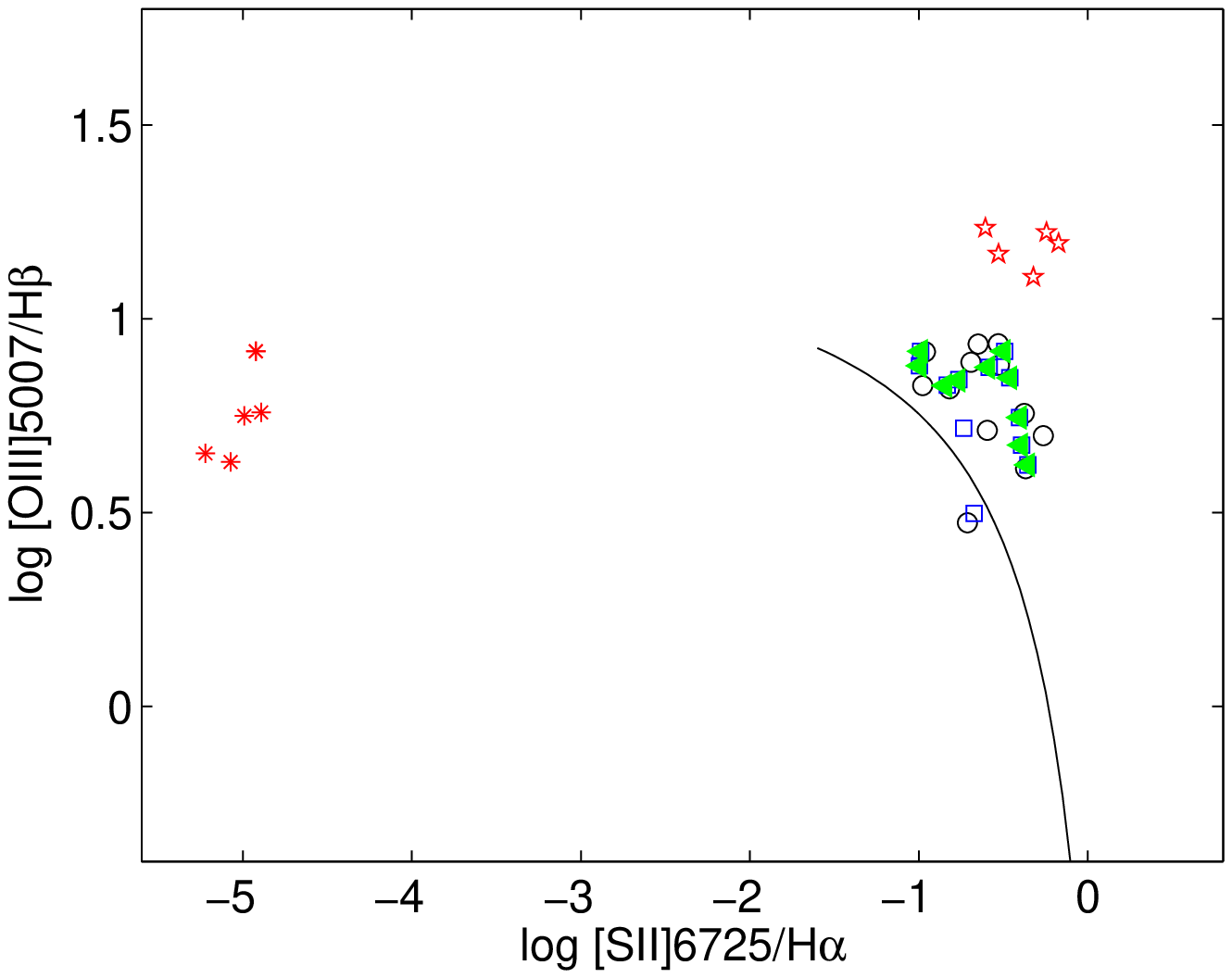}
\caption{BPT diagrams [\ion{O}{iii}]/\Hb vs [\ion{N}{ii}]/\Ha  (upper panel) and [\ion{O}{iii}]/\Hb\ vs [\ion{S}{ii}]/\Ha  (bottom panel) for line ratios presented in Table~\ref{tab:linerat_comp}.   Black open circles are the observed data; blue open squares are the results of
RD models; green triangles represent averaged results; red asterisks represent matter-bounded SD models while red stars show the shock dominated region 
(radiation-bounded models).}
\label{fig:bptdiag}
\end{figure}

Figure~\ref{fig:bptdiag} shows the location of the R1-R8 regions selected for our analysis in the traditional BPT diagrams   [\ion{N}{ii}]/\Ha vs [\ion{O}{iii}]/\Hb (upper panel) and [\ion{S}{ii}]/\Ha vs [\ion{O}{iii}]/\Hb (bottom panel) proposed by \citet{vo+87}. In both  plots, the line that separates the AGN domain from the star-formation and \ion{H}{ii} regions is taken from \citet{vo+87}. Note that only the average spectrum from region R2b is on the border line between AGN and star-forming region. All other observed ratios are in the AGN-dominated region.

\begin{figure}
\includegraphics[width=4.1cm]{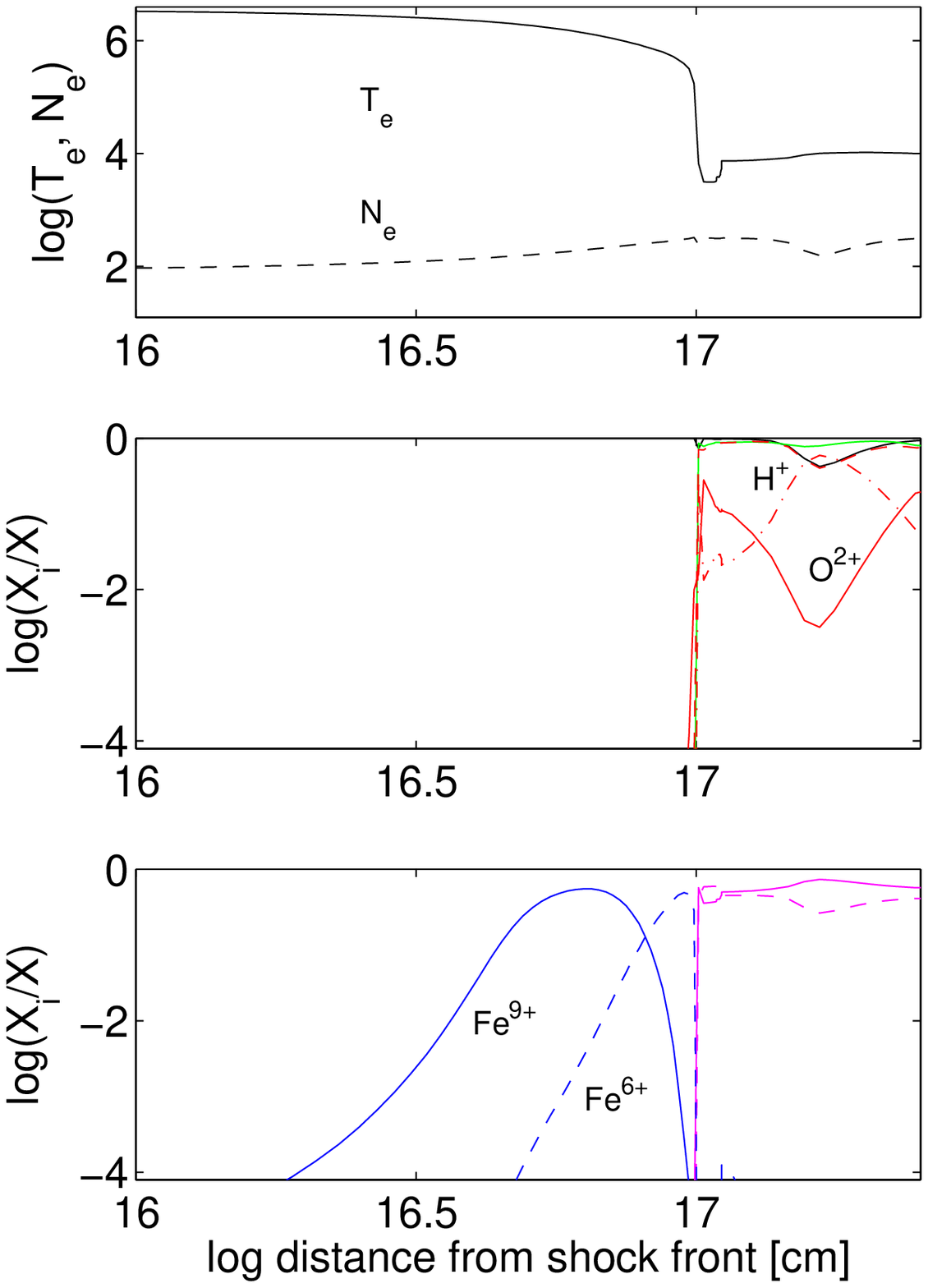}
\includegraphics[width=4.1cm]{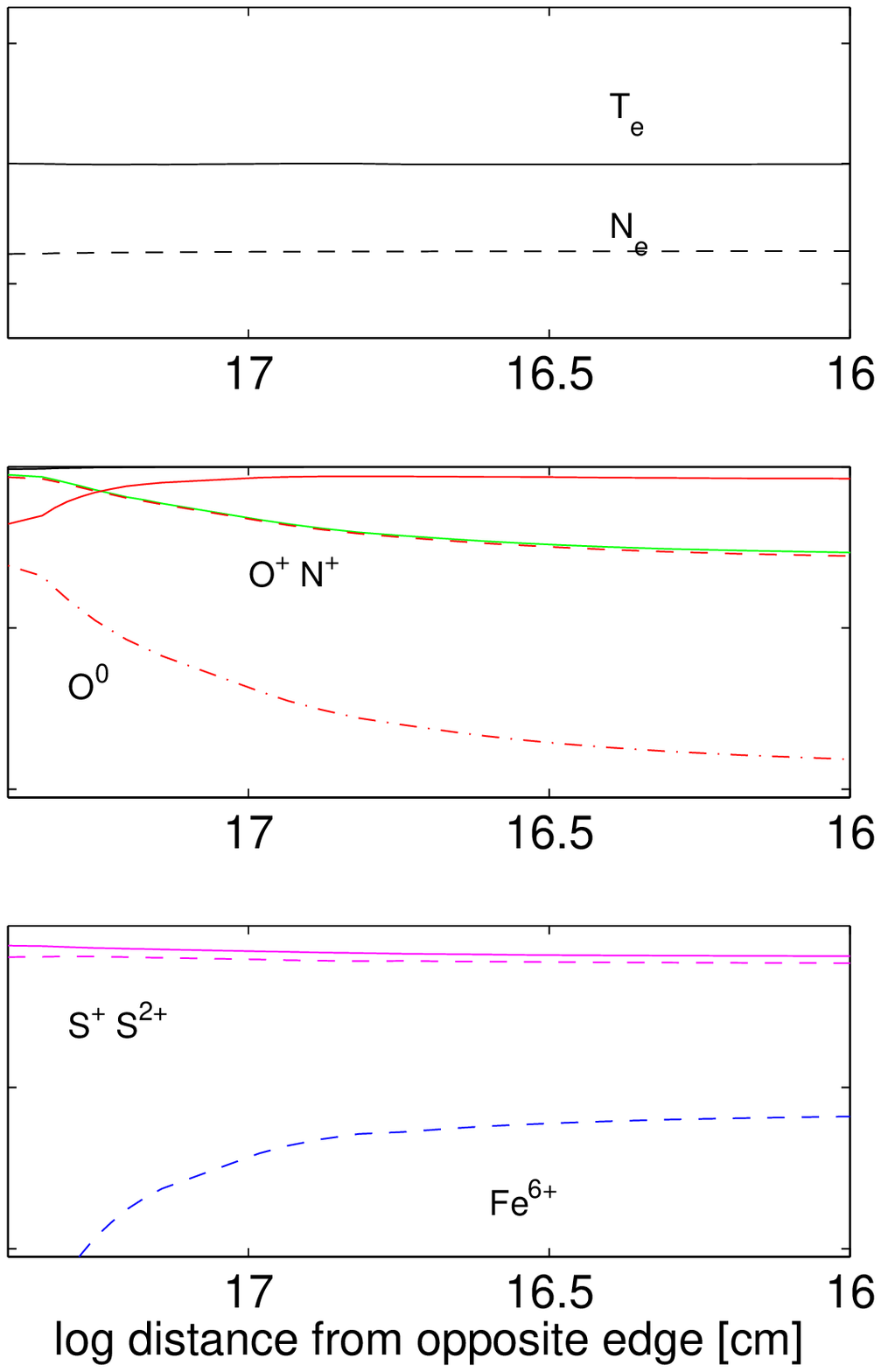}
\caption{The profiles of $T_{\rm e}$, $n_{\rm e}$, and of the fractional abundance of the most significant
ions throughout the clouds in region R1a calculated by a RD model.}
\label{fig:TeNeprofilesRD}
\end{figure}

\begin{figure}
\centering 
\includegraphics[width=8.0cm]{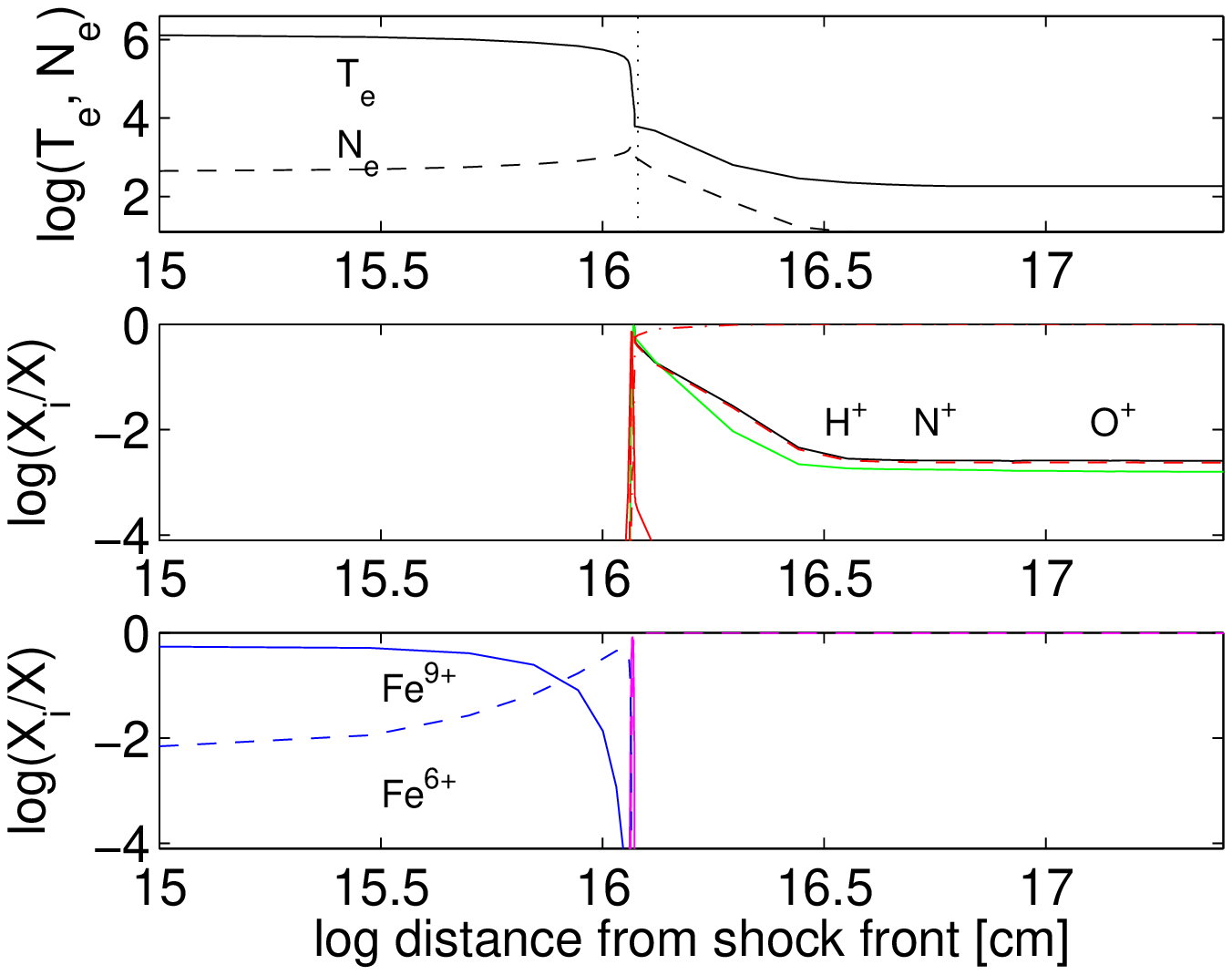}
\caption{The profiles of $T_{\rm e}$, $n_{\rm e}$, and of the fractional abundance of the most significant
ions throughout the clouds in region R1b calculated by a SD model.}
\label{fig:TeNeprofilesSD}
\end{figure}

Each one of the R1-R8 region  spectra  cover lines from different ionisation levels.  
Relatively high values of [Fe\,{\sc vii}]/\Hb and [O\,{\sc i}]/\Hb can be hardly fitted by a single model. 
We have chosen to reproduce the spectrum in each region by the
contribution of two representative models. One, shock dominated (SD), is calculated considering that the flux from the
active centre (AC) is zero, i.e. it does not reach the emitting gas. The other model, radiation dominated (RD), assumes that the 
radiation from the AC dominates although shocks are also at work. A  perfect fit to the spectra of the different 
regions is not  expected  because too 
many different physical conditions coexist throughout the observed regions. 
RD models are radiation bounded, therefore, they are adapted to reproduce all the line ratios in each spectrum 
except  lines from the high ionisation levels. In fact,  relatively high temperatures of the emitting gas
adapted to high [\ion{Fe}{vii}] and [\ion{Fe}{x}] line emission fluxes appear   in the region near the shock front downstream
even for the RD models, depending on the shock velocity.
The temperature   downstream decreases throughout the gaseous clouds  following the cooling rate  
which yields gas recombination.
In a large region of gas  the temperature is maintained  $\geq$ 10$^4$ K by radiation from the AC and by secondary 
radiation. Therefore lines from medium and low ionisation levels (e.g. \Hb)  are strong  and
the  calculated  [\ion{Fe}{vii}]/\Hb and [\ion{Fe}{X}]/\Hb ratios are low compared with the observed ones.
On the other hand, relatively high
[Fe\,{\sc x}]/\Hb and [Fe\,{\sc vii}]/\Hb, [Ar\,{\sc x}]/\Hb and  [Ar\,{\sc v}]/\Hb are better reproduced by  
matter-bounded SD models.  
As an example,  we show in Figs.~\ref{fig:TeNeprofilesRD} and \ref{fig:TeNeprofilesSD}
the profiles for  electron  temperature, electron  density and 
fractional abundances of the most significant ions throughout a RD and a SD cloud (mRD1a and mSD1b in
Table~\ref{tab:fwhmR1a}, respectively)  ejected outwards from the AC.

In Fig.~\ref{fig:TeNeprofilesRD} the  nebula is divided into two halves. The shock front is on the left side of the 
left panel, while the edge reached by the photoionising flux from the AC is  on  the right of the right panel,  
opposite to the shock front.
In the right panel of Fig.~\ref{fig:TeNeprofilesRD} the X-axis scale  is logarithmic  and reverse in order to show 
with the same precision the physical conditions  at the two edges of the  cloud. 
The shock dominated and the radiation dominated
sides of the emitting nebula  are bridged by  secondary radiation, 
produced by the slabs of gas heated by the shock, leading to a large zone of gas  at T$\sim$10$^4$K.  
In the SD  dominated clouds (Fig.~\ref{fig:TeNeprofilesSD}), the internal edge  opposite to the shock front 
 shows the cool gas recombined  region.
The downstream region throughout the  cloud  is interrupted before  recombination of high-level ions
such as Fe$^{6+}$ and Fe$^{9+}$ and also
to avoid  a strong \Hb line intensity, which could reduce the  [Fe\,{\sc vii}]/\Hb and the other high-ionisation 
line ratios.
Moreover, a pure shock model radiation-bounded  would lead to  relatively high [O\,{\sc ii}]/\Hb line ratios. 
In the top panel of Fig.~\ref{fig:TeNeprofilesSD} we have marked by a dotted vertical line
the distance from the shock front at which the calculations stop for model mSD1b.
The choice of the shock velocity in SD models is constrained  by the observed [Fe\,{\sc vii}]/[Fe\,{\sc x}]  and  
[Fe\,{\sc vii}]/\Hb  line ratios.

Some elements (e.g.  S and N) can be depleted from the gaseous phase because they are  trapped into
dust grains and into molecules. 
The grains are sputtered throughout the shock front and downstream depending on the shock velocity.
Therefore, the abundances of the heavy elements relative to the solar value can  change from model to model.
For the other elements (He, C, O, Ne, Mg, Si, Ar, Cl, and Fe) solar abundances relative to H \citep{asplund_2009} 
were  adopted.

\subsection{Calculation details} \label{sect:calculation}

The main input parameters adopted  to model the 12 integrated emission line regions marked in Fig.~\ref{fig:regions} 
are those that lead  to the best  fit  of calculated  to  observed line ratios  and continuum emission fluxes.
{\sc suma} accounts for photoionisation and heating by   primary and secondary  radiation and for
collisional processes due to shocks.

The input parameters such as  the shock velocity \Vs, the atomic
preshock density \n0, and the preshock
magnetic field \B0 (for  all models \B0~=~10$^{-4}$ Gauss is adopted)
define the hydrodynamical field.
They  are combined in the compression equation,  which is resolved
throughout each slab of gas
in order to obtain the density profile  across the emitting clouds. The AGN photoionising continuum radiation is 
represented by a power-law   radiation flux $F$  in  number of photons cm$^{-2}$\,s$^{-1}$\, eV$^{-1}$ at the Lyman limit
and  spectral indices  $\alpha_{\rm UV}$~=~-1.5 and $\alpha_X$~=~-0.7.
The primary radiation source does not depend on the host galaxy physical conditions but it affects the surrounding 
gas.   
This  region  is not considered as a unique cloud, but as a  sequence of plane-parallel slabs (up to 300) with different 
geometrical thickness calculated automatically following the temperature gradient.
The secondary diffuse radiation is emitted from the slabs of gas heated  by both the radiation flux reaching the gas 
and the shock.
Primary and secondary radiation are calculated by radiation transfer.
The calculations initiate at the shock front where the gas is compressed and adiabatically thermalised, reaching a maximum
temperature in the immediate post-shock region {\it T} $\sim$ 1.5$\times 10^5$ (\Vs/100 \kms)$^2$.
{\it T} decreases downstream,  leading to gas recombination.  The cooling rate is calculated in each slab.
The line and continuum emitting  regions throughout the galaxy cover  an ensemble of fragmented clouds.
The geometrical thickness $D$ of the clouds is an input parameter that is calculated consistently with the physical 
conditions and element abundances of the emitting gas. 

The fractional abundances of the ions are calculated resolving the ionisation equations
for each element (H, He, C, N, O, Ne, Mg, Si, S, Ar, Cl, Fe) for each ionisation level.
Then, the calculated line ratios, integrated throughout the cloud geometrical width, are compared with the
observed ones. The calculation process is repeated, changing  the input parameters until the observed data are reproduced
by  model results,  at maximum within 10-20 percent
for the strongest line ratios and within 50 percent for the weakest ones.

\subsection{Modelling results}

In  Table~\ref{tab:linerat_comp} the average observed line ratios (marked with the sufix "obs") are presented.  They are compared with RD and SD model results
in the following two rows. In the next row the weighted sum of the calculated line ratios are reported.
The models were constrained by the [\ion{O}{iii}]/\Hb, [\ion{Fe}{vii}]/\Hb, [\ion{N}{ii}]/\Hb, [\ion{S}{ii}]/\Hb, and [\ion{O}{ii}]/\Hb line ratios.
The models are described in Table~\ref{tab:inputpars}. The last column of that table shows the \Hb fluxes calculated at the nebula (\Hbc). It allows us to  calculate the line flux intensities for each line using the line ratios of Table~\ref{tab:linerat_comp} and eventually add up the RD and SD model contributions  in each region by the relative weights.

Dividing the total flux  calculated for each line by the \Hbc~ weighted sum,  we obtain the results for the total line ratios. They can be compared to the observed data in Table~\ref{tab:linerat_comp}.
The total \Hb line flux calculated at the nebula and
the relative weights $w$ adopted to add up the RD and SD  model results are given in the two bottom lines of Table~\ref{tab:weights}, respectively.
The R2b and R4 spectra are  reproduced  by only the RD model.

As mentioned above, the SD models that appear in Table~\ref{tab:linerat_comp} are matter-bounded in order to obtain the
[\ion{Fe}{vii}]/\Hb and [\ion{Fe}{x}]/\Hb line ratios  suitable to the averaged models. In the bottom 5 lines of Table~\ref{tab:linerat_comp} we report 
the line ratios calculated by  the by the corresponding radiation-bounded SD models that we obtained using  a cloud geometrical thickness $D >$0.1~pc for all of them.
Model m104 was used to fit the spectra of R1a, R3, R5a, R5b and R6a while model m101 better reproduces R1b and R6b. Model m100 fits R2a, whereas models m107 and m108 are suitable to the R7 and R8 regions.
Radiation-bounded SD models are calculated in order to  show the shock dominated
region  in BPT diagrams for relatively high shock velocities \Vs$\geq$ 100 \kms.  
Lower velocities are more adapted to star-formation and HII regions.

Our results show that not always low O/H, N/H and S/H  translate into low [\ion{O}{iii}]/\Hb,
[\ion{N}{ii}]/\Hb, and [\ion{S}{ii}]/\Hb ratios in the BPT diagrams. The geometrical thickness of the emitting cloud plays a key role in the modelling.

%In Table~\ref{tab:linerat_comp} the observed line ratios for the different analysed regions are presented. For each 
%region, the first line shows  the observed line ratios.
%They are compared with the best RD and SD model results
%in the following two rows. In the next row, the weighted sum of the calculated line ratios is reported.
%The models were constrained by the [O\,{\sc iii}]/\Hb, [Fe\,{\sc vii}]/\Hb, [N\,{\sc ii}]/\Hb, [S\,{\sc ii}]/\Hb and 
%[O\,{\sc ii}]/\Hb line ratios. The input parameters of the models are listed in Table~\ref{tab:inputpars}. The last column of that table shows the \Hb fluxes calculated at the nebula (\Hbc). They allow to derive the line flux intensities from the line ratios and eventually add up the RD and SD model contributions in each region taking into account the relative weights. Dividing the total flux  calculated for each line by the \Hbc~ weighted sum,  we obtain the final model results for every line ratio that is then compared to the observed data in Table~\ref{tab:linerat_comp}. The total \Hb line flux calculated at the nebula and the relative weights $w$ adopted to add up the RD and SD  model  results are given in the last two bottom lines of Table~\ref{tab:weights}. Note that the R2b and  R4 spectra are  reproduced  by only the RD model.

The results  obtained by modelling the observation data in the different regions  are commented below.

\begin{enumerate}
\item
Table~\ref{tab:linerat_comp} indicates that  for all the observed  regions,  [Fe\,{\sc vii}]/[Fe\,{\sc x}]~$>$~1. 
Model results show  that this constraint  is achieved by \Vs $\leq$~550 \kms.
The good fit to the observed [Fe\,{\sc vii}]/\Hb confirms our hypothesis that  SD models are  radiation bounded. 
[Fe\,{\sc x}] lines are weaker and indeed seldom observed.

\item Different relative weights are  adopted in different regions (See Table~\ref{tab:weights} to sum the SD and RD
cloud contributions). In regions R1a and R6a, the [Ar\,{\sc x}]~$\lambda$5533/\Hb is higher by a factor of $>$ 100 
than in the other regions. At the relatively high velocities observed  from the [Ar\,{\sc x}] line profile, 
this line can be blended with [Cl\,{\sc iii}]~$\lambda$5539.  [Cl\,{\sc iii}]/\Hb results 0.02 and 0.025 in the R1a 
and R6a regions, respectively.

\item Regarding the preshock density, we have found that RD models are constrained by the 
[O\,{\sc ii}]~$\lambda$7320/\Hb 
line ratios, which are  $\leq$0.1 in all the observed regions. This threshold is achieved by clouds of relatively  
low  \n0,  $\sim$20-60 \cm3. In Table~\ref{tab:weights}, \Te and \ne in each region  are calculated from the 
average observed data reported in Table~\ref{tab:linerat_comp}.  It can be seen that \ne
are roughly proportional to \n0 for  RD
  model results,  recalling that in the downstream region throughout the clouds
  the gas is compressed by factors of 5-10, depending on the shock velocity and the magnetic field.
  SD models are less constrained by the choice of \n0 because in matter-bounded models the
  [O\,{\sc ii}]~$\lambda$7320/\Hb line ratio is always $<$~1.
  The preshock densities adopted in this work  agree with  those 
  obtained by fitting the Circinus galaxy observed spectra by \citet{contini_1998}.

\item We have reproduced  the O, Ar and Fe line ratios to \Hb by solar abundances relative to H 
  (6.6 $\times$ 10$^{-4}$, 3.27 $\times$ 10$^{-6}$ and 3.2 $\times$  10$^{-5}$, respectively).
   The N/H  solar relative  abundance (10$^{-4}$)  was   altered by factors of  0.4-1.8  to
  improve the fit of the corresponding line ratios. The S/H relative abundances  were found lower than solar (2 $\times$ 10$^{-5}$) in all the observed regions, due to the presence of dust. 

\item The \Hbc flux calculated 
  by SD models  results lower by  factors between $<$~10 and $>$~100 than for RD models
  because SD models are matter bounded.
  High  fragmentation of the clouds may be explained by turbulence   created by  shocks  at the shock-fronts.

\item The photoionisation flux from the AC is relatively low throughout the observed regions,  similar to $F$ 
  found for LINERS.
  In particular, it declines in the external regions R7 and R8. 
  
\end{enumerate}

\begin{table*} 
 \centering 
\caption{Comparison of the calculated line ratios to \Hb with the observed  average in the different regions.} \label{tab:linerat_comp} 
 \begin{tabular}{ccccccccccccccc} \hline  \hline 
\ Region    &[O\,{\sc iii}]&[Ar\,{\sc x}] &[Fe\,{\sc vii}] &[O\,{\sc i}]  &[S\,{\sc iii}]& [Fe\,{\sc x}]& \Ha & [N\,{\sc ii}]& [S\,{\sc ii}]& [Ar\,{\sc v}]&[Ar\,{\sc iii}]& [O\,{\sc ii}]& [S\,{\sc iii}] \\
\           &5007  &  5533& 6087   & 6300 & 6312 & 6375 & 6563&6583  &  6717& 7006 & 7136  & 7320 &9069& \\ \hline
\ R1a (obs) &10.31 &0.02  &0.12    &0.23  & 0.04 &0.02  &3.10 &2.13  & 0.63 & 0.07 & 0.27  & 0.06 &0.45\\
\ mRD1a     &10.0  &3E-4  &0.022   &0.1   & 0.03 &0.005 &2.92 &2.    &0.76  &0.007 & 0.3   & 0.09 &0.6 \\
\ mSD1a     &11.   &0.03  &13.3    &0.0   &0.002 &0.9   &4.2  &0.003 &0.0   &0.9   & 0.02  &0.03  &0.0 \\
\ mR1a      &10.    &5.E-4&0.11    &0.1   &0.03  &0.01  &2.93 &2.    &0.75  &0.013 &0.3    &0.09  &0.59 \\
\ R1b (obs) &8.8   & -    &0.22    &0.12  & 0.03 &-     &3.1  & 1.32 & 0.47 & 0.07 & 0.2   &0.06  &0.35 \\
\ mRD1b     &9.3   & 3.E-4&0.024   &0.1   & 0.02 &0.006 &2.9  & 1.8  & 0.5  &0.006 & 0.35  & 0.1  &0.4 \\
\ mSD1b     &7.5   &6.4E-4&15.6    & 0.0  & 6.E-4& 0.014&4.9  & 0.001& 0.0  &1.2   & 0.011 & 0.01 &0.002\\
\  mR1b     &9.27  &3.E-4 &0.24    & 0.1  &0.02  &0.006 &2.93 &1.8   &0.5   &0.02  &0.34   &0.1   &0.4 \\
\ R2a (obs) &8.97  &-     &0.22    &0.07  &-     &-     &3.23 &0.87  &0.34  &0.09  &0.21   &-     &0.37\\
\ mRD2a     &9.    &3E-5  &0.027   &0.05  &0.028 &6E-5  & 2.92& 0.80 &0.43  &0.004 & 0.37  &0.1   &   0.6 \\
\ mSD2a     &7.65  &0.09  &12.3    & 1.E-5& 0.002& 2.   &3.9  & 0.005& 5E-5 &0.07  & 0.012 & 0.02 &0.02 \\
\ mR2a      &8.97  &0.0014&0.22    &0.05  &0.027 &0.03  &2.94 &0.79  &0.42  &0.005 &0.36   &0.1   &0.06  \\
\ R2b (obs) &3.97  &-     & -      &0.17  &-     &-     &3.09 &1.1   &0.60  &-     &0.19   &0.37  &  -  \\
\ mRD2b     &4.2   &1.3E-5&0.017   &0.2   &0.007 &3.E-6 &2.97 &1.2   &0.63  &0.003 &0.18   &0.27  &0.46 \\
\ R3 (obs)  &10.96 &-     &0.22    &0.13  &-     &-     &3.38 &0.97  &0.37  &0.09  &021    &0.14  &0.32 \\
\ mRD3      &10.1  &6.5E-5&0.05    &0.01  &0.04  &1.E-4 &2.97 &0.9   &0.3   &0.01  &0.38   &0.13  &0.9  \\
\ mSD3      &11.   &0.03  &13.3    & 0.0  & 0.02 & 0.9  &4.2  & 0.003&0.0   &0.94  & 0.013 & 0.034&0.0   \\
\ mR3       &10.1  &4.E-4 &0.22    &0.01  &0.04  &0.01  &2.98 &0.9   &0.3   &0.02  &0.37   &0.128 &0.89 \\
\ R4 (obs)  &6.88  &-     &0.03    &0.2   &0.02   &-    & 3.11&2.45  &0.79  &0.02  &0.114  & 0.05 &0.22 \\
\ mRD4      &6.97  &4E-5  &0.03    &0.05  &0.03  &0.0   &2.93 &2.2   &0.54  &0.003 &0.3    &0.09  &0.9  \\
\ R5a (obs) &10.12 &-     &0.08    &0.32  &0.04  & -    &3.10 &3.25  &0.93  &0.04  &0.25   &0.11  &0.50 \\
\ mRD5a     &9.4   &1E-4  &0.02    &0.24  &0.03  &0.001 &2.9  &2.6   &1.0   &0.005 &0.3    &0.11  &0.6  \\
\ mSD5a     &11.   &0.03  &13.3    &0.0   &0.002 &0.9   &4.2  &0.003 &0.0   &0.94  &0.02   &0.034 &0.0  \\
\ mR5a      &9.4   &2.4E-4&0.082   &0.24  &0.03  &0.005 &2.9  &2.6   &1.    &0.009 &0.3    &0.11  &0.6  \\
\ R5b (obs) &11.49 &-     &0.09    &0.20  &0.04  &0.01  &3.12 &2.47  &0.7   &0.04  &0.3    &0.06  &0.58 \\
\ mRD5b     &11.0  &3E-6  &0.044   &0.01  &0.04  &0.006 &2.93 &2.    &0.3   &0.01  &0.33   &0.06  &0.9  \\ 
\ mSD5b     &11.0  &0.03  &13.3    & 0.0  & 0.002& 0.9  &4.2  & 0.003& 0.0  &0.94  & 0.02  & 0.034&0.0  \\
\ mR5b      &11.   &1E-4  &0.09    &0.01  &0.04  &0.01  &2.93 &2.    &0.3   &0.013  &0.33   &0.06 &0.89 \\
\ R6a (obs) &11.5  &0.04  &0.15    &0.62  &0.03  &0.13  &3.12 &3.37  &0.92  &0.03  &0.25   &0.11  &0.44 \\
\ mRD6a     &11.0  &3E-4  &0.02    &0.4   &0.035 &0.005 &2.92 &2.2   &0.94  &0.007 & 0.35  &0.1   &0.7  \\
\ mSD6a     & 11.0 &0.03  &13.3    &0.0   &0.002 &0.9   &4.2  &0.003 &0.0   &0.94  &0.02   &0.034 &0.0   \\
\ mR6a      &11.   &6.E-4 &0.16    &0.4   &0.034 &0.014 &2.94 &2.2   &0.93  &0.02  &0.35   &0.1   &0.69  \\
\ R6b (obs) &7.6   &-     &0.06    &0.46  &0.02  &0.02  &3.11 &4.02  &1.31  &-     &0.18   &0.09  &0.21 \\
\ mRD6b     &7.42  &1.2E-4&0.02    &0.34  &0.02  &0.002 &2.94 &3.6   &1.16  &0.003 &0.28   &0.12  &0.5  \\
\ mSD6b     &7.5   &6.E-4 &15.6    & 0.0  & 6.E-4& 0.014&4.9  & 0.002& 0.0  &1.26  & 0.011 & 0.015&0.002\\
\ mR6b      &7.42  &1.2E-4&0.051   &0.34  &0.02  &0.002 &2.94 &3.6   &1.16  &0.006 &0.28   &0.12  &0.5  \\
\ R7 (obs)  &5.48  &-     &-       &0.43  &-     &-     &3.8  &3.93  &1.63  &-     &0.16   &0.09  &0.25  \\
\ mRD7      &5.6   &3E-4  &0.023   &0.4   &0.02  &0.005 &2.95 &3.6   &1.3   &0.003 &0.25   &0.09  &0.4   \\
\ mSD7      &5.7   &0.14  &11.8    &0.0   &9E-4  &3.0   &3.9  &2.2E-3&3.3E-5&0.67  &0.09   &0.07   &0.003 \\
\ mR7       &5.6   &8.E-4 &0.06    &0.4   &0.02   &0.02 &2.96 &3.6  &1.3   &0.005 &0.25   &0.09   &0.4 \\
\ R8 (obs)  &6.66  &-     &0.09    &0.59  &-     &-     &3.74 &5.28  &2.04  &-     &0.26   &0.19  &0.40    \\
\ mRD8      &6.3   &1.3E-4&0.023   &0.64  &0.02  &0.002 &2.96 &4.7   &1.2   &0.004 &0.23   &0.18  &0.4    \\
\ mSD8      &6.    &0.04  &12.8    &0.0   &6E-4  &1.2   &4.   &0.003 &2.4E-5&0.7   &0.09   &0.01  &0.005  \\
\ mR8       &6.3   &2.E-4 &0.09    &0.64  &0.02   &0.008  &3.   &4.7   &1.2   &0.007 &0.23   &0.18  &0.40 \\ \hline
\ m100      &17.1  &7.e-3 &1.07    &0.42  &0.1   &0.17  &5.0  &3.3   &2.38  &0.08  &0.15    &4.3    &0.3\\
\ m101     &22.9  &4e-5  &0.91    &0.44  &0.11  &8e-4  &5.85 &4.2   &1.45  &0.12  &0.2     &7.25   &0.3\\
\ m104      &19.6  &3.e-3 &1.12    &0.44  &0.1   &0.079 &5.5  &0.39  &1.63  &0.1   &0.2     &6.     &0.34\\
\ m107      &20.9  &0.014 &1.13    &0.27  &0.13  &0.29  &4.86 &3.7   &3.26  &0.09  &0.19    &3.59   &0.45\\
\ m108      &22.3  &4.-3  &1.24    &0.34  &0.11  &0.12  &5.27 &6.0   &3.0   &0.1   &0.2     &3.89   &0.4 \\ \hline
 \end{tabular}

\end{table*}

\begin{table} 
\scriptsize
 \centering 
\caption{Input parameters of the models for each of the regions analysed.} \label{tab:inputpars} 
 \begin{tabular}{lcccccccccccccccc} \hline  \hline 
	 \         &\Vs   &\n0   &$D$   &$F$               &N/H        &  S/H    & \Hbc$^1$  \\      
\         & \kms & \cm3 & pc   &units$^2$         & 10$^{-4}$ & 10$^{-4}$   & \ergcs \\ \hline 
\ mRD1a   &500   & 20   &0.14  &2.                 &1.         &0.12  & 3.3E-3     \\
\ mSD1a   &270 & 180     &0.0015&-                 &1.         &0.1     &2.6E-5 \\
\ mRD1b   &500   &20    &0.13  &1.7               &1.         &0.08    & 3.3E-3  \\
\ mSD1b   &280   &200   &0.0005&-                 &1.         &0.1     &1.9E-5  \\
\ mRD2a   &200   &40    &0.05  &1.2               &0.4        & 0.1    &1.8E-3  \\   
\ mSD2a   &300   &100   &0.004 &-                 &1.         &0.12    &1.8E-5   \\
\ mRD2b   &180   &60    &0.03  &1.2               &0.4        &0.06    &2.7E-3 \\
\ mRD3    &200   &45    &0.023 &1.3               &0.8        &0.15    &1.0E-3   \\
\ mSD3    &270 & 180     &0.0015&-                 &1.         &0.1     &2.6E-5 \\
\ mRD4    &200   &40    &0.043 &1.0               &1.0        &0.13    &1.1E-3   \\
\ mRD5a   &260   &41    &0.067 &2.7               &1.4        &0.13    &4.4E-3   \\
\ mSD5a   &270 & 180     &0.0015&-                 &1.         &0.1     &2.6E-5 \\
\ mRD5b   &300   &30    &0.05  &1.5               &1.8        &0.16    &1.5E-3\\
\ mSD5b   &270 & 180     &0.0015&-                 &1.         &0.1     &2.6E-5 \\
\ mRD6a   &550   &20    &0.17  &2.7               &1.         &0.13     &4.8E-3 \\
\ mSD6a   &270 & 180     &0.0015&-                 &1.         &0.1     &2.6E-5 \\
\ mRD6b   &280   &38    &0.067 &2.                &1.         &0.12    &3.8E-3  \\
\ mSD6b   &280   &200   &0.0005&-                 &1.         &0.1     &1.9E-5  \\
\ mRD7    &350   &20    &0.17  &0.86              &1.        & 0.13    &2.E-3  \\
\ mSD7    &350   &20    &0.03  &-                 &1.         &0.13    &3.6E-6  \\
\ mRD8    &250   &20    &0.26  &0.7               &1.3        &0.1     &1.4E-3  \\
\ mSD8    &250   &20    &0.04  &-                 &1.3        &0.1  &2.3E-6       \\ \hline

 \end{tabular}

$^1$ \Hb calculated at the nebula;
$^2$ in 10$^9$ photon cm$^{-2}$ s$^{-1}$ eV$^{-1}$ at the Lyman limit

\label{tab:inputpars} 
\end{table}

\begin{table*} 
 \centering 
\caption{Medium values of $T_{\rm e}$~ and $n_{\rm e}$, total \Hbc~ and weights $w$ calculated for the different regions modelled with {\sc suma}.} \label{tab:weights}
 \begin{tabular}{lccccccccccccc} \hline  \hline 
\                   &R1a     & R1b   &R2a    &R2b   &R3      &R4   &R5a    &R5b    &R6a    &R6b   &R7    &R8   \\ \hline   
\ \Te (K) $^1$      &118884  &10750  &-      &-     &-       &12692&11800  &11065  &10500  &-     &-     &-    \\
\ \Ne (\cm3) $^2$   &190     &136    &179    &100   &406     &100  &234    &242    &516    &100   &220   &126  \\ 
\ \Hb$_{tot}$ $^3$  &3.33E-3&3.35E-3 &1.83E-3&2.7E-3&1.13E-3 &1.1E-3&4.4E-3&1.5E-3 &4.85E-3&4.82E-3&2.E-3 &1.4E-3 \\ 
\  w                &1      &2.5     &1.6    &-     &0.5     &-     &0.8   &0.2    &2      &1      &0.2   &3.2   \\ \hline
\end{tabular} 
 
$^1$ \Te is obtained from [S\,{\sc iii}]$\lambda$9069/[S\,{\sc iii}]$\lambda$6312  
$^2$ \Ne is obtained from [S\,{\sc ii}]$\lambda$6716/[S\,{\sc ii}]$\lambda$6731;  
$^3$ in \ergcs

\end{table*}

%##########################################################
%################## DICUSAO FINAL   ############
%##########################################################

\subsection{Discussion}

High energy emission lines with an X-ray emission counterpart is indicative of photoionisation by a central source as the dominant ionisation process \citep{ardila_2011}. In the work of \citet{ferguson_1997}, 
the minimum and maximum distances suitable  to find high ionisation lines from a cloud photoionised by the central 
source are presented. For example, for the [Fe\,{\sc vii}]~$\lambda$6087  line, the models predict an upper limit of 110 pc from the AGN. Here, we found that this line in Circinus extends to distances of up to 700 pc, ruling out the possibility of being produced by the central source alone. Moreover, the extended coronal emission is coincident with the central portion of the ionisation cone. Kinematics evidence show that this HIG is out of the galaxy plane.
The predicted size of the emission regions of [Fe\,{\sc x}] and [Ar\,{\sc x}] are even smaller than that of [Fe\,{\sc vii}] under a central source photoionisation scenario: They are limited to 20 pc and 1 pc, respectively.  However, we clearly detect both (see Fig.\ref{fig:fluxos_2}) at distances larger than 140 pc in the nuclear region, and up to 370 pc in the extended outflow. This result further supports our hypothesis that an additional mechanism must be at work to power that coronal emission.

The photoionisaton code {\sc suma} \citep{contini_2011}, which couples the effect of  photoionisation by the central source and shocks,   show that the [Fe\,{\sc vii}]~$\lambda$6087 line  can be produced away from the AC when the shock  velocities are $>$~200 ~km\,s$^{-1}$. We measure from MUSE values much higher than that, as can be seen in Figure 
\ref{fig:momentos_alta1}. Note though, that values of shock velocities and gas velocities do not necessarily need to match. 
Figure \ref{fig:temperaturas} shows evidence of a higher electron temperature at the edges of the [Fe\,{\sc vii}] clouds and that the temperature of the red component appears to be lower than that of the blue component. This is in line with the concept of a possible precursor 
gas. In this case we are assuming that the gas did not have time to cool down and we see the effect of the precursor. Following  the hypothesis  that a precursor is photoionising the gas of 
the red component in Figure \ref{fig:temperaturas}, the shocked gas  in the region of the blue component 
has not yet reached the region of the red component. It may be suggestive of a jet that has already passed and cooled. Apparently, the same effect as that we are seeing in the blue component will be found for the red component in a few thousand years.

\section{Concluding remarks}

Detailed analysis of MUSE data  allowed us to study, for the first time in the literature, the high-ionised outflow detected by means of the [Fe\,{\sc vii}]~$\lambda$6087 line in the Circinus Galaxy.  The emission fills the central portion of the prominent NW ionisation cone and is clearly aligned to the radio jet. The coronal gas extends at least up to 700~pc from the nucleus although we do not discard that it extends further out.  Moreover, we report on the first confirmed detection of extended [Fe\,{\sc x}]~$\lambda$6374 gas in Circinus. That emission is produced by an extended cloud of several tens of parsecs size, clearly resolved
with MUSE and visible up to $\sim$350 pc from the nucleus. The region where the extended   emission is observed coincides in
position with the brightest portion of the extended [Fe\,{\sc vii}] gas. 

We examined the behaviour of the different emission line flux distributions relative to H$\beta$. The results show that most of the high-ionisation lines are enhanced in the extended [Fe\,{\sc vii}] region. In contrast, low-ionisation lines are weak relative to H$\beta$ in the region dominated by the extended coronal emission, becoming stronger outside the region filled with the high-ionisation gas. This result suggests an additional mechanism at work favouring the coronal line production. 

Kinematic evidence as well as the examination of the physical conditions of the high-ionised outflow indicates that it is out of the galaxy plane, following non-circular motions. All evidence suggests that it is likely produced by a radio jet that passed through the region, inflating the gas and producing expanding bubbles, seen in the form of approaching and receding gas shells.  These structures display
velocities of a few hundred ~km\,s$^{-1}$, spatially coincident with prominent hard X-ray emission detected by Chandra. Density and temperature sensitive line ratios show that the extended high-ionisation gas is characterised 
by temperatures of up to 25000~K and electron density  \ne $> 10^4$ cm$^{-3}$. We do not discard some contribution from stellar winds, given that molecular outflows have been detected  with ALMA in projection with the ionisation cone. 

The above scenario is further supported by means of ionisation models that couple the effects of central source photoionisation plus shocks. The good fit to the observed [Fe\,{\sc vii}]/H$\beta$ confirms our hypothesis that shock-dominated clouds  must be present in the ionisation cone of Circinus. These clouds are responsible for the bulk of the extended coronal emission. Shock velocities with V$_{\rm s}~>~$200~km\,s$^{-1}$ are required for the production of such lines, in agreement with the observed gas kinematics.

Overall, our analysis find compelling evidence that the innermost portion of the ionisation cone in Circinus is jet-driven. It also demonstrates the role of low-power radio jets in a low-luminosity AGN, being able to produce outflow rates of order of 0.4~M$\odot$~yr$^{-1}$, similar to the ones observed in sources of higher luminosity. Detailed analysis of similar sources are needed in order to quantify the relevance of the kinetic channel in the feedback of active galactic nuclei.
 
\section*{Acknowledgements}

We are grateful to the anonymous Referee for useful comments and suggestions to improve this manuscript.
ARA acknowledges Conselho Nacional de Desenvolvimento Cient\'{\i}fico e Tecnol\'ogico (CNPq) for partial support to this work through grant 312036/2019-1.

\section*{Data availability}
The  data  underlying  this  article  are  available  in  the  European Southern Observatory  archive.  The  data  can  be  obtained  in  raw quality through the MUSE raw data query form (\url{http://archive.eso.org/wdb/wdb/eso/muse/form}). Science quality data can be obtained from the Spectral data products query form http://archive.eso.org/wdb/wdb/adp/phase3spec tral/form?collectionname=MUSE.


%%%%%%%%%%%%%%%%% APPENDICES %%%%%%%%%%%%%%%%%%%%%

\appendix


\begin{table} 
\scriptsize
 \centering
\caption{FWHM measured in region R1a  ($ \Delta$X/$\Delta$Y: 8"-12"/10"-14"). } \label{tab:fwhmR1a}
\begin{tabular}{lccccccccccccc} \hline  \hline 
\                 & FWHM$^{1,2}$ & max FWHM$^{1,2}$ & max FWHM$^{2}$ \\  \hline 
 \  [O\,{\sc iii}]$\lambda$5007 &       227.09 &      283.11 &      487.05  \\
 \    HeII$\lambda$5412 &       201.78 &      205.36 &      607.22  \\
 \  [Ar\,{\sc x}]$\lambda$5533 &       284.48 &      284.48 &      559.74  \\
 \ [Fe\,{\sc vii}]$\lambda$6087 &       196.50 &      237.36 &      541.28  \\
 \    [O\,{\sc i}]$\lambda$6300 &       185.51 &      239.16 &      522.33  \\
 \  [S\,{\sc iii}]$\lambda$6312 &       176.84 &      185.48 &      405.93  \\
\      [O\,{\sc i}]$\lambda$6364 &       204.99 &      204.99 &      517.25  \\
 \   [Fe\,{\sc x}]$\lambda$6375 &       272.87 &      272.87 &      530.82  \\
 \   \Ha  $\lambda$6563 &       201.95 &      274.45 &      502.65  \\
 \   [N\,{\sc ii}]$\lambda$6583 &       175.23 &      229.31 &      466.03  \\
 \     HeI$\lambda$6678 &       250.34 &      263.44 &      494.12  \\
 \   [S\,{\sc ii}]$\lambda$6716 &       196.58 &      268.94 &      419.48  \\
 \   [Ar\,{\sc v}]$\lambda$7006 &       173.00 &      202.94 &      432.77  \\
 \  [Ar\,{\sc iii}]$\lambda$7136 &      181.78 &      248.67 &      400.67  \\
 \   [O\,{\sc ii}]$\lambda$7320 &       216.15 &      251.08 &      647.57  \\
 \  [S\,{\sc iii}]$\lambda$9069 &       160.61 &      182.38 &      354.02  \\ \hline  
\end{tabular}

$^1$ average. $^2$ in km\,s$^{-1}$.\\ 
\end{table}

\begin{table} \scriptsize 
 \centering 
\caption{FWHM measured in region R1b  $ \Delta$X/$\Delta$Y: 12"-15"/14"-17".} \label{tab:fwhmR1b}
 \begin{tabular}{lccccccccccccc} \hline  \hline 
\                 & FWHM$^{1,2}$ & max FWHM$^{1,2}$ & max FWHM$^{2}$ \\  \hline 

\  [O\,{\sc iii}]$\lambda$5007       &       238.07 &      298.36 &      654.85  \\ 
\ HeII$\lambda$5412          &       160.34 &      160.34 &      170.82  \\ 
\  [Ar\,{\sc x}]$\lambda$5533        &       208.82 &      208.82 &      425.95  \\ 
\ [Fe\,{\sc vii}]$\lambda$6087       &       208.44 &      237.79 &      429.15  \\ 
\    [O\,{\sc i}]$\lambda$6300       &       161.98 &      165.97 &      380.53  \\ 
\  [S\,{\sc iii}]$\lambda$6312       &       143.07 &      143.07 &      170.48  \\ 
\      [O\,{\sc i}]$\lambda$6364       &          -   &         -   &          -   \\ 
\   [Fe\,{\sc x}]$\lambda$6375       &       235.34 &      235.34 &      530.82  \\ 
\   \Ha $\lambda$6563          &       206.53 &      275.76 &      502.65  \\ 
\   [N\,{\sc ii}]$\lambda$6583       &       184.74 &      207.37 &      434.19  \\ 
\     HeI$\lambda$6678       &       164.22 &      164.22 &      212.97  \\ 
\   [S\,{\sc ii}]$\lambda$6716       &       203.46 &      273.07 &      491.36  \\ 
\   [Ar\,{\sc v}]$\lambda$7006       &       175.96 &      185.29 &      383.71  \\ 
\   [Ar\,{\sc iii}]$\lambda$7136     &       177.54 &      200.93 &      462.73  \\ 
\   [O\,{\sc ii}]$\lambda$7320       &       104.33 &      104.33 &      116.20  \\ 
\  [S\,{\sc iii}]$\lambda$9069       &       216.16 &      216.16 &      298.11  \\ 
\hline  \end{tabular} 
 \\ 
$^1$ average. $^2$ in km\,s$^{-1}$.\\ 
\end{table}

\begin{table} \scriptsize 
 \centering 
\caption{FWHM measured in region R2a  $ \Delta$X/$\Delta$Y: 18"-20"/14"-16".} \label{tab:fwhmR2a}
 \begin{tabular}{lccccccccccccc} \hline  \hline 
\                 & FWHM$^{1,2}$ & max FWHM$^{1,2}$ & max FWHM$^{2}$ \\  \hline 

\  [O\,{\sc iii}]$\lambda$5007       &       211.04 &      247.87 &      542.23  \\ 
\ HeII$\lambda$5412          &          -   &         -   &          -   \\ 
\  [Ar\,{\sc x}]$\lambda$5533        &       210.92 &      210.92 &      317.83  \\ 
\ [Fe\,{\sc vii}]$\lambda$6087       &       164.46 &      164.46 &      232.47  \\ 
\    [O\,{\sc i}]$\lambda$6300       &       124.02 &      124.02 &      124.02  \\ 
\  [S\,{\sc iii}]$\lambda$6312       &          -   &         -   &          -   \\ 
\      [O\,{\sc i}]$\lambda$6364       &          -   &         -   &          -   \\ 
\   [Fe\,{\sc x}]$\lambda$6375       &       199.80 &      199.80 &      272.80  \\ 
\   \Ha $\lambda$6563          &       197.09 &      257.99 &      502.65  \\ 
\   [N\,{\sc ii}]$\lambda$6583       &       193.77 &      203.19 &      299.39  \\ 
\     HeI$\lambda$6678       &          -   &         -   &          -   \\ 
\   [S\,{\sc ii}]$\lambda$6716       &       179.84 &      199.06 &      309.16  \\ 
\   [Ar\,{\sc v}]$\lambda$7006       &       159.28 &      159.28 &      221.86  \\ 
\   [Ar\,{\sc iii}]$\lambda$7136     &       172.76 &      175.44 &      330.17  \\ 
\   [O\,{\sc ii}]$\lambda$7320       &          -   &         -   &          -   \\ 
\  [S\,{\sc iii}]$\lambda$9069       &       118.10 &      118.10 &      161.98  \\ 
\hline  \end{tabular} 
 \\ 
$^1$ average. $^2$ in km\,s$^{-1}$.\\ 
\end{table}

\begin{table} \scriptsize 
 \centering 
\caption{FWHM measured in region R2b  $ \Delta$X/$\Delta$Y: 17"-19"/17"-19".} \label{tab:fwhmR2b}
 \begin{tabular}{lccccccccccccc} \hline  \hline 
\                 & FWHM$^{1,2}$ & max FWHM$^{1,2}$ & max FWHM$^{2}$ \\  \hline 

\  [O\,{\sc iii}]$\lambda$5007       &       231.31 &      276.70 &      414.29  \\ 
\ HeII$\lambda$5412          &          -   &         -   &          -   \\ 
\  [Ar\,{\sc x}]$\lambda$5533        &       419.93 &      419.93 &      419.93  \\ 
\ [Fe\,{\sc vii}]$\lambda$6087       &          -   &         -   &          -   \\ 
\    [O\,{\sc i}]$\lambda$6300       &       130.53 &      130.53 &      169.56  \\ 
\  [S\,{\sc iii}]$\lambda$6312       &          -   &         -   &          -   \\ 
\      [O\,{\sc i}]$\lambda$6364       &          -   &         -   &          -   \\ 
\   [Fe\,{\sc x}]$\lambda$6375       &       317.62 &      317.62 &      431.50  \\ 
\   \Ha $\lambda$6563          &       212.09 &      302.18 &      502.65  \\ 
\   [N\,{\sc ii}]$\lambda$6583       &       145.83 &      145.83 &      172.41  \\ 
\     HeI$\lambda$6678       &          -   &         -   &          -   \\ 
\   [S\,{\sc ii}]$\lambda$6716       &       139.08 &      145.08 &      222.67  \\ 
\   [Ar\,{\sc v}]$\lambda$7006       &          -   &         -   &          -   \\ 
\   [Ar\,{\sc iii}]$\lambda$7136     &       117.63 &      117.63 &      147.07  \\ 
\   [O\,{\sc ii}]$\lambda$7320       &       103.63 &      103.63 &      103.63  \\ 
\  [S\,{\sc iii}]$\lambda$9069       &          -   &         -   &          -   \\ 
\hline  \end{tabular} 
 \\ 
$^1$ average. $^2$ in km\,s$^{-1}$.\\ 
\end{table}

\begin{table} \scriptsize 
 \scriptsize
 \centering
\caption{FWHM measured in region R3 $ \Delta$X/$\Delta$Y: 20"-25"/16"-19".} \label{tab:fwhmR3}
 \begin{tabular}{lccccccccccccc} \hline  \hline 
\                 & FWHM$^{1,2}$ & max FWHM$^{1,2}$ & max FWHM$^{2}$ \\  \hline 

\  [O\,{\sc iii}]$\lambda$5007       &       242.93 &      311.40 &      654.85  \\ 
\ HeII$\lambda$5412          &          -   &         -   &          -   \\ 
\  [Ar\,{\sc x}]$\lambda$5533        &       165.09 &      165.09 &      239.91  \\ 
\ [Fe\,{\sc vii}]$\lambda$6087       &       217.67 &      217.67 &      541.28  \\ 
\    [O\,{\sc i}]$\lambda$6300       &       137.28 &      137.28 &      254.49  \\ 
\  [S\,{\sc iii}]$\lambda$6312       &          -   &         -   &          -   \\ 
\      [O\,{\sc i}]$\lambda$6364       &          -   &         -   &          -   \\ 
\   [Fe\,{\sc x}]$\lambda$6375       &       167.85 &      167.85 &      247.35  \\ 
\   \Ha $\lambda$6563          &       223.04 &      315.90 &      502.65  \\ 
\   [N\,{\sc ii}]$\lambda$6583       &       159.52 &      161.39 &      287.12  \\ 
\     HeI$\lambda$6678       &          -   &         -   &          -   \\ 
\   [S\,{\sc ii}]$\lambda$6716       &       153.09 &      164.37 &      431.63  \\ 
\   [Ar\,{\sc v}]$\lambda$7006       &       137.14 &      137.14 &      203.06  \\ 
\   [Ar\,{\sc iii}]$\lambda$7136     &       150.73 &      155.35 &      305.29  \\ 
\   [O\,{\sc ii}]$\lambda$7320       &       109.65 &      109.65 &      164.40  \\ 
\  [S\,{\sc iii}]$\lambda$9069       &       175.48 &      175.48 &      256.08  \\ 
\hline  \end{tabular} 
 \\ 
$^1$ average. $^2$ in km\,s$^{-1}$.\\ 
\end{table}

\begin{table} \scriptsize 
 \centering 
\caption{FWHM measured in region R4  $ \Delta$X/$\Delta$Y: 4"-7"/8"-11".} \label{tab:fwhmR4}
 \begin{tabular}{lccccccccccccc} \hline  \hline 
\                 & FWHM$^{1,2}$ & max FWHM$^{1,2}$ & max FWHM$^{2}$ \\  \hline 

\  [O\,{\sc iii}]$\lambda$5007       &       194.89 &      217.95 &      317.95  \\ 
\ HeII$\lambda$5412          &       213.49 &      213.49 &      325.61  \\ 
\  [Ar\,{\sc x}]$\lambda$5533        &       236.47 &      236.47 &      344.75  \\ 
\ [Fe\,{\sc vii}]$\lambda$6087       &       224.46 &      237.48 &      541.28  \\ 
\    [O\,{\sc i}]$\lambda$6300       &       170.21 &      214.54 &      355.62  \\ 
\  [S\,{\sc iii}]$\lambda$6312       &       198.70 &      198.70 &      426.15  \\ 
\      [O\,{\sc i}]$\lambda$6364       &          -   &         -   &          -   \\ 
\   [Fe\,{\sc x}]$\lambda$6375       &       204.29 &      204.29 &      437.36  \\ 
\   \Ha $\lambda$6563          &       169.76 &      221.52 &      306.84  \\ 
\   [N\,{\sc ii}]$\lambda$6583       &       167.60 &      217.07 &      307.90  \\ 
\     HeI$\lambda$6678       &       160.63 &      186.11 &      470.34  \\ 
\   [S\,{\sc ii}]$\lambda$6716       &       143.90 &      173.15 &      241.78  \\ 
\   [Ar\,{\sc v}]$\lambda$7006       &       185.58 &      188.31 &      314.73  \\ 
\   [Ar\,{\sc iii}]$\lambda$7136     &       146.03 &      184.46 &      282.23  \\ 
\   [O\,{\sc ii}]$\lambda$7320       &       270.55 &      336.79 &      647.57  \\ 
\  [S\,{\sc iii}]$\lambda$9069       &       125.74 &      140.00 &      184.42  \\ 
\hline  \end{tabular} 
 \\ 
$^1$ average. $^2$ in km\,s$^{-1}$.\\ 
\end{table}

\begin{table} \scriptsize 
 \centering 
\caption{FWHM measured in region R5a  $ \Delta$X/$\Delta$Y: 11"-14"/5"-7".} \label{tab:fwhmR5a}
 \begin{tabular}{lccccccccccccc} \hline  \hline 
\                 & FWHM$^{1,2}$ & max FWHM$^{1,2}$ & max FWHM$^{2}$ \\  \hline 

\  [O\,{\sc iii}]$\lambda$5007       &       246.70 &      322.52 &      458.39  \\ 
\ HeII$\lambda$5412          &       194.59 &      194.59 &      285.53  \\ 
\  [Ar\,{\sc x}]$\lambda$5533        &       157.38 &      157.38 &      174.62  \\ 
\ [Fe\,{\sc vii}]$\lambda$6087       &       196.67 &      201.44 &      295.66  \\ 
\    [O\,{\sc i}]$\lambda$6300       &       197.52 &      262.98 &      486.07  \\ 
\  [S\,{\sc iii}]$\lambda$6312       &       139.75 &      144.02 &      209.41  \\ 
\      [O\,{\sc i}]$\lambda$6364       &          -   &         -   &          -   \\ 
\   [Fe\,{\sc x}]$\lambda$6375       &       210.16 &      210.16 &      381.42  \\ 
\   \Ha$\lambda$6563          &       226.96 &      317.83 &      489.60  \\ 
\   [N\,{\sc ii}]$\lambda$6583       &       213.06 &      293.41 &      482.82  \\ 
\     HeI$\lambda$6678       &       187.75 &      187.75 &      352.43  \\ 
\   [S\,{\sc ii}]$\lambda$6716       &       195.42 &      256.15 &      368.11  \\ 
\   [Ar\,{\sc v}]$\lambda$7006       &       201.52 &      201.52 &      401.66  \\ 
\   [Ar\,{\sc iii}]$\lambda$7136     &       151.77 &      168.39 &      458.91  \\ 
\   [O\,{\sc ii}]$\lambda$7320       &       222.81 &      247.85 &      647.57  \\ 
\  [S\,{\sc iii}]$\lambda$9069       &       134.21 &      140.29 &      207.92  \\ 
\hline  \end{tabular} 
 \\ 
$^1$ average. $^2$ in km\,s$^{-1}$.\\ 
\end{table}

\begin{table} \scriptsize 
 \centering 
\caption{FWHM measured in region R5b  $ \Delta$X/$\Delta$Y: 15"-19"/5"-8".} \label{tab:fwhmR5b}
 \begin{tabular}{lccccccccccccc} \hline  \hline 
\                 & FWHM$^{1,2}$ & max FWHM$^{1,2}$ & max FWHM$^{2}$ \\  \hline 

\  [O\,{\sc iii}]$\lambda$5007       &       241.61 &      310.98 &      654.85  \\ 
\ HeII$\lambda$5412          &       185.20 &      185.20 &      488.50  \\ 
\  [Ar\,{\sc x}]$\lambda$5533        &       292.35 &      292.35 &      447.36  \\ 
\ [Fe\,{\sc vii}]$\lambda$6087       &       177.53 &      179.65 &      325.64  \\ 
\    [O\,{\sc i}]$\lambda$6300       &       204.50 &      268.16 &      522.33  \\ 
\  [S\,{\sc iii}]$\lambda$6312       &       165.16 &      173.30 &      522.33  \\ 
\     [O\,{\sc i}]$\lambda$6364       &          -   &         -   &          -   \\ 
\   [Fe\,{\sc x}]$\lambda$6375       &       256.48 &      256.48 &      530.82  \\ 
\   \Ha $\lambda$6563          &       238.23 &      332.96 &      502.65  \\ 
\   [N\,{\sc ii}]$\lambda$6583       &       224.92 &      311.84 &      502.65  \\ 
\     HeI$\lambda$6678       &       207.98 &      207.98 &      426.86  \\ 
\   [S\,{\sc ii}]$\lambda$6716       &       193.17 &      255.07 &      491.36  \\ 
\   [Ar\,{\sc v}]$\lambda$7006       &       155.43 &      159.66 &      387.87  \\ 
\   [Ar\,{\sc iii}]$\lambda$7136     &       172.32 &      203.44 &      462.73  \\ 
\   [O\,{\sc ii}]$\lambda$7320       &       164.27 &      179.51 &      647.57  \\ 
\  [S\,{\sc iii}]$\lambda$9069       &       161.72 &      185.80 &      363.64  \\ 
\hline  \end{tabular} 
 \\ 
$^1$ average. $^2$ in km\,s$^{-1}$.\\ 
\end{table}

\begin{table} \scriptsize 
 \centering 
\caption{FWHM measured in region R6a  $ \Delta$X/$\Delta$Y: 0"-1"/0"-1".} \label{tab:fwhmR6a}
 \begin{tabular}{lccccccccccccc} \hline  \hline 
\                 & FWHM$^{1,2}$ & max FWHM$^{1,2}$ & max FWHM$^{2}$ \\  \hline 

\  [O\,{\sc iii}]$\lambda$5007       &       266.54 &      362.76 &      393.56  \\ 
\ HeII$\lambda$5412          &       165.43 &      165.43 &      203.56  \\ 
\  [Ar\,{\sc x}]$\lambda$5533        &       187.49 &      187.49 &      194.74  \\ 
\ [Fe\,{\sc vii}]$\lambda$6087       &       261.17 &      391.28 &      416.86  \\ 
\    [O\,{\sc i}]$\lambda$6300       &       195.78 &      267.54 &      300.10  \\ 
\  [S\,{\sc iii}]$\lambda$6312       &       124.02 &      124.02 &      124.02  \\ 
\      [O\,{\sc i}]$\lambda$6364       &       162.66 &      201.91 &      222.50  \\ 
\   [Fe\,{\sc x}]$\lambda$6375       &       196.16 &      196.16 &      201.39  \\ 
\   \Ha $\lambda$6563          &       236.59 &      350.79 &      367.16  \\ 
\   [N\,{\sc ii}]$\lambda$6583       &       210.90 &      301.26 &      312.09  \\ 
\     HeI$\lambda$6678       &       282.17 &      448.87 &      494.12  \\ 
\   [S\,{\sc ii}]$\lambda$6716       &       207.79 &      295.13 &      307.59  \\ 
\   [Ar\,{\sc v}]$\lambda$7006       &       172.40 &      202.81 &      232.23  \\ 
\   [Ar\,{\sc iii}]$\lambda$7136     &       219.19 &      326.07 &      361.89  \\ 
\   [O\,{\sc ii}]$\lambda$7320       &       146.09 &      153.17 &      184.16  \\ 
\  [S\,{\sc iii}]$\lambda$9069       &       164.44 &      238.50 &      243.81  \\ 
\hline  \end{tabular} 
 \\ 
$^1$ average. $^2$ in km\,s$^{-1}$.\\ 
\end{table}

\begin{table} \scriptsize 
 \centering 
\caption{FWHM measured in region R6b  $ \Delta$X/$\Delta$Y: 3"-4"/4"-5".} \label{tab:fwhmR6b}
 \begin{tabular}{lccccccccccccc} \hline  \hline 
\                 & FWHM$^{1,2}$ & max FWHM$^{1,2}$ & max FWHM$^{2}$ \\  \hline 

\  [O\,{\sc iii}]$\lambda$5007       &       264.12 &      357.91 &      383.07  \\ 
\ HeII$\lambda$5412          &          -   &         -   &          -   \\ 
\  [Ar\,{\sc x}]$\lambda$5533        &       579.34 &      579.34 &      579.34  \\ 
\ [Fe\,{\sc vii}]$\lambda$6087       &       200.24 &      200.24 &      206.85  \\ 
\    [O\,{\sc i}]$\lambda$6300       &       147.08 &      170.14 &      247.64  \\ 
\  [S\,{\sc iii}]$\lambda$6312       &          -   &         -   &          -   \\ 
\      [O\,{\sc i}]$\lambda$6364       &       122.44 &      122.44 &      122.44  \\ 
\   [Fe\,{\sc x}]$\lambda$6375       &       243.09 &      243.09 &      459.43  \\ 
\   \Ha $\lambda$6563          &       193.78 &      269.57 &      307.16  \\ 
\   [N\,{\sc ii}]$\lambda$6583       &       187.13 &      250.60 &      348.42  \\ 
\     HeI$\lambda$6678       &          -   &         -   &          -   \\ 
\   [S\,{\sc ii}]$\lambda$6716       &       195.90 &      273.22 &      335.07  \\ 
\   [Ar\,{\sc v}]$\lambda$7006       &          -   &         -   &          -   \\ 
\   [Ar\,{\sc iii}]$\lambda$7136     &       196.58 &      285.60 &      462.73  \\ 
\   [O\,{\sc ii}]$\lambda$7320       &       245.82 &      348.97 &      459.17  \\ 
\  [S\,{\sc iii}]$\lambda$9069       &       208.45 &      208.45 &      226.24  \\ 
\hline  \end{tabular} 
 \\ 
$^1$ average. $^2$ in km\,s$^{-1}$.\\ 
\end{table}

\begin{table} \scriptsize 
 \centering 
\caption{FWHM  measured in region R7  $ \Delta$X/$\Delta$Y: 0"-5"/15"-20".} \label{tab:fwhmR7}
 \begin{tabular}{lccccccccccccc} \hline  \hline 
\                 & FWHM$^{1,2}$ & max FWHM$^{1,2}$ & max FWHM$^{2}$ \\  \hline 

\  [O\,{\sc iii}]$\lambda$5007       &       249.75 &      327.65 &      654.85  \\ 
\ HeII$\lambda$5412          &          -   &         -   &          -   \\ 
\  [Ar\,{\sc x}]$\lambda$5533        &       267.58 &      267.58 &      505.81  \\ 
\ [Fe\,{\sc vii}]$\lambda$6087       &          -   &         -   &          -   \\ 
\    [O\,{\sc i}]$\lambda$6300       &       131.75 &      132.91 &      219.62  \\ 
\  [S\,{\sc iii}]$\lambda$6312       &          -   &         -   &          -   \\ 
\      [O\,{\sc i}]$\lambda$6364       &          -   &         -   &          -   \\ 
\   [Fe\,{\sc x}]$\lambda$6375       &       269.31 &      269.31 &      530.82  \\ 
\   \Ha$\lambda$6563          &       204.41 &      280.92 &      502.65  \\ 
\   [N\,{\sc ii}]$\lambda$6583       &       155.42 &      180.21 &      375.20  \\ 
\     HeI$\lambda$6678       &       136.23 &      136.23 &      177.78  \\ 
\   [S\,{\sc ii}]$\lambda$6716       &       152.87 &      184.34 &      491.36  \\ 
\   [Ar\,{\sc v}]$\lambda$7006       &          -   &         -   &          -   \\ 
\   [Ar\,{\sc iii}]$\lambda$7136     &       133.55 &      136.27 &      462.73  \\ 
\   [O\,{\sc ii}]$\lambda$7320       &       114.38 &      114.38 &      175.10  \\ 
\  [S\,{\sc iii}]$\lambda$9069       &       104.90 &      104.90 &      176.47  \\ 
\hline  \end{tabular} 
 \\ 
$^1$ average. $^2$ in km\,s$^{-1}$.\\ 
\end{table}

\begin{table} \scriptsize
 \centering 
\caption{FWHM measured in region R8  $ \Delta$X/$\Delta$Y: 15"-25"/0"-4".} \label{tab:fwhmR8}
 \begin{tabular}{lccccccccccccc} \hline  \hline 
\                 & FWHM$^{1,2}$ & max FWHM$^{1,2}$ & max FWHM$^{2}$ \\  \hline 

\  [O\,{\sc iii}]$\lambda$5007       &       254.62 &      323.29 &      654.85  \\ 
\ HeII$\lambda$5412          &          -   &         -   &          -   \\ 
\  [Ar\,{\sc x}]$\lambda$5533        &       185.23 &      185.23 &      371.87  \\ 
\ [Fe\,{\sc vii}]$\lambda$6087       &       143.02 &      143.02 &      143.02  \\ 
\    [O\,{\sc i}]$\lambda$6300       &       168.02 &      177.55 &      380.05  \\ 
\  [S\,{\sc iii}]$\lambda$6312       &          -   &         -   &          -   \\ 
\      [O\,{\sc i}]$\lambda$6364       &          -   &         -   &          -   \\ 
\   [Fe\,{\sc x}]$\lambda$6375       &       218.10 &      218.10 &      530.82  \\ 
\   \Ha $\lambda$6563          &       223.30 &      294.64 &      502.65  \\ 
\   [N\,{\sc ii}]$\lambda$6583       &       173.35 &      189.60 &      294.05  \\ 
\     HeI$\lambda$6678       &          -   &         -   &          -   \\ 
\   [S\,{\sc ii}]$\lambda$6716       &       178.47 &      211.43 &      491.36  \\ 
\   [Ar\,{\sc v}]$\lambda$7006       &       211.18 &      211.18 &      295.03  \\ 
\   [Ar\,{\sc iii}]$\lambda$7136     &       161.44 &      163.10 &      323.07  \\ 
\   [O\,{\sc ii}]$\lambda$7320       &       104.71 &      104.71 &      122.15  \\ 
\  [S\,{\sc iii}]$\lambda$9069       &        91.48 &       91.48 &      126.39  \\ 
\hline  \end{tabular} 
 \\ 
$^1$ average. $^2$ in km\,s$^{-1}$.\\ 

\end{table}

\end{document}